\documentclass[utf8]{FrontiersinHarvard} 
\usepackage{url,hyperref,lineno}
\usepackage{txfonts,natbib,graphicx,amsmath,multirow,xspace}
\usepackage{appendix}
\usepackage{xcolor}
\usepackage{comment}
\usepackage{microtype}
\usepackage[onehalfspacing]{setspace}

\definecolor{gold}{RGB}{218,165,32}
\definecolor{pink}{RGB}{212,76,133}
\definecolor{denim}{RGB}{60,82,145}
\definecolor{purple}{RGB}{128,0,128}

\newcommand{\msun}{$M_{\odot}$\xspace}

\newcommand{\fluxcgs}{ergs~cm$^{-2}$~s$^{-1}$\xspace}
\newcommand{\lumcgs}{ergs~s$^{-1}$\xspace}
\newcommand{\rin}{\ensuremath{R_{\rm in}}\xspace}
\newcommand{\rg}{\ensuremath{R_{g}}\xspace}
\newcommand{\risco}{\ensuremath{R_{\mathrm{ISCO}}\xspace}}

\newcommand{\nustar}{\textit{NuSTAR}\xspace}

\newcommand{\swift}{\textit{Swift}\xspace}

\newcommand{\nicer}{\textit{NICER}\xspace}

\newcommand{\rxte}{\textit{RXTE}\xspace}

\newcommand{\xmm}{\textit{XMM-Newton}\xspace}
\newcommand{\hexp}{\textit{HEX-P}\xspace}

\newcommand{\fdcoeq}{\ensuremath{
   \mathrm{FDCUT}(E) = A\ E^{-\Gamma}\
   \frac{1}{1 + \mathrm{e}^{(E-\ecut)/\efold}}
}}

\newcommand{\npexeq}{\ensuremath{
   \mathrm{NPEX}(E) = (AE^{-\Gamma_{1}} + BE^{+\Gamma_{2}})
   \mathrm{e}^{-E/\efold}
}}

\newcommand{\efold}{\ensuremath{E_{\rm{fold}}}}
\newcommand{\ecut}{\ensuremath{E_{\rm{cut}}}}

\def\keyFont{\fontsize{8}{11}\helveticabold }
\def\firstAuthorLast{Ludlam {et~al.}} 

\def\Authors{R.~M.~Ludlam$^{1,*}$, C.~Malacaria$^{2}$, E.~Sokolova-Lapa$^{3}$, F.~Fuerst$^{4}$, P.~Pradhan$^{5}$, A.~W.~Shaw$^{6}$, K.~Pottschmidt$^{7,8}$, S.~Pike$^{9}$, G.~Vasilopoulos$^{10}$, J.~Wilms$^{3}$, J.~A.~Garc\'ia$^{11,12}$, K.~Madsen$^{11}$, D. Stern$^{13}$, C.~Maitra$^{14}$, M.~Del Santo$^{15}$, D.~J. Walton$^{16}$, M.~C.~Brumback$^{17}$, and J.~van den Eijnden$^{18}$}

\def\Address{ \footnotesize
$^{1}$Department of Physics $\&$ Astronomy, Wayne State University, 666 West Hancock Street, Detroit, MI 48201, USA

$^{2}$ International Space Science Institute, Hallerstrasse 6, 3012
Bern, Switzerland

$^{3}$ Dr. Karl Remeis-Observatory and Erlangen Centre for Astroparticle Physics, Friedrich-Alexander Universität Erlangen-Nürnberg, Sternwartstr. 7, 96049 Bamberg, Germany

$^{4}$ Quasar Science Resources S.L for European Space Agency (ESA), European Space Astronomy Centre (ESAC), Camino Bajo del Castillo
s/n, 28692 Villanueva de la Ca\~{n}ada, Madrid, Spain

$^{5}$ Embry Riddle Aeronautical University, 3700 Willow Creek Road, Prescott, AZ 86301, USA

$^{6}$ Department of Physics and Astronomy, Butler University, 4600 Sunset Ave, Indianapolis, IN, 46208, USA

$^{7}$ CRESST and CSST, University of Maryland, Baltimore County, 1000 Hilltop Circle, Baltimore, MD 21250, USA

$^{8}$ Astroparticle Physics Laboratory, NASA Goddard Space Flight Center, Greenbelt, MD 20771, USA

$^{9}$ Department of Astronomy and Astrophysics, University of California, San Diego, 9500 Gilman Dr., La Jolla, CA 92093, USA

$^{10}$ Department of Physics, National and Kapodistrian University of Athens, University Campus Zografos, GR 15783, Athens, Greece

$^{11}$ X-ray Astrophysics Laboratory, NASA Goddard Space Flight Center, Greenbelt, MD 20771, USA

$^{12}$ California Institute of Technology, Pasadena, CA 91125, USA

$^{13}$ Jet Propulsion Laboratory, California Institute of Technology, 4800 Oak Grove Dr., Pasadena, CA 91109, USA

$^{14}$ Max-Planck-Institut f{\"u}r extraterrestrische Physik, Gie{\ss}enbachstra{\ss}e 1, 85748 Garching, Germany

$^{15}$ INAF, Istituto di Astrofisica Spaziale e Fisica Cosmica di Palermo, Via U. La Malfa 153, 90146 Palermo, Italy 

$^{16}$ Centre for Astrophysics Research, University of Hertfordshire, College Lane, Hatfield, AL10 9AB, UK

$^{17}$ Department of Physics, Middlebury College, Middlebury, VT 05753, USA

$^{18}$ Department of Physics, University of Warwick, Coventry CV4 7AL, UK
}

\def\corrAuthor{R.~M.~Ludlam}

\def\corrEmail{renee.ludlam@wayne.edu}

\begin{document}
\onecolumn
\firstpage{1}

\title {The High Energy X-ray Probe (HEX-P): A New Window into Neutron Star Accretion} 

\author[\firstAuthorLast ]{\Authors} 
\address{} 
\correspondance{} 

\extraAuth{}

\maketitle
\begin{abstract}
Accreting neutron stars (NSs) represent a unique laboratory for probing the physics of accretion in the presence of strong magnetic fields ($B\gtrsim 10^8$\,G). Additionally, the matter inside the NS itself exists in an ultra-dense, cold state that cannot be reproduced in Earth-based laboratories. Hence, observational studies of these objects are a way to probe the most extreme physical regimes. Here we present an overview of the field and discuss the most important outstanding problems related to NS accretion. We show how these open questions regarding accreting NSs in both low-mass and high-mass X-ray binary systems can be addressed with the \textit{High-Energy X-ray Probe} (\hexp) via simulated data.
In particular, with the broad X-ray passband and improved sensitivity afforded by a low X-ray background, \hexp will be able to 1) distinguish between competing continuum emission models; 2) provide tighter upper limits on NS radii via reflection modeling techniques that are independent and complementary to other existing methods; 3) constrain magnetic field geometry, plasma parameters, and accretion column emission patterns by characterizing fundamental and harmonic cyclotron lines and exploring their behavior with pulse phase; 4) directly measure the surface magnetic field strength of highly magnetized NSs at the lowest accretion luminosities; as well as 5) detect cyclotron line features in extragalactic sources and probe their dependence on luminosity in the super-Eddington regime in order to distinguish between geometrical evolution and accretion-induced decay of the magnetic field.  In these ways \hexp will provide an essential new tool for exploring the physics of NSs, their magnetic fields, and the physics of extreme accretion.

\tiny
\keyFont{ \section{Keywords:} Neutron stars; Accretion; High-mass X-ray binary stars; Low-mass X-ray binary stars; Magnetic fields; High energy astrophysics; X-ray astronomy; X-ray telescopes} 
\end{abstract}

\section{Introduction}
\subsection{Accreting Neutron Stars}
\begin{figure}
\centering
\includegraphics[width=\textwidth,trim=2 6 0 2,clip]{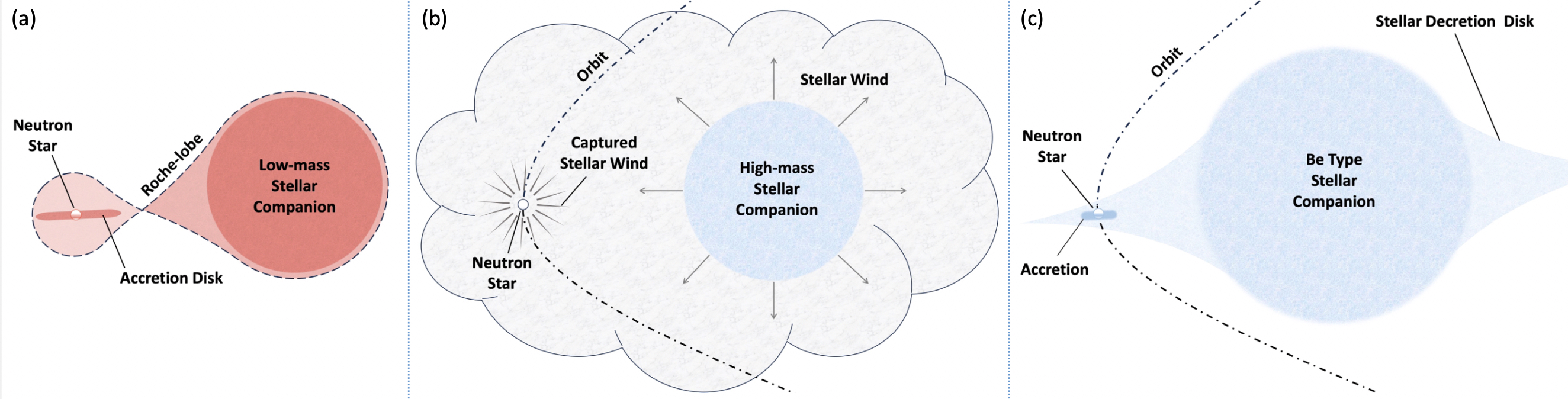}
\caption{Simple schematic diagrams of several types of X-ray binary systems: (a) a LMXB system where the NS accretes from a stellar companion via Roche-lobe overflow, (b) a NS in a HMXB that accretes from material captured from stellar winds that are launched from the companion, and (c) a NS accreting from the decretion disk of a Be type stellar companion. }
\label{fig:XRB}
\end{figure}

Accretion is a ubiquitous process in the Universe, from the formation of stars and planets to supermassive black holes at the center of galaxies. X-ray binary systems are composed of a compact object (CO), either a black hole (BH) or neutron star (NS), that accretes from a stellar companion. These systems can be further categorized based on the mass of the stellar companion: low-mass and high-mass (see \citealt{Longair2011} for a more detailed review). Figure~\ref{fig:XRB} shows a simple schematic for accretion onto NSs in several types of binary systems. Low-mass X-ray binaries (LMXBs) have a $\lesssim1$\,\msun companion that transfers matter via Roche-lobe overflow to form an accretion disk around the compact object. High-mass X-ray binaries (HMXBs) have a more massive ($\gtrsim5$\,\msun), early-type stellar companion and accretion onto the CO can occur in various ways, including periodic Roche lobe overflow (such as when the CO passes through the decretion disk of a Be type star), or capture of material ejected through dense stellar winds. We note that there is a rare class of LMXBs that accrete from the slow winds of a late-type stellar companion and exhibit properties akin to HMXBs (see \citealt{bozzo22} and references therein)\footnote{Although we do not discuss this special class of objects further, some of the science cases we explore (e.g., cyclotron line studies) are applicable to these sources as well.}. 

NSs are the densest known objects with a surface in the Universe. The matter inside a NS exists in an ultra-dense, cold state that cannot be replicated in terrestrial laboratories. The only way to discern how matter behaves under these conditions is by determining the equation of state (EoS). The EoS sets the mass and radius of the NS through the Tolman-Oppenheimer-Volkoff equations \citep{Tolman1934,Tolman1939, Oppenheimer1939} and astronomical measurements of NS masses and radii are therefore crucial for determining which theoretical EoS models are viable (see, e.g., \citealt{Lattimer2011}). In particular, accretion onto NSs probes the properties and behavior of matter in the presence of a strong magnetic field ($B\sim10^{8-9}$\,G for LMXBs and $B\sim10^{12}$\,G for HMXBs; see \citealt{caballero12}). 

In order to fully encapsulate the accretion emission from these systems, a broad X-ray passband is necessary (see \S2 for LMXBs and \S3 for HMXBs). Currently, \nustar \citep{harrison13} is the only focusing hard X-ray telescope with a passband from 3--80\,keV. Energies below 3\,keV have to be supplemented with other X-ray telescopes, such as \nicer \citep{Gendreau12}, \swift \citep{Gehrels04}, or \xmm \citep{xmm}, to construct a broad energy passband. While observations can be scheduled to occur during the same observing period, often the data are not strictly simultaneous due to the different orbits of the missions resulting in various degrees of overlap. Since NS binary systems are highly variable, sometimes on time-scales comparable to or shorter than the typical exposure itself, simultaneous observations over a broad X-ray passband are invaluable for studying these systems.

\subsection{The High Energy X-ray Probe: HEX-P}
The \textit{High Energy X-ray Probe} (\hexp; Madsen et al.\ 2023, \textit{in prep.}) is a probe-class mission concept that offers sensitive broad-band X-ray coverage (0.2--80\,keV) with exceptional spectral, timing and angular capabilities. It features two high-energy telescopes (HETs) that focus hard X-rays and one low-energy telescope (LET) that focuses lower-energy X-rays.

The LET consists of a segmented mirror assembly coated with Ir on monocrystalline silicon that achieves an angular resolution of $3.5''$, and a low-energy DEPFET detector, of the same type as the Wide Field Imager (WFI; \citealt{wfi}) onboard \textit{Athena} \citep{Nandra13}. It has $512 \times 512$ pixels that cover a field of view of $11.3'\times 11.3'$. The LET has an effective passband of 0.2--25\,keV, and a full frame readout time of 2\,ms, which can be operated in a 128 and 64 channel window mode for higher count-rates to mitigate pile-up and faster readout. Pile-up effects remain below an acceptable limit of ${\sim}1\%$ for fluxes up to ${\sim}100$\,mCrab in the smallest window configuration. Excising the core of the PSF, a common practice in X-ray astronomy, will allow for observations of brighter sources, with a typical loss of up to ${\sim}60\%$ of the total photon counts.

The HET consists of two co-aligned telescopes and detector modules. The optics are made of Ni-electroformed full shell mirror substrates, leveraging the heritage of \textit{XMM-Newton}, and coated with Pt/C and W/Si multilayers for an effective passband of 2--80\,keV. The high-energy detectors are of the same type as flown on \nustar, and they consist of 16 CZT sensors per focal plane, tiled $4\times4$, for a total of $128 \times 128$ pixel spanning a field of view of $13.4' \times 13.4'$.

This paper highlights interesting existing open questions about NSs and accretion in strong magnetic fields, and demonstrates \hexp's unique ability to address these with the current best estimate (CBE) mission capabilities. All simulations presented here were produced with a set of response files that represents the observatory performance based on CBEs as of Spring 2023 (see Madsen et al.\ 2023, \textit{in prep.}). The effective area is derived from raytracing calculations for the mirror design including obscuration by all known structures. The detector responses are based on simulations performed by the respective hardware groups, with an optical blocking filter for the LET and a Be window and thermal insulation for the HET. The LET background was derived from a GEANT4 simulation \citep{Eraerds2021} of the WFI instrument, and the HET background was derived from a GEANT4 simulation of the \nustar instrument. Both assume \hexp is in an L1 orbit. The broad X-ray passband and superior sensitivity will provide a unique opportunity to study accretion onto NSs across a wide range of energies, luminosity, and dynamical regimes.

\section{Low-mass X-ray Binaries}
\subsection{Background}
Persistently accreting NS LMXBs are divided into two types based upon characteristic shapes that are traced out in hardness-intensity diagrams and color-color diagrams \citep{hasinger89}: `Z' sources and `atoll' sources. Z sources trace out a Z-shaped pattern with three distinct branches: the horizontal, normal, and flaring branch (HB, NB, and FB, respectively), with the FB being the softest spectral state. They can be further divided into two subgroups, Sco-like and Cyg-like, based upon how much time they spend in the different branches. Sco-like sources spend little to no time in the HB and extended periods of time in the FB ($>12$\,hours), whereas Cyg-like sources have a strong HB and spend little time in the FB \citep{kuulkers97}. Atoll sources are either observed in the soft `banana' state or hard `island' state. A difference in mass accretion rate is thought to be the primary driver between the behavioral differences between atoll and Z sources. Atoll sources probe a lower range in Eddington luminosity (${\sim} 0.01$--$0.5\,L_{\mathrm{Edd}}$) in comparison to the near-Eddington Z sources (${\sim} 0.5$--$1.0\,L_{\mathrm{Edd}}$: \citealt{vanderklis05}). Further evidence that the average mass accretion rate is responsible for the observed difference between the two classes comes from observing transient NS LMXBs that have transitioned between the two classes during outburst (e.g., XTE\,J1701$-$462: \citealt{homan10}).

NS LMXBs occupy a number of spectral states which vary considerably in terms of the models and spectral parameters needed to describe the continuum emission \citep{barret01}. In the very hard state, the spectrum is dominated by Comptonization that can be modeled with an absorbed power-law component with a photon index $\Gamma\sim1$ \citep{ludlam16, parikh17} or two thermal Comptonization components assuming two distinct populations of seed photons from different plasma temperatures \citep{fiocchi19}. In the hard state, the spectrum is dominated by a hard Comptonization component with $\Gamma=1.5$--2.0 and a soft thermal component arising from a single temperature blackbody component or multi-color disk blackbody with a temperature\,$\lesssim1$\,keV \citep{barret00, church01}.  In the soft state, the spectrum becomes thermally dominated with weakly Comptonized emission. Model choices for the thermal and Comptonization components in the soft state vary in the literature leading to two classical descriptions. The ``Eastern'' model, after \cite{mitsuda89}, uses a multi-color disk blackbody in combination with a Comptonized blackbody component, while the ``Western'' model, after \cite{white88},  uses a single-temperature blackbody and a Comptonized disk component. \cite{Lin07} devised a hybrid model for hard and soft state spectra based upon \textit{RXTE} observations of two transient atoll systems that resulted in a coherent picture of the spectral evolution (e.g., the thermal components follow the expected $L_{x}\propto T^{4}$ relation). In this model, the soft state assumes two thermal components (i.e., a single-temperature blackbody and a disk blackbody) and weak Comptonization accounts for the power-law component. This hybrid double thermal model has been used in many NS LMXBs studies (e.g., \citealt{cackett08, cackett09, Lin10}), though not exclusively. For instance, a recent study using thermal Comptonization from a blackbody and a multi-color accretion disk blackbody (akin to the ``Eastern" model) to model the X-ray spectra of the atoll 4U~1820-30 found good agreement with the observed jet variability in this system \citep{marino23}. In the absence of multi-wavelength data to support a choice of continuum model, it is difficult to ascertain which prescription of the spectra is appropriate. Due to the soft spectral shape in these states, the source spectrum typically becomes background dominated above 30\,keV even when observed with \nustar.  A broad X-ray passband with large effective collecting area and low X-ray background that can observe a large number of sources at various luminosity levels and spectral states is needed to further our understanding of these sources.

\begin{figure}
\centering
\includegraphics[width=0.7\textwidth,trim=10 10 0 0,clip]{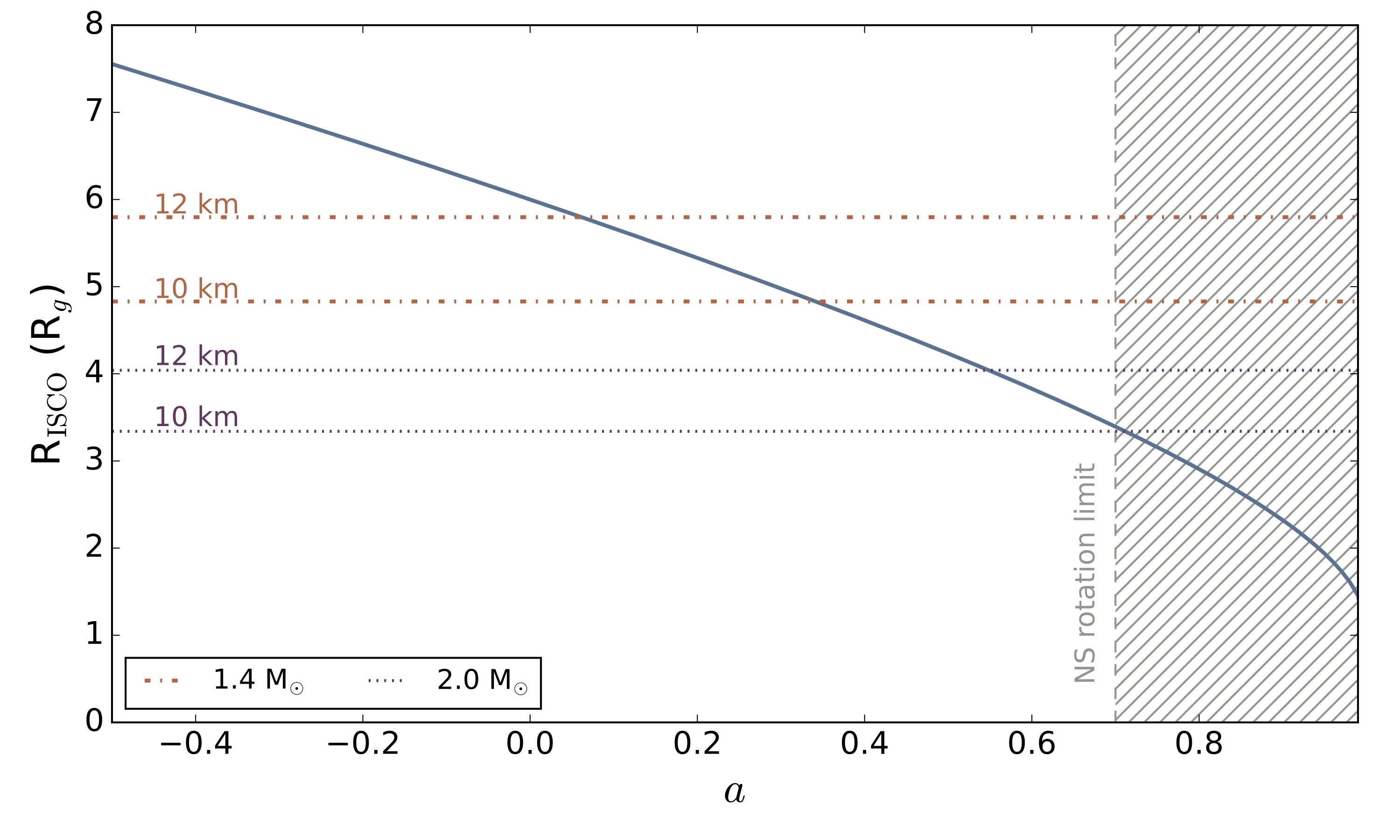}
\caption{Position of the innermost stable circular orbit (ISCO) in units of gravitational radii versus the dimensionless spin parameter $a$ of the CO. The hatched region indicates where the NS surpasses the rotational limit and would break apart. The horizontal dot-dashed lines indicate the corresponding values of 10\,km and 12\,km radius for a canonical NS mass of 1.4\,\msun. The horizontal dotted lines indicate the same radius values for a more massive NS of 2.0\,\msun.}
\label{fig:isco}
\end{figure}

The accretion disks in these systems can be externally illuminated by hot electrons in the corona \citep{Sunyaev1991} or from the thermal emission of the NS or boundary layer region \citep{popham2001}. 
We note that the exact geometry of the corona is unclear (see \citealt{Degenaar18} for some possible geometries), but X-ray polarization measurements with {\it IXPE} are beginning to shed light on coronal orientation and presence of boundary layer regions in some systems (e.g., \citealt{Jayasurya23, Ursini23, Cocchi23, Farinelli23}); building up our understanding of the accretion geometry when coupled with X-ray energy spectral studies.
Regardless of the hard X-ray source, the disk reprocesses the hard X-ray emission and re-emits discrete emission lines superimposed upon a reprocessed continuum known as the `reflection' spectrum \citep{Ross2005,Garcia2010}. The spectrum is then relativistically blurred due to the extreme environment close to the NS, where accreting material reaches relativistic velocities as it falls into the NS's deep gravitational well \citep{Fabian1989,Dauser2013}. Studying reflection in NS LMXBs allows for properties of NSs and the disk itself to be measured, such as the NS magnetic field strength \citep{ibragimov2009, cackett09, ludlam17}, extent of the boundary layer \citep{popham2001, king16, Ludlam21}, and NS radius \citep{cackett08, ludlam17a, Ludlam22}.  Of particular interest is both the inner disk radius, \rin, and dimensionless spin parameter, $a=cJ/GM^2$ (which is the mass normalized total angular momentum $J$ of the CO); the latter is typically fixed in current studies due to limitations in data quality. The spin parameter has important consequences for both accreting NSs and BHs (see Connors et al.\ 2023, \textit{in prep.}, and Piotrowska et al.\ 2023, \textit{in prep.}, for \hexp science with BH X-ray binaries and supermassive BHs, respectively). The spin sets the location of the innermost stable circular orbit (ISCO) where a higher spin corresponds to a smaller ISCO  (\citealt{bardeen72}). Consequently, the position of the inner disk radius for higher spin values cannot be replicated by lower spin if the data is of sufficient quality to recover \rin accurately. The majority of NSs in LMXBs have spin $a \lesssim 0.3$ \citep{galloway2008, miller2011} and the difference in the location of the ISCO decreases from 6 gravitational radii (\rg~$=GM/c^2$) at $a=0$ to 4.98\,\rg at $a=0.3$ (see Fig.~\ref{fig:isco}). Therefore, targeting an accreting NS with a higher spin, as indicated through a high measured spin frequency, can decrease the upper limit on NS radii obtained from reflection modeling by roughly 2--3 kilometers. 

In order to capture the entire reflection spectrum (i.e., the low-energy O~VIII and iron~(Fe)~L emission lines near 1\,keV \citep{madej14, ludlam18, Ludlam21}, the Fe~K emission lines at 6.4--6.97~keV, and the Compton backscattering hump above $15$\,keV), determine the appropriate continuum model, and measure the absorption column along the line of sight, a broad X-ray passband is necessary. Hence, the broad X-ray sensitivity provided by the LET and HETs on \hexp is crucial for studying accreting NS LMXBs.

\subsection{Simulated Science Cases for LMXBs}
\label{subsec:lmxb_sim}

All simulations in this section were conducted in \textsc{xspec} \citep{arnaud96} via the `fakeit' command, which draws photons from a randomized seed distribution,  with V07 of the \hexp response files selecting an 80\% PSF correction, assuming the data would be extracted from a 15 arcsec region for the HET and 8 arcsec region for the LET, as well as the anticipated background for the telescope at L1. The simulated spectra were grouped via \textsc{grppha} to have a minimum of 25 counts per bin to allow for the use of $\chi^{2}$ statistics. The flux for each of the following simulations can be found in Table~S1 in the Supplemental Materials.

\subsubsection{Distinguishing between continuum models}

To demonstrate \hexp's ability to distinguish between different continuum models, we base our simulations on the models used to fit simultaneous \nicer and \nustar observations of the accreting atoll 4U~1735$-$44 published in \cite{ludlam20}. Model 1 is based on the hybrid double thermal continuum model of \cite{Lin07} while Model 2 is based on the ``Eastern'' model of \cite{mitsuda89}. Both models provide an adequate description of the data in the 0.3--30\,keV energy band (the \nustar observations were background dominated above 30\,keV) when neglecting the reflection spectrum, but data at higher energies would distinguish between these two continuum models. We take the continuum model and parameter values for Model 1 and Model 2 from Table 1 of \cite{ludlam20} as input for the \hexp simulation; the exact values can be found in Table S2 of the Supplementary Materials. The overall models\footnote{The definition of each model component and their parameters can be found in the \textsc{xspec} manual: https://heasarc.gsfc.nasa.gov/xanadu/xspec/manual/node128.html} are as follows: (a) Model 1: \textsc{tbabs}*(\textsc{diskbb}+\textsc{bbody}+\textsc{pow}) and (b) Model 2: \textsc{tbabs}*(\textsc{diskbb}+\textsc{nthcomp}).  
Note that we fix the absorption column index to a value between the ones used in Model 1 and Model 2, $N_{\rm H}=4\times10^{21}$\,cm$^{-2}$, to focus on the difference in the spectral shape due to the model component choices rather than a difference in absorption column between the models. We chose an exposure time of 20\,ks, roughly equivalent to the exposure time of the \nustar observation in the aforementioned study.

\begin{figure}
\begin{center}
\includegraphics[width=0.7\textwidth,trim=10 0 0 0,clip]{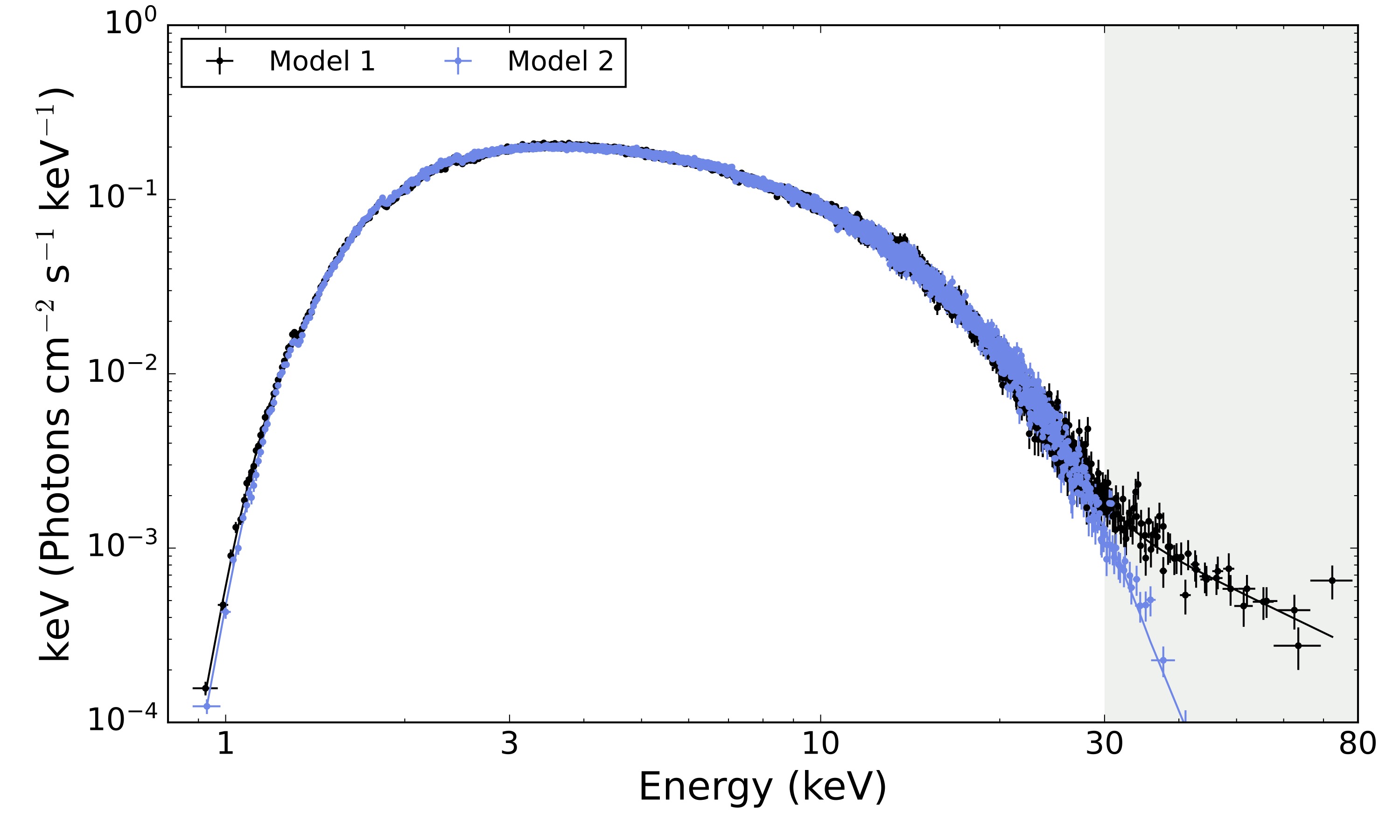}
\caption{Comparison of simulated 20\,ks \hexp (LET+2 HETs) observation for two different continuum models typically used to describe NS LMXB soft state spectra: a double thermal continuum model with weak Comptonization (Model 1: black) and an accretion disk with blackbody Comptonization (Model 2: blue). The models diverge above 30\,keV where currently available missions quickly become background dominated for the same exposure time. To emphasize this point, the same S/N near 30 keV could be achieved with \nustar in an exposure time of 96.6\,ks. The broad X-ray passband and improved sensitivity of \hexp will differentiate the models.}
\label{fig:cont_comp}
\end{center}
\end{figure}

The simulated \hexp spectra remain above the background in the HET bands up to 80\,keV, though they dive into the LET background below 1\,keV. This could be remedied by increasing the exposure time, though, notably, the spectrum above 30\,keV is where the models diverge. The reduced $\chi^{2}$ is $\leq1.03$ when fitting each simulation with its own input model. Attempting to fit each simulated spectrum with the opposing model description provides an inadequate fit  with $\Delta\chi^{2}>1000$ for the same number of degrees of freedom ($d.o.f.$). Figure~\ref{fig:cont_comp} shows the \hexp spectrum for the Model 2 input of a thermal disk with a Comptonized blackbody from a boundary layer region. For direct comparison, we show a simulated spectrum using Model 1 (the double thermal continuum model with weak Comptonization) with parameter values from fitting the simulated spectrum of Model 2 below 30 keV, since these models perform equally well within this energy regime. This highlights that though these models are nearly identical below 30 keV, they diverge at higher energies inaccessible by current operating missions. The weakly Comptonized emission above 30 keV cannot be replicated by Model 2, and demonstrates the potential for a broadband X-ray mission like \hexp to discern between competing continuum models that would otherwise equally describe the data below 30\,keV.

\subsubsection{Neutron star radius constraints from relativistic reflection modeling}

\begin{table}[t!]
\caption{Input model parameters for simulating reflection model data with \hexp. Values are taken from Table 3: RNS1 of \citet{Ludlam22} for \mbox{Cygnus~X-2} in the NB and updated using the public version of {\sc relxillNS} (v2.2). Additionally, we show one case of the recovered model parameters from a simulated \hexp spectrum after performing a Markov Chain Monte Carlo (MCMC) analysis with a burn-in of $10^6$ and chain length of $5\times10^5$ to emulate a standard analysis of the data. Errors are reported at the $1\sigma$ confidence level, though the errors for \rin and $a$ are drawn from the bivariate distribution between the two due to their correlated nature.}
\label{tab:reflsimpar}

\begin{center}
\begin{tabular}{lccc}
\hline
Model Component & Parameter & Input & Recovered for $a=0.3$\\
\hline
{\sc tbfeo} & $N_{\rm H}$ ($10^{22}$ cm$^{-2}$)& 0.39 & 0.39$^{\dagger}$\\
& O & 1.16 & 1.16$^{\dagger}$\\
& Fe & 1.00 & 1.00$^{\dagger}$\\
{\sc diskbb} & $T_{\rm in}$ (keV) & 1.74 & $1.74\pm0.01$\\
& norm & 117 & $117\pm1$\\
{\sc powerlaw} & $\Gamma$ & 3.49 & $3.49\pm0.01$\\
& norm & 6.0 & $5.9\pm0.1$\\
{\sc rexillns} & $q$ & 2.21 & $2.21\pm0.04$\\
& $a$ & 0, 0.17, 0.30 & $0.3\pm0.2$\\
& $i$ ($^{\circ}$) & 65 & 65$^{\dagger}$\\
& \rin (\risco) & 1.00 & $1.02_{-0.02}^{+0.10}$\\
& $R_{\rm out}$ (\rg)& 1000 & 1000$^{\dagger}$\\
& $kT_{\rm bb}$ (keV) & 2.49 &$2.49\pm0.01$\\
& $\log(\xi/\mathrm{[erg\,cm\,s}^{-1}])$ & 2.00 &$ 2.04 _{- 0.10 }^{+ 0.03}$\\
& $A_{\rm Fe}$ & 5.9 & $ 4.9 _{- 0.8 }^{+ 0.7 }$ \\
& $\log(N [\rm cm^{-3}])$ & 18 & $17.9\pm0.1$\\
& $f_{\rm refl}$ & 1.0 & $1.0\pm0.1$\\
& norm (10$^{-3}$) & 3.23 & $3.2\pm0.1$\\
\hline
{$^{\dagger}$~=~fixed} & $\chi^2/d.o.f.$ & & $3132.3/3102$\\
\end{tabular}
\end{center}
\end{table}

To demonstrate that \hexp would improve the radius constraints obtained by reflection modeling of NSs, we base our simulations on the results of a recent investigation of Cygnus~X-2 that utilized simultaneous data from \nicer and \nustar \citep{Ludlam22}. This source is of particular interest since it has a dynamical NS mass measurement of $M_{\rm NS}=1.71\pm0.21$\,\msun \citep{casares10}. The source was analyzed in the NB, HB, and the vertex between those branches. Fixing the spin at $a=0$, \citet{Ludlam22} found that the inner disk radius remained close to \risco. In the current literature it is customary to find studies in which the spin of the NS has been fixed while fitting models to the data 
given the highly degenerate nature of \rin and $a$. We set up our simulations using the Cygnus~X-2 data in the NB modeled with the latest public release of the self-consistent reflection model tailored for thermal emission illuminating the accretion disk, \textsc{relxillNS} (v2.2: \citealt{Garcia22}). The input parameters for the simulation\footnote{Note that there was an update to the publicly available \textsc{relxill} model's definition of the reflection fraction parameter for non-lamppost type models. As a result the values we use do not exactly match those of \cite{Ludlam22}, which used a proprietary developmental version of \textsc{relxillNS}.  We have verified that the main results are consistent within errors when fitting the \nustar spectrum in the NB.} are provided in Table~\ref{tab:reflsimpar}. We leave \rin at 1\,\risco\ for all simulations while varying the input spin since the disk remained consistent with this value while the source was in the various branches. We chose three spin values as test cases: non-rotating ($a=0$), the highest value expected for a NS LMXB ($a=0.3$), and $a=0.17$ which is an approximation based on the spin frequency of the source \citep{mondal18}. We use an exposure time of 100\,ks and investigate how well \rin and spin can be recovered independently.

\begin{figure*}
\includegraphics[width=0.99\textwidth,trim=20 10 10 0,clip]{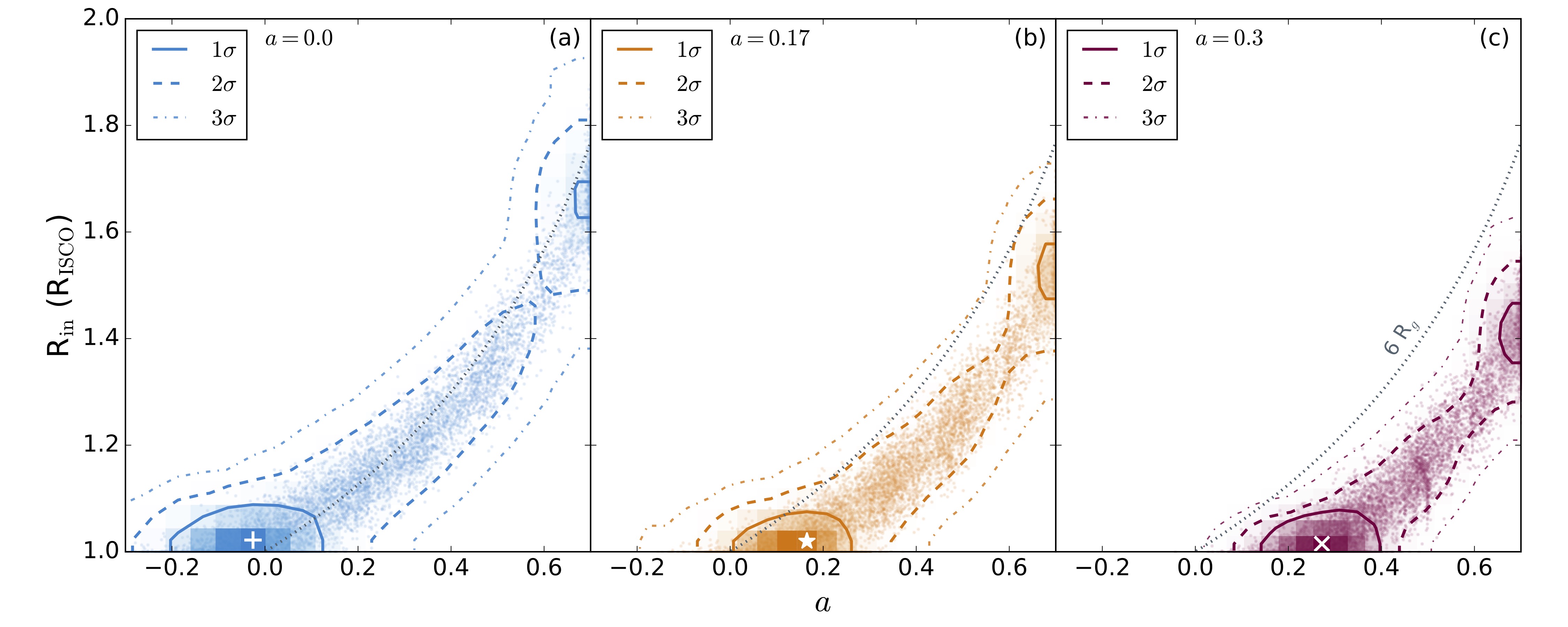}
\caption{Posterior distribution of $a$ and \rin for $10^{4}$ iterations of simulating \hexp spectra for three spin values: (a) $a=0.0$, (b) $a=0.17$, and (c) $a=0.3$. The $1\sigma$, $2\sigma$, and $3\sigma$ contours are shown. The white points indicate the highest probability value. The dotted grey line indicates a constant line of 6\,\rg which corresponds to the ISCO for $a=0$. The distribution tightens at high values of spin as relativistic effects become stronger. However, the distributions show that the data are able to trace out unique values of inner disk radii by trending towards lower values of \rg at higher $a$. }
\label{fig:Rin_a}
\end{figure*}

Given the degenerate nature of the parameters of interest and that each spectrum simulated is created via a randomized photon generation, the results obtained can vary with each iteration. We therefore simulate $10^{4}$ spectra per spin value to determine the likelihood of constraining \rin and $a$. The LET data were modeled in the 0.3--15\,keV energy band while the HET data were modeled in the 2--80\,keV band. We impose an upper limit on the spin ($a\leq0.7$) when fitting the simulated spectra so as to not surpass the rotational break-up limit of a NS \citep{Lattimer04}. We fit each spectrum and then perform error scans to ensure that each iteration reached a minimum goodness of fit of $\chi^2/\mathrm{d.o.f.}\leq1.1$ prior to obtaining values for inner disk radius and spin from each simulated spectrum, thus building up the posterior distribution shown in Figure~\ref{fig:Rin_a}.

\begin{figure*}[t!]
\centering
\includegraphics[width=0.98\textwidth,trim=1 0 0 0,clip]{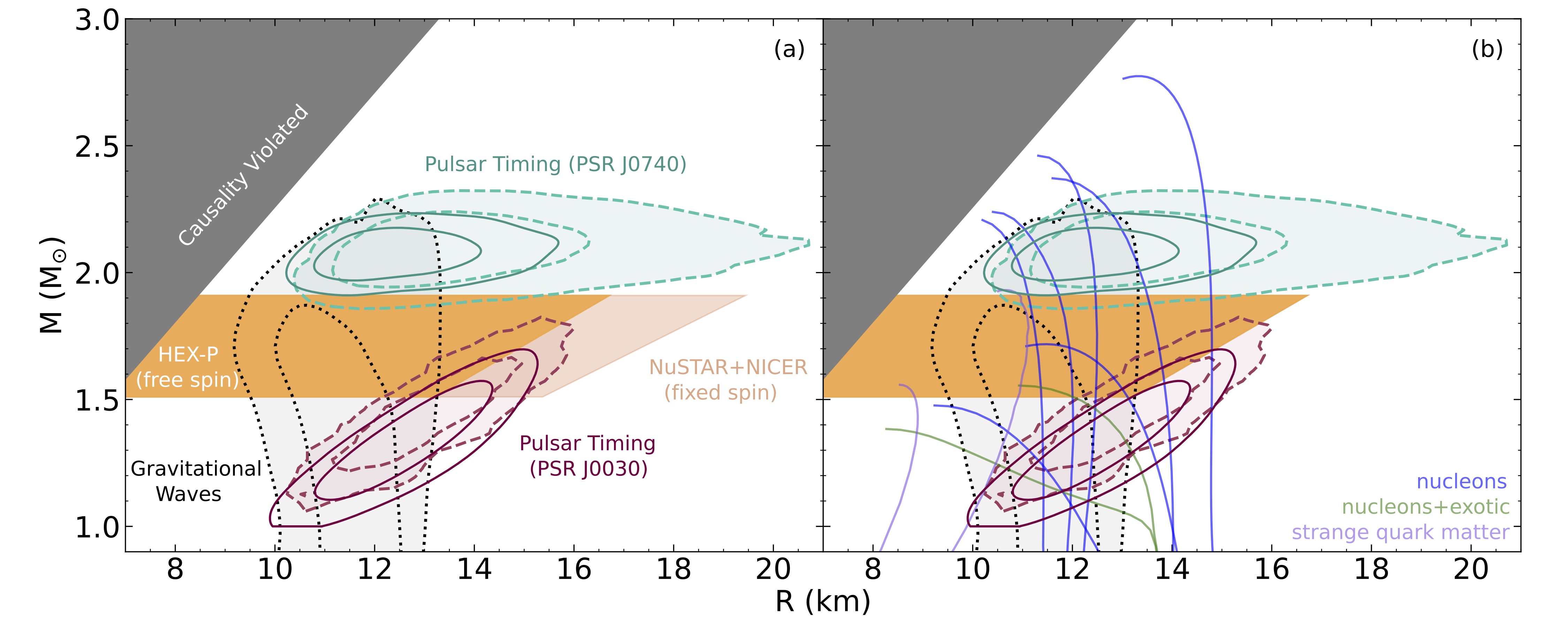}
\caption{Mass and radius constraints from reflection modeling compared to NS gravitational wave events and \nicer pulsar light curve modeling. The darker solid orange region indicates the improved radius constraints for reflection modeling of \hexp data based on Cygnus X-2 for the case of high $a$. Note that both \rin and $a$ are free parameters when fitting the \hexp data, while the \nustar plus \nicer analysis fixes $a=0$ (lighter solid orange region in panel a; \citealt{Ludlam22}). The solid gray region indicates where causality is violated (i.e., the sound speed within the NS exceeds the speed of light). Pulsar light curve modeling of \nicer data for PSR J0740+6620 is indicated in teal (dashed lines: \citealt{miller21}; solid lines: \citealt{Riley21}). Maroon indicates the results for light curve modeling of PSR~J0030+0452 (dashed lines: \citealt{miller19}; solid lines: \citealt{Riley19}). The black dotted line denotes the mass-radius constraints from the combined GW170817 \citep{abbott19} and GW190425 \citep{ligo20} signatures using a piece-wise polytropic model as reported in \citet{Raaijmakers21}. Confidence contours correspond to the 68\% and 95\% credible regions.  Panel (b) shows select EoS models from \cite{Lattimer01} to demonstrate the behavior of different internal compositions on the mass-radius plane.}
\label{fig:MRcomp}
\end{figure*}

Errors are drawn from the bivariate posterior distribution due to their correlated nature and reported at the $1\sigma$ confidence level.  
We find that for an input of $a=0$, we recover $a=-0.03_{-0.17}^{+0.15}$ and \rin~$=1.02_{-0.02}^{+0.07}$\,\risco. For an input of $a=0.17$, we recover $a=0.17_{-0.16}^{+0.09}$ and \rin~$=1.02_{-0.02}^{+0.05}$\,\risco. Lastly, for an input of $a=0.30$, we recover $a=0.27\pm0.13$ and \rin~$=1.02_{-0.02}^{+0.05}$\,\risco. The ability to recover the input parameters improves at higher spin as the relativistic effects become stronger. Recall that the position of the inner disk radius for higher spin values cannot be replicated by lower spin, thus if the data is of sufficient quality to recover \rin accurately this will become apparent. For example, an input of $\rin=1\ \risco$ for a spin of $a=0$ corresponds to $6\ \rg$. This value can be mimicked by a higher spin and \rin beyond 1\,\risco\ (e.g., $6\ \rg$ is equal to $1.2\ \risco$ for $a=0.3$). However, an input of $\rin=1\ \risco$ for $a=0.3$ corresponds to $4.98\ \rg$ and cannot be replicated by a lower spin so long as the upper limit recovered when fitting the data returns a value of $\rin<1.2\ \risco$. We note that Cygnus~X-2 is a bright LMXB (${\sim}0.7$\,Crab) and would thus need to be observed in a configuration that would reduce pile-up effects in the LET. However, we obtain consistent results at the $1\sigma$ level when performing the same exercise with just the HETs, which do not suffer from pile-up, though the $3\sigma$ level can no longer rule out a non-rotating NS as is the case when the LET is included.
Hence the combined observing power of the HETs and LET provide an improvement over the current capabilities of existing missions.

Figure~\ref{fig:MRcomp} shows the improvement provided by \hexp for reflection modeling of spinning NS LMXBs in comparison to simultaneous \nicer and \nustar results with fixed spin. Additionally, the current best constraints from gravitational wave events of binary NS mergers and pulsar light curve modeling demonstrate how the various methods can work in concert to narrow down the allowable region for the EoS. Each method has its own underlying systematic uncertainties, so the firmest EoS constraints will require multiple measurement approaches. Utilizing the dynamical NS mass estimate of Cygnus X-2 ($M_{\rm NS}=1.71\pm0.21$\,\msun: \citealt{casares10}), \citet{Ludlam22} reported an upper limit on the radius of $R_{\rm NS}=19.5$\,km for $M_{\rm NS}=1.92$\,\msun and $R_{\rm NS}=15.3$\,km for $M_{\rm NS}=1.5$\,\msun. The lower limit on spin and upper limit on \rin return a conservative upper limit on the NS radius.
From the \hexp simulations we calculate an upper limit of $R_{\rm NS}=16.7$\,km for $M_{\rm NS}=1.92$\,\msun and $R_{\rm NS}=13.2$\,km for $M_{\rm NS}=1.5$\,\msun for an input of $a=0.3$. This demonstrates the power of \hexp to study rotating NSs with reflection studies. For comparison, we conducted the same simulation with \nustar response files. For the highest spin value of $a=0.3$, \nustar recovers $a=0.67_{-0.54}^{+0.03}$ and \rin~$=1.02_{-0.02}^{+0.16}$\,\risco. The conservative upper limit on the NS radius from the location of the inner disk radius is larger than 6\,\rg and therefore does not provide an improvement over current reflection studies that assume the spin is fixed at $a=0$. Performing this test with the combination of \nicer and \nustar marginally improves the recovered spin value to $a=0.68_{-0.50}^{+0.02}$, but with a larger upper limit on \rin~$=1.02_{-0.02}^{+0.18}$\,\risco, which is still $>1$\,km larger than the constraints obtained from \hexp  (i.e., $R_{\rm NS}=18.5$\,km for $M_{\rm NS}=1.92$\,\msun and $R_{\rm NS}=14.4$\,km for $M_{\rm NS}=1.5$\,\msun). 

We demonstrate that these results do not depend on drawing from a posterior distribution of simulated spectra by conducting Markov Chain Monte Carlo (MCMC) analysis on one of the simulated \hexp spectra with an input spin of $a=0.3$. The results of this test are shown in Table~\ref{tab:reflsimpar} with the inclination fixed at the median value from optical and X-ray studies \citep{orosz99, Ludlam22}. The lower limit on spin of $a=0.1$ and upper limit on \rin~$=1.12$\,\risco\, still provides an improvement over current \nicer and \nustar analyses, which fix $a=0$. The MCMC analysis finds an upper limit of $R_{\rm NS}=$~18\,km for $M_{\rm NS}=$~1.92\,\msun and $R_{\rm NS}=$~14.1\,km for $M_{\rm NS}=$~1.5\,\msun.  

For completeness we note that the Kerr metric is used to describe the space-time close to the NS in the relativistic reflection model. As the angular momentum (i.e., spin) increases, the NS can become oblate, thus causing deviations in space-time from the Kerr metric due to an induced quadrupole moment. The exact induced deviation depends upon the EoS of the NS, but this has an effect $<10$\%  \citep{sibgatullin98} in the anticipated range of spin parameters for NS LMXBs ($a\lesssim0.3$).  Given that our $1\sigma$ errors for \rin are larger than this deviation from the Kerr metric, the \hexp spectra would not be sensitive enough to measure this effect. However, our conservative upper limits on NS radius are still at larger radii than the deviation in the \risco\ and hence are not in conflict even if we are insensitive to an induced quadrupole moment. Therefore, reflection studies of accreting NS LMXBs with \hexp would provide improved upper limits on NS radii in comparison to current studies.

\section{High-mass X-ray Binaries}\label{sec:hmxb}
\subsection{Background}\label{subsec:hmxbback}

\begin{figure}[t!]
\begin{center}
\includegraphics[width=0.95\textwidth]{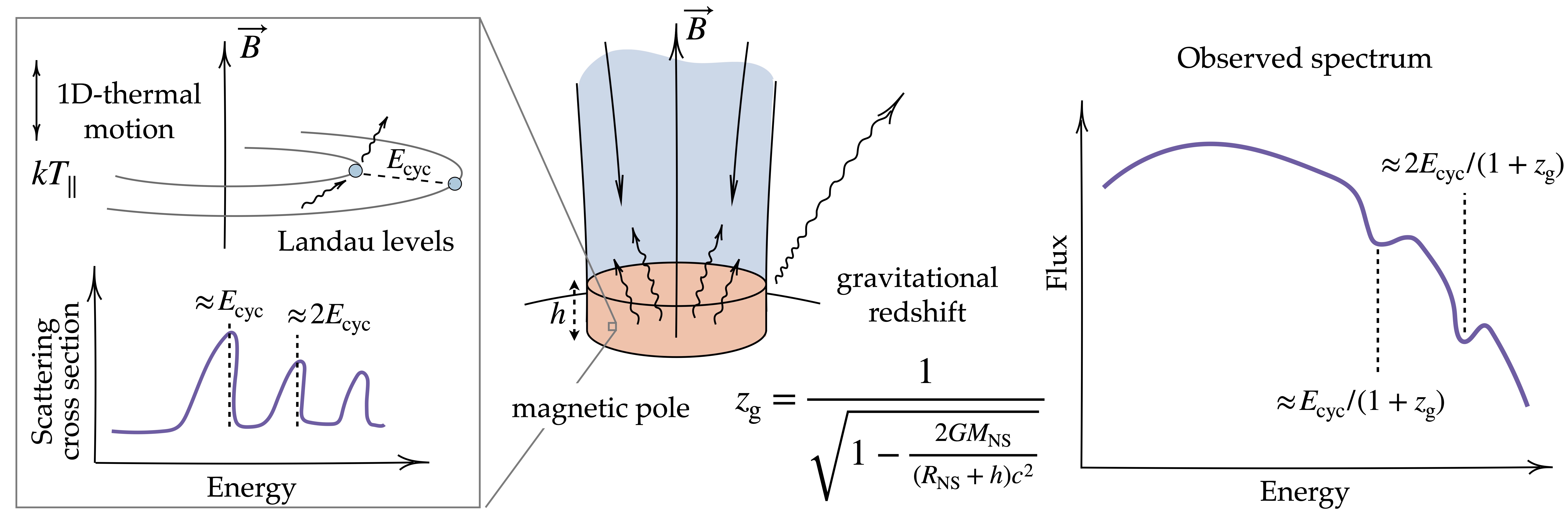}
\caption{Simplified depiction of the formation of cyclotron lines in the spectra of accreting highly magnetized neutron stars due to photon scattering off electrons, whose motion perpendicular to the field lines is restricted by Landau quantization. The latter leads to strong cyclotron resonances in the Compton scattering cross section, broadened by the electron thermal motion parallel to the magnetic field. The centroid energies of the cyclotron lines in observed spectra approximately correspond to the redshited energies of the resonances (with some correction for the scattering redistribution effects during the lines' formation). The gravitational redshift is dependent on the height of the line-forming region, $h$, above the surface of the neutron star.}
\label{fig:cycf}
\end{center}
\end{figure}

NSs in HMXBs are typically highly magnetized, with a dipolar magnetic field strength of ${\sim}10^{12}\,$G. The pressure of the TeraGauss magnetic field disrupts the inflow of the accreting matter, channelling it onto the magnetic poles \citep{DaviesPringle1981}. There, the kinetic energy of the matter is released as radiation in a highly anisotropic manner. Since the NS rotates, the sources are visible as X-ray pulsars. Studying this emission is crucial to understanding the physics of plasma accretion and the interaction of radiation with high magnetic fields that are orders of magnitude stronger than those achievable in laboratories on Earth. Due to their extreme surface magnetic field strength, quantum effects take place at the site of emission. In particular, the electron motion in the direction perpendicular to the magnetic field lines becomes quantized. This affects the electron scattering cross section, leading to a resonance at an energy that is proportional to the magnetic field strength \citep{Meszaros1992}. As a result, the spectra of X-ray pulsars (XRPs) can contain absorption line-like features called cyclotron resonant scattering features (CRSFs), or cyclotron lines, at an energy of 
\begin{equation}
E_{cyc}\approx\frac{n}{(1+z_g)}11.6\,B_{12}\,\rm keV, 
\end{equation}
where $z_g$ is the gravitational redshift (typically 0.3 for standard NS mass and radius) and $B_{12}$ is the NS (polar) magnetic field strength in units of $10^{12}\,$G, while $n$ is the number of Landau levels involved\footnote{In the following, the cyclotron line energy correspondent to $n=1$ is labeled as \textit{fundamental}, while for $n>1$ lines are labeled as the $n_i$ \textit{harmonic}, with $i=n$.}. 
Cyclotron lines represent the most direct way to probe the NS magnetic field on or close to the NS surface. Up to now, cyclotron lines have been confirmed in the spectra of about 40 sources \citep{Staubert2019}. Figure~\ref{fig:cycf} shows a schematic description of the formation of cyclotron harmonics.

When observed at typical outburst luminosity of $\gtrsim10^{36}\,$erg\,s$^{-1}$, the spectra of XRPs have been described with phenomenological models (a power-law with high-energy cutoff) modified by interstellar absorption, an Fe~K$\alpha$ emission line around 6.4\,keV, low-temperature blackbody components and, at times, a component called the 10\,keV feature \citep[][and references therein]{Mushtukov2022}. However, at lower luminosity accretion regimes, the spectrum often exhibits a transition to
a double-hump shape \citep{Tsygankov2019} which has recently been explained in terms of a low-energy thermal component peaking at ${\sim}5\,$keV and a high-energy Comptonized component peaking at ${\sim}35\,$keV representing the broadened red wing of the cyclotron line \citep{Sokolova-Lapa2021, Mushtukov2021}. A dip between the two components was sometimes interpreted as a cyclotron line \citep[see, e.g.,][for the case of X~Per]{Doroshenko2012}, but is likely a continuum feature, as was shown for sources with a known fundamental cyclotron line at higher energies after the transition to a low-luminosity state \citep[see, e.g.,][]{Tsygankov2019}.

These low luminosity states of accretion characterized by two-hump spectra are generally associated with braking of the accretion flow in the NS atmosphere (\citealt{Mushtukov2021}, \citealt{Sokolova-Lapa2021}, although a model assuming an extended collisionless shock instead was recently proposed by \citealt{Becker2022}). Rapid deceleration of the plasma in the atmosphere proceeds mainly by Coulomb collisions and leads to the formation of an overheated outermost layer with electron temperatures of ${\sim}30\,\mathrm{keV}$. Such high temperatures significantly enhance Comptonization and together with resonant redistribution lead to the formation of the high-energy excess in the spectra around the cyclotron line. When the mass-accretion rate onto the poles is increased, the pressure of the emitted radiation starts affecting the dynamics of the accretion flow. The radiation becomes capable of decelerating the flow above the surface, reducing direct heating of the atmosphere and giving rise to a radiation-dominated shock \citep{Basko1976}. The onset of this regime is typically associated with a critical luminosity \citep[$L_{\rm crit}{\sim}10^{37}\,\mathrm{erg}\,\mathrm{s}^{-1}$:][]{Basko1976,Becker2012, Mushtukov2015} and growth of the accretion column as an emitting structure. Modeling emission from accretion columns is a very challenging problem due to its dynamic nature, multi-dimensional geometry, and the necessity of including general relativistic effects. On the other hand, studying the low-luminosity regime with emission from the heated atmosphere of the polar cap allows easier access to fundamental parameters of accreting NSs (see \S~\ref{sec:lowl} for more details).

Cyclotron lines exhibit variations with accretion luminosity, pulse phase, and over secular time scales. Variation of the cyclotron line centroid energy $E_{\rm cyc}$ with X-ray luminosity $L_X$ can be exploited to gain insight into the accretion regime.  For example, low-luminosity XRPs often show a positive correlation between $E_{\rm cyc}$ and $L_X$, while high-luminosity XRPs show a negative correlation (see, e.g., \citealt{Klochkov2011}). The two trends are assumed to be separated by the critical luminosity and thus are related to the transition between the different accretion regimes described above.
On the other hand, studying the pulse-phase dependence of the $E_{\rm cyc}$ (as well as the line depth and width) can shed light upon the geometrical configuration of the NS \citep{Nishimura2011}. Moreover, some sources show evidence of a pulse phase-transient cyclotron line; i.e., a CRSF that only appears at certain pulse phases \citep{Klochkov2008, Kong2022}. Finally, secular variation of the cyclotron line has been attributed to a geometric reconfiguration of the polar field or to a magnetic field burial due to continued accretion \citep{Staubert2020,Bala2020}.

One of the biggest challenges for the detection and characterization of cyclotron lines is determining a robust broadband continuum model.
The centroid energy of the line as well as its other parameters can show significantly different best-fit values when modelled with different continua. Due to this model-dependence, the very presence of a cyclotron line has been questioned at times \citep{DiSalvo1998, Doroshenko2012, Doroshenko2020} as has its luminosity dependence \citep{mueller2013}. A similar argument also holds for the continuum best-fit values. The ``true'' continuum model can only be inferred if data cover a sufficiently wide energy passband to constrain absorption affecting the softer X-ray energies and the various spectral components at harder X-ray energies \citep{Sokolova-Lapa2021, Malacaria2023}. Moreover, properly constraining the broadband continuum is necessary when multiple cyclotron line harmonics are present, as in the cases of 4U 0115+63 \citep{Heindl2004} and V 0332+53 \citep{Pottschmidt2005}.
In addition, observing the broadband continuum from accreting XRPs can help constraining physical parameters of the NSs thanks to the recent development of physically motivated spectral models (e.g., \citealt[and references therein]{Farinelli16, Becker2022}).
This is especially important at low-luminosity, where the accretion flow free-falls on the NS surface and our understanding of the physical mechanisms is more uncertain concerning, e.g., how the flow transitions from radiation-dominated to gas-dominated, or the detailed production of seed photons from cyclotron, bremsstrahlung, and blackbody mechanisms at the site of impact.
\hexp\ will bring crucial contributions also in this field, as it will be able to probe broadband spectral emission from intrinsically dim sources whose required observing exposure with current facilities would be prohibitive.

Last but not least measuring phase-dependent spectral components with the highest sensitivity is required in order to develop consistent physical models for the emission process that include X-ray polarization properties. Recent \textit{IXPE} observations of accreting pulsars like Cen X-3, Vela X-1, EXO~2030+375 and others provided new constraints as they showed low, energy-dependent polarization degrees of $<$10\% \citep{Tsygankov2022,Forsblom2023, Malacaria23b}. Radiative transfer models of highly magnetized plasmas predict considerably higher polarization degrees of 50--80\% \citep{Meszaros88,Caiazzo2021,Sokolova-Lapa2023}. Initial candidates for explaining the difference highlight the potential importance of the temperature structure of the plasma in the accretion column, of angle-dependent beaming, and of the propagation of X-rays in the magnetosphere for lowering the polarization degree. Furthermore it has been demonstrated that the X-ray continuum and cyclotron line profile can be expected to directly show subtle, complex imprints of polarization effects \citep{Sokolova-Lapa2023}. This is an intriguing possible reason for the occasionally observed and notoriously difficult to constrain ``10\,keV feature'' mentioned above, as well as for the fact that some accreting pulsars do not show cyclotron lines.

Advancing our understanding of HMXBs in general and cyclotron line sources in particular therefore requires broadband spectra with high 
sensitivity throughout the relevant energy passband, with a limited background as provided by focusing X-ray facilities, and medium spectral energy resolution \citep{Wolff2019}.  Thanks to its leap in observational capabilities, \hexp\ will be capable of overcoming all above-mentioned challenges and push our understanding of accretion onto XRPs forward. In the next section, we simulate a few exemplary cases highlighting the gains enabled by \hexp with respect to currently available X-ray facilities for which the investigated cyclotron lines centroid energy, accretion regime, or the necessary exposure time prevent a comprehensive study of the physical mechanisms at work.

\subsection{Simulated Science Cases for HMXBs}

The following simulations show \hexp's potential to tackle different cyclotron line science cases using several example sources.  Each source has been chosen to represent a specific science case, either because it allows us to observe a certain accretion regime and luminosity range or due to its specific cyclotron line properties, such as the existence of $n>1$ harmonics. We begin with the prototypical persistent cyclotron line sources Cen X-3 and Vela X-1, which show moderate to high fluxes and luminosities, and allow for exquisitely detailed and sensitive parameter constraints with \hexp. Then we focus on transient accreting pulsars for which \hexp will provide unprecedented access to low fluxes, allowing study of extreme luminosity regimes, e.g., for GX 304-1 in quiescence or for super-Eddington outbursts of extragalactic sources like SMC X-2. Similar to \S~\ref{subsec:lmxb_sim}, all simulations were performed via the `fakeit' command in  \textsc{xspec} \citep{arnaud96} or \texttt{ISIS} \citep{Houck}, employing version v07 of the \hexp response files, selecting an 80\% PSF correction, assuming a $15''$ extraction region for the HETs and $8''$ for the LET, as well as taking the expected background at L1 into account. The flux for each of the following simulations can be found in Table~S1 in the Supplemental Materials.

\subsubsection{Constraining magnetic field geometry from cyclotron line harmonics} 
\label{sec:cenx3}

As discussed earlier, cyclotron lines result from transitions between quantized Landau energy levels of electrons in the presence of a magnetic field \citep{Meszaros1992}. Measuring the energy, width, and strength of cyclotron lines in HMXBs is crucial for understanding the underlying physics of these systems. The energy of cyclotron lines provides insights into the line-forming region in the accretion column. The width of cyclotron lines provides valuable information about the geometry, temperature distribution, and plasma conditions within the accretion column (see e.g., \citealt{Becker2012,Staubert2014} and references therein). Broadening of cyclotron lines can be influenced by factors such as electron thermal motion, turbulence, and relativistic effects. In the last few decades, comprehensive analyses of HMXBs have directly  unveiled many important properties of these systems (see, e.g., \citealt{Staubert2019,pradhan2021} and references therein). In addition to the fundamental, higher harmonics of the cyclotron line are also sometimes present in the X-ray spectrum. They arise due to higher-order interactions between X-rays and the Landau levels. By detecting and analyzing fundamental cyclotron lines and their higher harmonics, one can determine the magnetic field strength in different regions of the accretion column. Furthermore, the strength or intensity of fundamental cyclotron lines and their higher harmonics relative to the fundamental line provide insight into the scattering efficiency and the fraction of scattered photons, improving our understanding of the emission processes and properties of the compact object (e.g., see \citealt{Alexander1991, Schwarm2017} and references therein).

\begin{figure}
    \centering
    \includegraphics[width=0.95\textwidth, trim=0 20 0 10,clip]{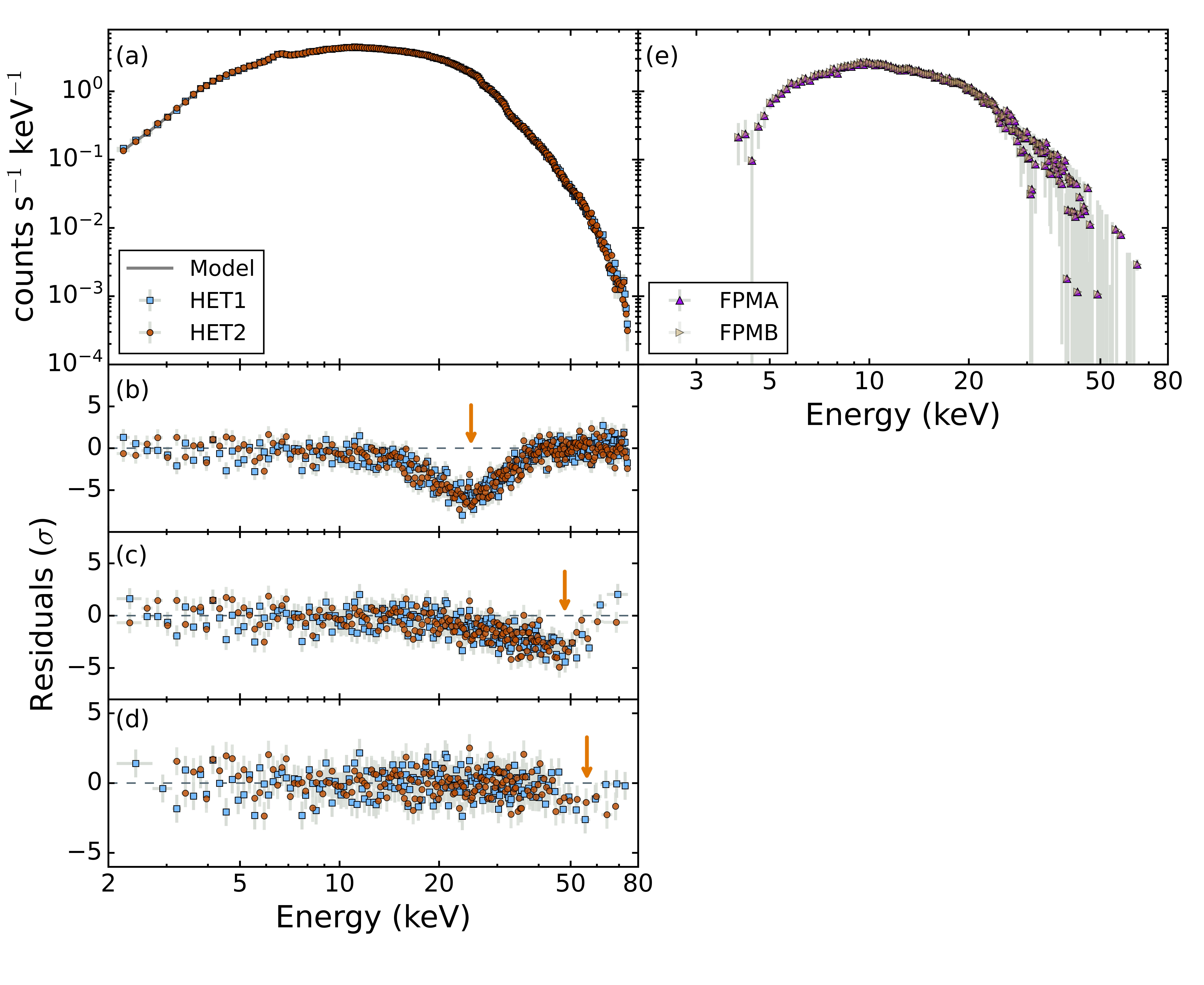}
    \caption{ Simulated 20\,ks HET spectra for Cen X-3, as well as the best-fit model are shown in panel (a). Panels (b), (c), and (d) show the residuals obtained when we set the strength of fundamental, $n_2$ harmonic, and $n_3$ harmonic to zero, respectively. We re-binned the data visually and provide arrows for the cyclotron line components to aid with clarity. For comparison, we provide a simulated \nustar spectrum for using the same model and exposure time which is shown in panel (e). The \nustar data are strongly background dominated at these flux levels, which makes constraining the higher energy harmonics difficult even at much longer exposure times (see text for more details).}
        \label{fig:cenx3}
\end{figure}

Given the limited passband and / or higher background of instruments like \rxte and \textit{Suzaku}, there have been few studies to date of cyclotron lines higher harmonics in HMXBs \citep[see Table 3 of][]{pradhan2021}. More recent instruments like \nustar and \textit{Insight}/HXMT provide a significant improvement, but as we look into the future, these studies will be revolutionized by a mission with a broad passband coupled with very low background like \hexp. In order to test this, we simulated a 20\,ks \hexp spectrum of Cen X-3 in \textsc{xspec}. Due to its bright nature and that Cen X-3 is a strong hard X-ray emitter (see Table S1 in the Supplemental Material), we concentrate on the HETs only which are devoid of pile-up although we have verified that the results below are consistent when including the LET. Cen X-3, an eclipsing HMXB, consists of an O star and a pulsar with a rotational period of $\sim$4.8\,s \citep{Chodil,Giacconi,Schreier}. The X-ray spectrum of this source exhibits multiple emission lines (Fe, Si, Mg, Ne: \citealt{iaria2005,tugay2009,naik2012,aftab2019, ferrin2021}), and the X-ray emission has been detected beyond 70\,keV. The fundamental cyclotron line occurs near 29\,keV \citep{tomar2021} with a possible $n_2$ harmonic at $\sim$47\,keV \citep{Yang2023} as seen from recent HXMT results. With the passband of \hexp extending up to 80\,keV and significantly less background, we investigated the possibility of detecting an $n_3$ harmonic in the source.

The baseline model for our simulation is \texttt{NPEX} (Equation \ref{eq:npex}; \citealt{M95, makishima1999}) and the values of model parameters are taken from Table~1 (ObsID P010131101602) of \citet{Yang2023}. The functional form of \texttt{NPEX} is given by
\begin{equation}\label{eq:npex}
\npexeq
\end{equation} where $\Gamma_{1}$ and $\Gamma_{2}$ are the negative and positive power-law indices, respectively. $\Gamma_{2}$, which approximates a Wien hump, was fixed at 2.0, $\Gamma_{1} = 0.8$, the folding energy \efold=6.9\,keV, and the absorption column $N_{\rm H} = 2.06 \times$ 10$^{22}$ cm$^{-2}$. 

The CRSFs are modeled with a multiplicative model of a line with a Gaussian optical depth profile with energy $E_i$, strength $d_i$, and width $\sigma_i$.  The fundamental line, $i=1$, is at $\sim$29\,keV with $d_1 \sim 1.4$ and $\sigma_1=7.6$ \,keV. The $n_2$ harmonic line, $i=2$, is at $\sim$47\,keV, with $d_2 \sim 2.3$ and $\sigma_2 = 9.7$\,keV. The 2-75\,keV flux is ${\sim}1\times$ 10$^{-8}\,\mathrm{erg}\,\mathrm{cm}^{-2}\,\mathrm{s}^{-1}$.
In order to showcase the unparalleled accuracy of \hexp in measuring magnetic fields correspondent to cyclotron line energies above 50\,keV, we incorporated a representative example into our model: we include an $n_3$ harmonic line at 70 keV ($i=3$), with the width ($\sigma_3$) and strength ($d_3$) set to match those of the $n_2$ harmonic line. By doing so, we investigate the ability of \hexp to effectively detect these parameters in the $n_3$ harmonic line, from which we can constrain the magnetic field strength and details about the magnetic field structure. 

A 20\,ks observation \hexp is able to constrain the energy of the $n_3$ harmonic well, to within 7\% (see Fig.~\ref{fig:cenx3}). We also applied an F-test to calculate the probability of chance improvement, with and without the $n_3$ harmonic line, and found that the probability of chance improvement when adding the $n_3$ harmonic line is below 0.06\%, confirming the robustness of the detection. We reiterate here that the clear improvement of \hexp over its predecessors in terms of broadband coverage and sensitivity makes it possible to investigate higher harmonic features -- which have clearly eluded \nustar (see e.g., \citealp{tomar2021}). 

Since cyclotron lines result from the resonant scattering of photons by electrons whose energies are quantized into Landau levels by the strong magnetic field, and the quantized energy levels of the electrons are harmonically spaced, one would naively expect the energy of the $n>1$ harmonics to be the integer multiples of the fundamental line. It has however been observed in various X-ray pulsars that this is generally not the case (see \citealt{Yang2023, Orlandini2012} and discussions within). The discrepancy can be understood by assuming a difference in line formation \textit{mechanisms} or in line-forming \textit{regions} for the fundamental and higher harmonics. In the former case, the fundamental would primarily arise from resonant scattering, while the higher harmonics involve additional effects like multiple scattering and photon spawning (see e.g., \citealt{Nishimura2003,Sch2007}). This, however, can not be the only reason because some systems show an an-harmonic spacing larger than predicted by this effect. One way to explain this difference is to take into account that the optical depths of the fundamental and the higher harmonics can be different if they are formed at different heights above the NS. The higher harmonics could be closer to the NS surface and the fundamental line situated at a height with weaker magnetic field strength (see, e.g., \citealt{Furst2018}). Another possibility is to consider a displacement of the magnetic dipole, which would also explain the energy difference of the two lines if the lines originate from the different poles of the NS \citep{Rodes2009}. Therefore, a significant phase dependence of the strength of the fundamental and higher harmonics is expected, which is not possible to probe with the current data sets. Here, too, \hexp can provide revolutionary new capabilities though (see section \ref{subsubsec:velax1}).

\begin{figure}[t!]
    \centering
    \includegraphics[width=0.6\columnwidth]{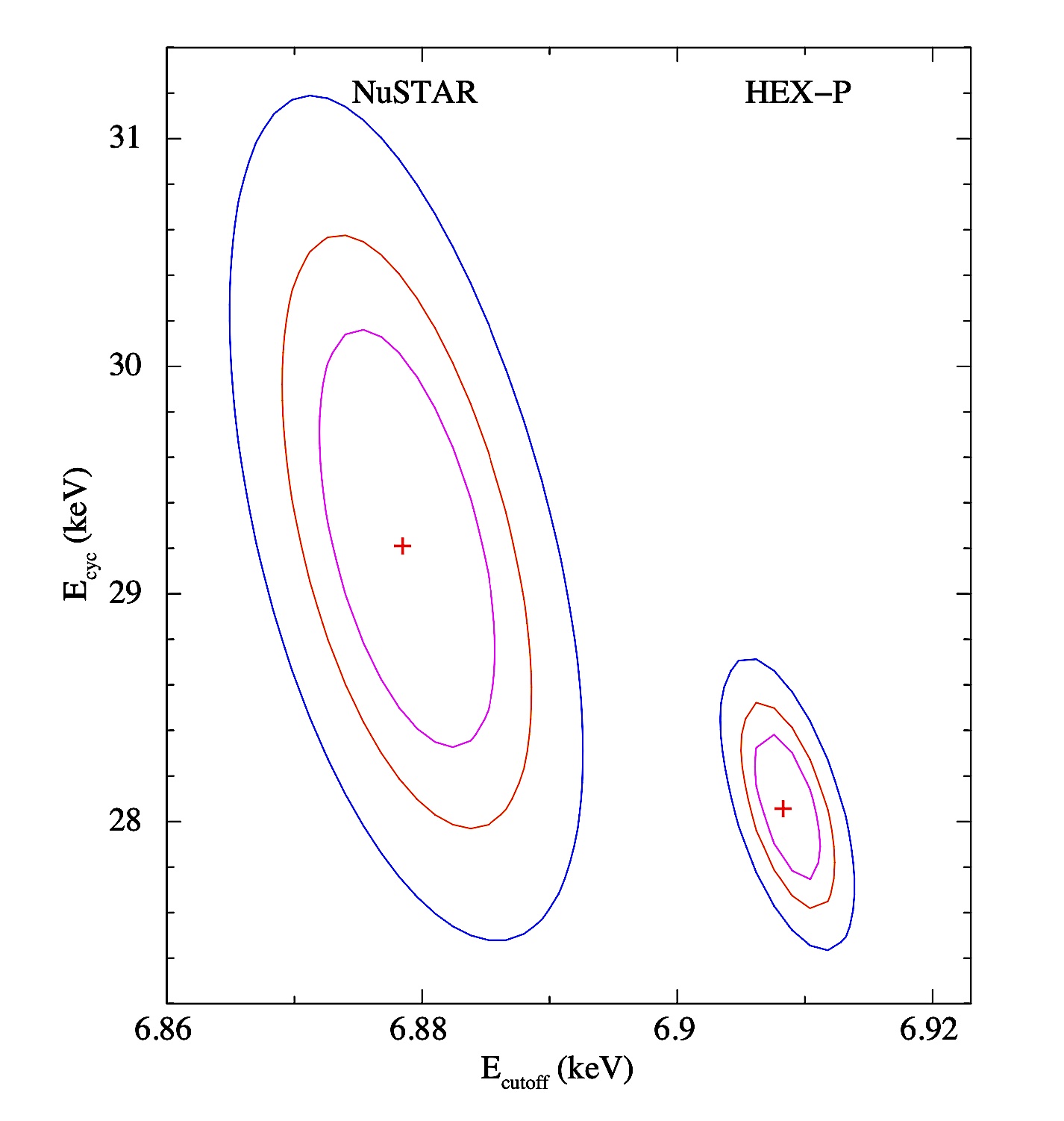}
    \caption{Contour plots (68\%, 90\% and 99\% confidence levels) for cutoff energy versus the fundamental cyclotron line energy for Cen X-3 using the same model and exposure for \nustar and \hexp. As evident from the plot, \hexp will provide better constraints on the continuum and cyclotron line parameters; thereby mitigating degeneracies between the two. Note that the cutoff energies are slightly shifted to aid visualization. }
        \label{contour-cenx3}
\end{figure}

Finally, note that while the $n_2$ harmonic of Cen X-3 at 47\,keV technically falls within the energy range covered by \nustar, meaningful constraints could not be derived with \nustar (see panel (e) of Fig.~\ref{fig:cenx3}) due to the dominant background in that energy range even when the exposure was set to 500\,ks. However, with \hexp, the parameters associated with the cyclotron line can be much better constrained, while also alleviating the degeneracies with the continuum. To emphasize the latter point, we present a contour plot in Figure~\ref{contour-cenx3} depicting the relationship between the fundamental cyclotron line energy and the cutoff energy for both \hexp and \nustar data for the same model and exposure of 20\,ks for Cen X-3. The plot clearly demonstrates that \hexp provides significantly more stringent constraints compared to \nustar. Such accurate measurements in short exposure times will enable detailed pulse phase-resolved spectra to explore the accretion physics in the accretion column in unprecedented detail, as discussed in the following section.

Note that Cen X-3 is highly variable, the X-ray flux varies by up-to two orders of magnitude even outside eclipses. Therefore, we also fit the actual \nustar data in bright flux state (ObsID 30101055002; exposure 21 ks), which has a flux an order of magnitude more than our simulations, and find the harmonics were not detected in the \nustar spectrum (see, \citep{tomar2021}). In order to make a comparison of \hexp vs \nustar for this bright state of Cen X-3, we simulated an HET spectrum using the \nustar model in bright state, while keeping the parameters of harmonics as above. We find that, even for this bright state of Cen X-3, \nustar will be able to obtain the same signal-to-noise as \hexp only if the exposure is increased by 20 times (i.e., 420\,ks).

\subsubsection{Constraining accretion column emission geometry through pulse phase-resolved spectroscopy} \label{subsubsec:velax1}

Vela X-1 is an archetypical HMXB. It has been well studied with all current and past X-ray missions \citep[see][for a recent review]{kretschmar21}. While not being exceptionally luminous (${\sim} 10^{36}$\,erg\,s$^{-1}$), its close distance ($d=1.9$\,kpc) and X-ray eclipses make it an ideal system to study NS magnetic fields and their interaction with the stellar wind of the companion (i.e., the mass donor). The X-ray spectrum of Vela X-1 shows two CRSFs, the fundamental at around 25\,keV and the $n_2$ harmonic around 55\,keV \citep[see][and references therein]{fuerst14}. These line energies are ideally covered by the \hexp energy band. However, in Vela X-1, the fundamental line around 25\,keV is much weaker and shallower than the $n_2$ harmonic line at 55\,keV. In fact, the 25\,keV line is so weak that there has been a long-standing discussion in the literature about its existence, which could only be settled once \nustar data were available \citep{fuerst14, kreykenbohm2002, Maitra2013}. The cyclotron lines in Vela~X-1 therefore represent a good test case to explore \hexp's sensitivity to broad and shallow spectral features.
 
We performed simulations to study how well \hexp can measure the energy, width, and depth of both CRSFs, in particular as a function of rotational phase of the NS. Variations of CRSF parameters as a function of phase constrain the magnetic field geometry and emission geometry of the accretion column \citep[see, e.g.][]{Iwakiri2019, Liu2020}.

\begin{figure}[t!]
    \centering
    \includegraphics[width=0.8\textwidth,trim=0 0 0 0,clip]{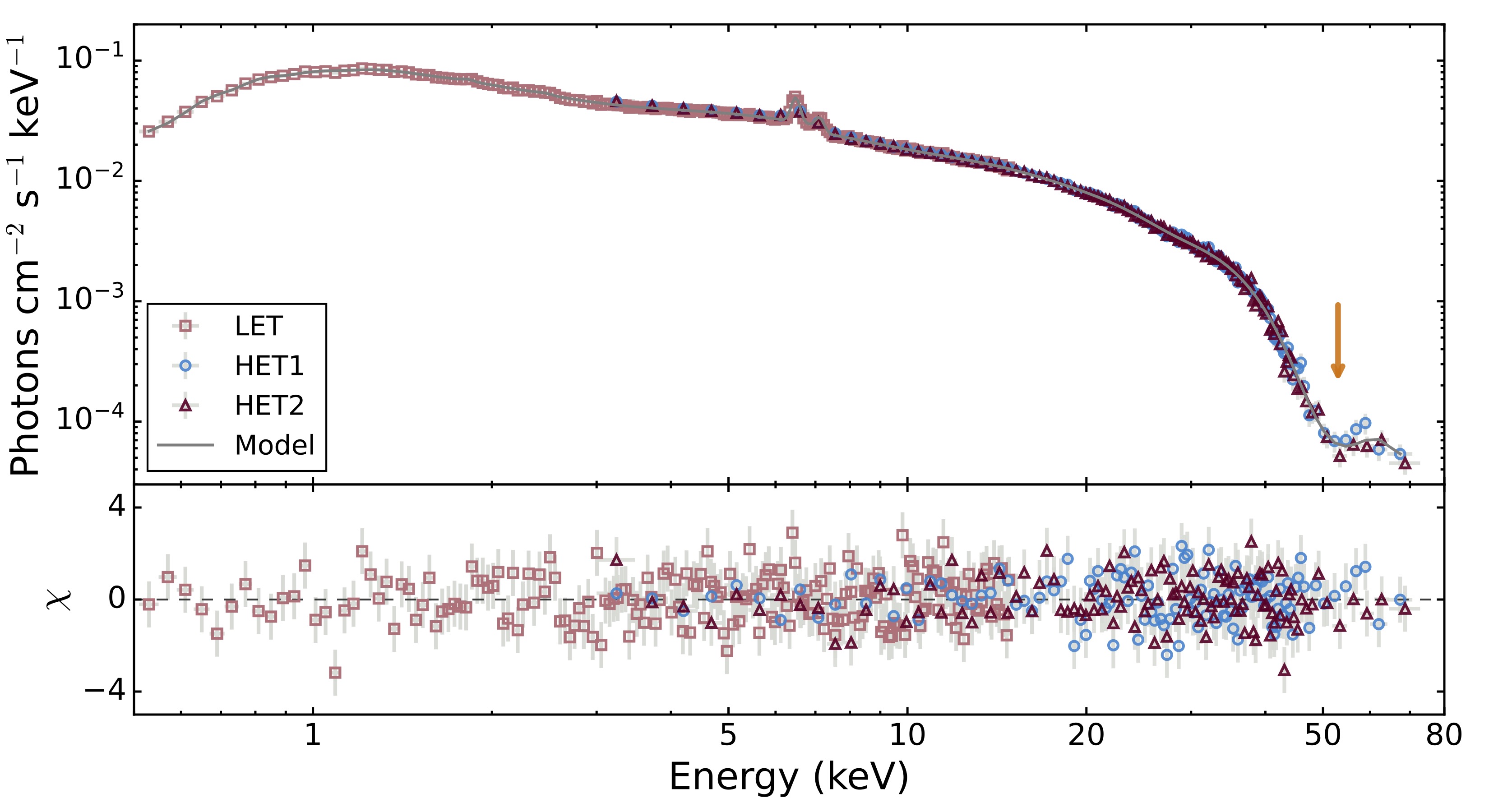}
    \caption{Simulated Vela X-1 spectrum for the bin covering phases 0--0.1 (see Fig.~\ref{fig:velax-1} for the spectral values in that bin). The $n_2$ harmonic cyclotron line is clearly visible as a strong dip at the highest energies as indicated by the arrow.}
    \label{fig:velax-1-spec}
\end{figure}

We base our simulations on the spectral fits of \nustar data presented by \citet{Diez2022}. In particular, the continuum is modeled by a powerlaw with an exponential cutoff at high energies (using the model ``FDcut'' in \textsc{xspec}, Equation \ref{eq:fdcut}; \citealt{tanaka86}), modified by neutral absorption column at low energies:
\begin{equation}\label{eq:fdcut}
\fdcoeq
\end{equation} 
where $\Gamma$ is the power-law index, \ecut\ is the cutoff energy, and \efold\ is the folding energy. In addition to the continuum model used by \citet{Diez2022} we also include soft X-ray emission lines at 6.4\,keV (Fe K$\alpha$), 2.4\,keV (S  K$\alpha$), 1.8\,keV (Si K$\alpha$), 1.4\,keV (Mg L$\alpha$), and 0.9\,keV (Ne IX), representing various atomic features in the spectrum \citep{Diez2023}. A simulated spectrum is shown in Figure~\ref{fig:velax-1-spec}.

As mentioned in \S\ref{sec:cenx3}, CRSFs are modeled using a multiplicative line model with a Gaussian optical depth profile described by its energy $E_i$, its strength $d_i$, and its width $\sigma_i$. Here the subscript $i$ denotes either the fundamental line around 25\,keV ($i=1$) or the $n_2$ harmonic line around 55\,keV ($i=2$). The width $\sigma_1$ is set to $0.5\times\sigma_2$ \citep{Diez2022}.
Instead of relying on the (rather uncertain; see, e.g., \citealt{Maitra2018}) current knowledge of the phase-resolved behavior of the CRSF parameters, we simulate that the four most relevant parameters ($E_1$, $E_2$, $d_1$, $d_2$) vary sinusoidally with random phase shifts to each other. This approach demonstrates the power of \hexp to resolve small changes in any of the parameters, even for relatively weak lines. In particular, we assume that the fundamental line $E_1$ varies by about $\pm$3.5\,keV as a function of phase, while the $n_2$ harmonic line energy $E_2$ varies by $\pm$5\,keV. The strength $d_1$ and $d_2$ vary by $\pm$0.3\,keV and $\pm$5.0\,keV, respectively.

In addition, we also allow the absorption column of the partial absorber to vary, with $\pm 5\times10^{22}$\,cm$^{-2}$ around the average value of $32.1\times10^{22}$\,{cm}$^{-2}$. While we do not necessarily expect that the absorption column will vary significantly as function of pulse phase, this variability highlights the capabilities of the \hexp/LET to measure small changes in absorbing columns on time-scales as short as 5\,ks. These variations have been observed in time-resolved spectroscopy and allow us to study the physical properties, like density and clump sizes of the accreted medium \citep{Diez2023}. At the same time, the changes in $N_\text{H}$ do not influence the high energy spectrum where the CRSFs are present.

\begin{figure}
    \centering
    \includegraphics[width=0.6\textwidth,trim=0 0 0 0,clip]{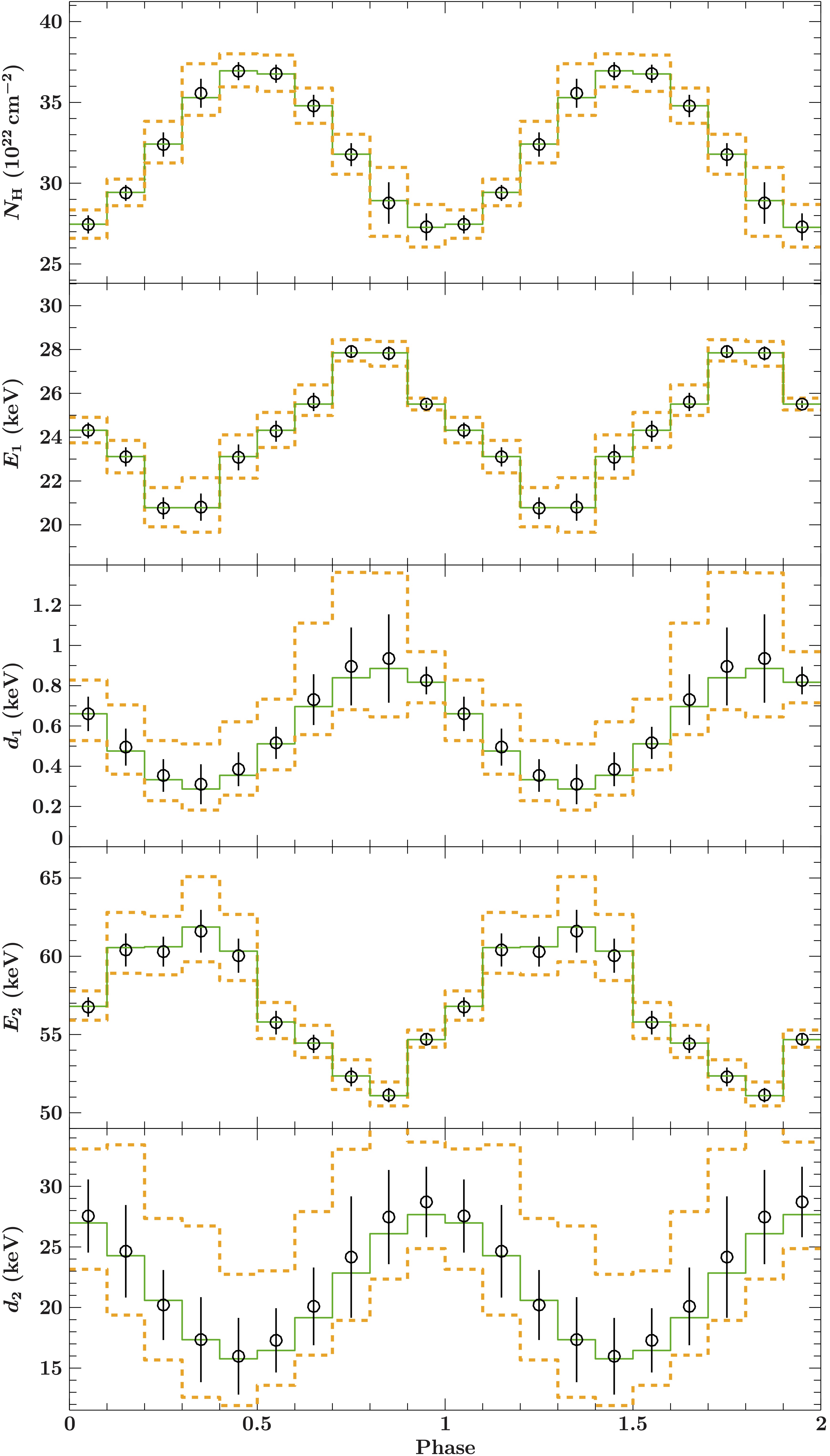}
    \caption{Absorption column and cyclotron line variability in Vela X-1, calculated from simulated phase-resolved \hexp\ spectra using the results of \citet{Diez2022} with an average flux of $6.7\times10^{-9}$\,erg\,s$^{-1}$\,cm$^{-2}$ in the 3--80\,keV band. The parameters from top to bottom are: absorption column $N_\text{H}$, fundamental line energy ($E_{1}$), fundamental line strength ($d_1$), $n_2$ harmonic line energy ($E_{2}$), and $n_2$ harmonic line strength ($d_2$). Note that the fundamental line width is set to half of the $n_2$ harmonic width. The black  uncertainties show the standard deviation for a sample of 100 simulations per phase bin. The orange dashed histogram indicates the expected 90\% uncertainties on each individual realization. The green line indicates the input values for each phase bin. For details about the simulation set-up see \S~\ref{subsubsec:velax1} and Table S3 in the Supplemental Material.}
    \label{fig:velax-1}
\end{figure}

For our simulations, we assume that LET will be operated in a fast readout mode to avoid pile-up and with negligible deadtime. We simulate spectra with 50\,ks exposure time each, and split the data into 10 phase-bins so that each one has an exposure time of 5\,ks. We simulate 100 individual spectra for each phase-bin and calculate the standard deviation (SD) of that sample as uncertainties for each parameter. We additionally calculate the 90\% and 10\%  percentile of the 90\% uncertainties of each realization as a realistic estimate of the expected uncertainties in a real 50\,ks observation. Using multiple realizations for each phase bin allows us to avoid issues with any one particular realization and shows that our uncertainties are properly sampled over the Poisson noise inherent to counting statistics. The results are shown in Figure~\ref{fig:velax-1}.

All parameters can be very well reconstructed, with the fundamental energy having uncertainties \mbox{$<0.5$\,keV} for phases where its strength is $>0.4$\,keV (i.e., phases 0.5--1.0). Even for weaker lines (despite having larger uncertainties of about $\pm 1$\,keV), the line can be still significantly detected. This is a significant improvement over \nustar, for which \citet{Furst2018} found that the energy was basically unconstrained for line strengths $<0.5$\,keV.

The $n_2$ harmonic line energy can be very well constrained (with uncertainties $<$2\,keV) for central energies of the line $<57$\,keV. Due to the exposure time of only 5\,ks per phase-bin and the low cutoff energy of  $\sim$25\,keV, the line appears at the very edge of the useful range of the spectrum. Therefore, higher energies become more difficult to constrain, as most of the line is outside the useful passband; however, we still find typical constraints of $\pm$3--4\,keV, even for a simulated line energy of 61\,keV. 

These results represent a significant improvement over previous phase-resolved studies of Vela X-1. For example, using \rxte, \citet{kreykenbohm99} found variations of the CRSF energies for both the fundamental and the $n_2$ harmonic line, but could only use 10~phase bins and still had average uncertainties of 2--3\,keV for the fundamental line, and 5--10\,keV for the $n_2$ harmonic. Recently, \citet{liu22} published results obtained on Vela~X-1 with \textit{Insight}/HXMT.  Using 16 phase-bins for a $\sim$100\,ks exposure, they found average uncertainties comparable to our \hexp simulations for the $n_2$ harmonic line, but with significantly larger uncertainties (a few keV) for the fundamental line. We also note that \citet{liu22} found the $n_2$ harmonic line at energies between 40--50\,keV, i.e., at much lower energies than simulated here and therefore in a part of the spectrum with a much higher signal-to-noise ratio.

\hexp observations would therefore be a big step forward in being able to obtain phase-resolved spectroscopy of bright HMXBs, where we can constrain the CRSF parameters, in particular the energy, much better than with existing instruments in shorter exposure times. This reduction in exposure time means that we can either observe more sources in less time or slice the phase-resolved spectra finer to obtain a more detailed look at the emission and magnetic field geometry of the accretion column \citep{Nishimura2003,Sch2007, Schwarm2017}.

\subsubsection{Constraining the surface magnetic field strength from quiescent observations of Be X-ray binaries}
\label{sec:lowl}

Details regarding the formation of cyclotron lines in the spectra of accreting NSs in HMXBs are highly debated in the literature.
The observed positive and negative correlations of the cyclotron energy with luminosity are often interpreted as a change of the dynamics of the accretion process in different accretion regimes, distinguished by a critical luminosity where a radiation-dominated shock forms in the column above the NS surface.
For high-luminosity regimes, when it is assumed that $L_\mathrm{X}>L_\mathrm{crit}$, several mechanisms for the formation of cyclotron lines have been suggested that explain the observed negative line energy versus luminosity correlation. These scenarios involve different locations for the line production: above the NS surface in an accretion column that is growing with increasing luminosity \citep[see, e.g.,][]{Becker2012} or in the illuminated atmosphere of the NS at lower magnetic latitudes, where the column radiation is reprocessed (as suggested by \citealt{Poutanen2013}; see, however, \citealt{Kylafis2021}).
For lower luminosities, $L_\mathrm{X}\lesssim L_\mathrm{crit}$, the formation of cyclotron lines is usually attributed to comparatively lower heights in the accretion column. In this case, a positive line energy versus luminosity correlation can be explained by the formation of a collisionless shock \citep[see, e.g.,][]{Rothschild2017, Vybornov2017} that is moving closer to the NS surface for higher luminosities or by the redshift due to bulk motion of the accretion flow \citep{Nishimura2014, Mushtukov2015}. However, to apply these models in a consistent way and to distinguish between their predictions, it is required to know the surface field at the magnetic pole of the NS. This can then serve as a reference to estimate, for example, the characteristic height of the line-forming region in the column or the velocity of the accretion flow near the surface.

Recent evidence of accretion in Be X-ray binaries in quiescence, i.e., with $L_\mathrm{X}\ll L_\mathrm{crit}$, and the simple emission region geometry expected in this case (i.e., a hot spot on the NS surface) together represent a unique opportunity to observe cyclotron lines at the energy corresponding to the surface value of the magnetic field.
From current observational examples it seems that for this accretion state to occur, the magnetic field of the source might have to be sufficiently high, $B\gtrsim5\times10^{12}\,\mathrm{G}$, and the spin period might have to be comparatively long,
$\gtrsim100\,\mathrm{s}$ \citep[as, e.g., for GX\,304$-$1 and GRO~J1008$-57$: ][]{Tsygankov2019a, Lutovinov2021}.
The low flux level, however, makes the detection of cyclotron lines at high energies very challenging.
GX\,304$-$1 is one of the first sources which unambiguously exhibited stable quiescent accretion \citep{Rouco2018} with a cyclotron line known from outburst observations \citep{Mihara2010, Malacaria2015}.
The corresponding flux level in the 2--$10\,\mathrm{keV}$ energy band, ${\sim}0.4\,\mathrm{mCrab}$, is below the sensitivity of \textit{Insight}/HXMT.
With \nustar we can access the high-energy emission, but are typically unable to constrain the turnover of the second hump of the characteristic double-hump spectrum (see \S\ref{subsec:hmxbback}) and the cyclotron line (\citealt[][Zainab et al., in prep.]{Tsygankov2019a}).

The best way to probe the above mentioned science cases is to observe the source during the quiescent accretion regime, i.e., at a low-luminosity state that is typically below the detection threshold for X-ray all sky monitors and accessible only with pointed observations subsequent to an X-ray outburst. 
Such a regime is also of importance since the simplified physics of plasma stopping at the nearly-static NS atmosphere allows for more detailed modeling of emission processes. The tenuous flow of matter stops at the NS atmosphere by Coulomb collisions, resulting in a high temperature gradient from ${\sim}30\,\mathrm{keV}$ at the top down to ${\sim}2\,\mathrm{keV}$ in the lower layers, separated by only ${\sim}10\,\mathrm{m}$ \citep[see, e.g.,][]{Sokolova-Lapa2021}. The intrinsic emission is mainly produced by magnetic bremsstrahlung \citep[for sufficiently low magnetic fields, cyclotron photons are produced as well from collisional excitations of electrons moving in bulk;][]{Mushtukov2021}, which is then modified by Compton scattering. This regime of accretion allows, for the first time, to combine modeling of the temperature and density structure of the emission region with a \textit{joint} simulation of the continuum and cyclotron line formation \citep{Mushtukov2021, Sokolova-Lapa2021}.

In this way, existing physically-motivated models principally provide access to information about the field strength at the poles of accreting highly magnetized NSs. However, the limitations of current high-energy missions do not permit us to constrain the corresponding cyclotron lines in the spectra. \hexp will open a new avenue for the accreting NSs community by allowing us to detect such features with sufficiently long exposures.

\begin{figure*}
    \centering
    \includegraphics[width=0.8\columnwidth]{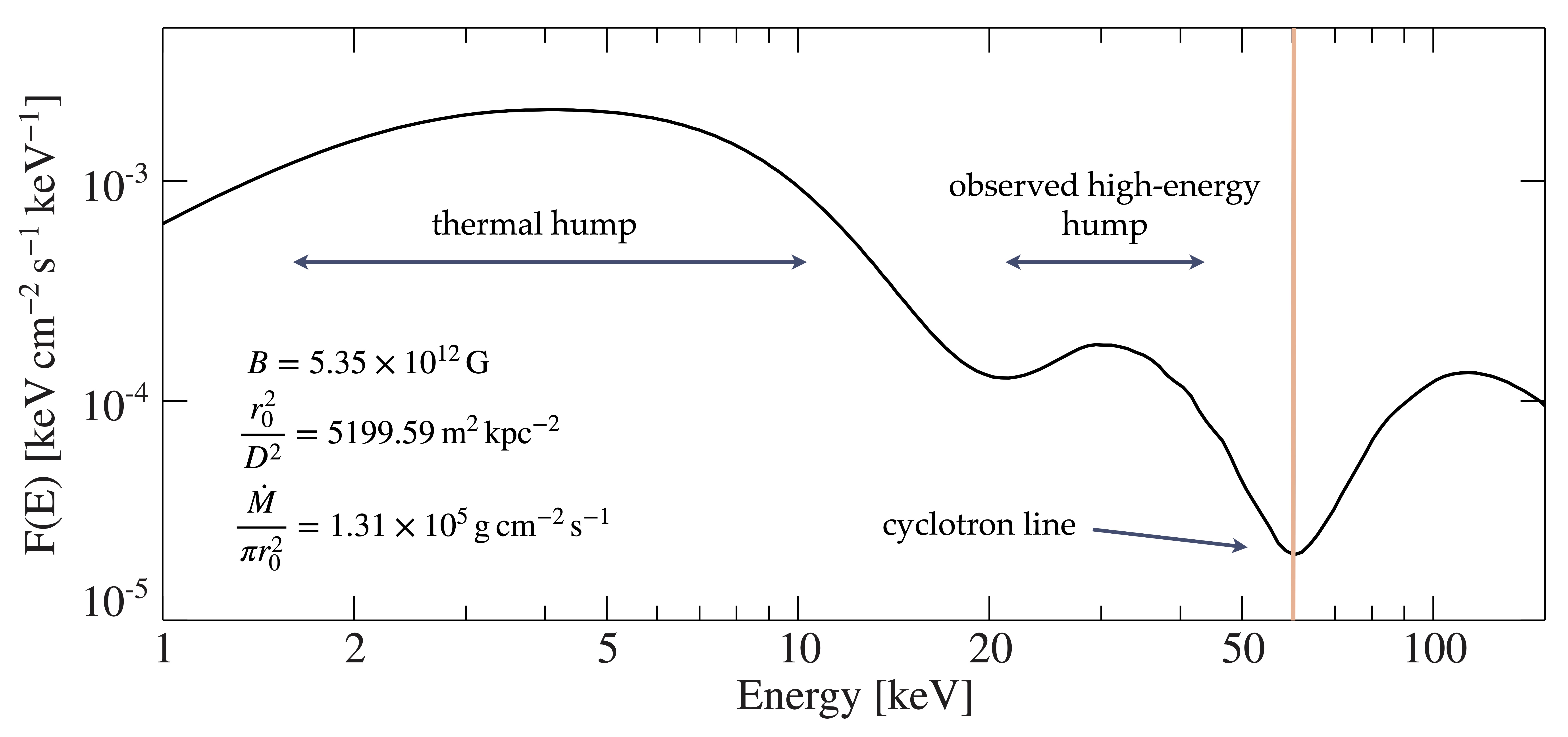}
    \caption{Model used to simulate the observation of the Be~X-ray binary GX\,304$-$1 in quiescence. The vertical orange line indicates the center of the cyclotron line. The cyclotron line and the continuum are calculated together using polarized radiative transfer simulations \citep{Sokolova-Lapa2021}. The low-energy ``thermal'' hump and the high-energy hump below the cyclotron line (the red wing of the cyclotron line) are typically observed as a ``two-hump'' spectrum from Be X-ray binaries in quiescence. }
    \label{fig:gxmod}
\end{figure*}

In order to demonstrate \hexp's capabilities to measure the surface magnetic field strength of a NS in quiescence, we simulate  observations of the quiescent state of the Be X-ray binary GX\,304$-$1, the first system for which a transition to the two-component spectrum was observed by \nustar \citep{Tsygankov2019a}.
We use the physical model \textsc{polcap} presented by \citet{Sokolova-Lapa2021}, which describes emission for the low-luminosity accretion regime. Here, we adopt an updated parametrization of the model (using the same underlying pre-calculated spectra; E.~Sokolova-Lapa, priv. comm.).
The parameters are the mass flux, $\dot{M}/\pi r_0^2$, where $\dot{M}$ is the mass-accretion rate and $r_0$ is the polar cap radius; the cyclotron energy, $E_\mathrm{cyc}$, corresponding to the polar magnetic field strength, $B$; and the normalization given in terms of $r_0^2/D^2$, where $D$ is the distance to the source.
We set the normalization and the mass flux based on the flux level and the shape of the spectrum as observed previously by \nustar.
The magnetic field strength is set to $B=5.35\times10^{12}\,\mathrm{G}$,  derived from the centroid energy of the cyclotron line observed during the previous outburst \citep[${\sim}50\,\mathrm{keV}$,][]{Jaisawal2016} and corrected by the gravitational redshift near the surface of the NS ($z_\mathrm{g}\approx0.24$, assuming  standard NS parameters).  
The corresponding spectrum calculated with the \textsc{polcap} model and the exact values of the parameters used for the simulations are shown in Figure~\ref{fig:gxmod}.
Due to internal averaging over the emission angles to obtain the total flux in the NS rest-frame \citep[see details in][]{Sokolova-Lapa2021}, the cyclotron line in the corresponding spectrum is located at ${\approx}60\,\mathrm{keV}$.
Taking into account the gravitational redshift, the cyclotron line in the observed spectrum is therefore expected at ${\approx}48\,\mathrm{keV}$.
We simulate a 60\,ks observation for all three \hexp\ instruments, combining the \textsc{polcap} model with \textsc{tbabs} \citep{Wilms2000} to account for interstellar absorption (i.e., \textsc{tbabs*polcap}), fixing $N_\mathrm{H}$ to the Galactic value of $1.1\times10^{22}\,\mathrm{cm}^{-2}$ in the direction of the source.
The resulting luminosity in the 1--80\,keV range is ${\sim}8\times10^{33}\,\mathrm{erg}\,\mathrm{s}^{-1}$.

\begin{figure*}
    \centering
    \begin{tabular}{p{0.48\textwidth} p{0.48\textwidth}}
        \vspace{0pt} \includegraphics[width=0.48\textwidth]{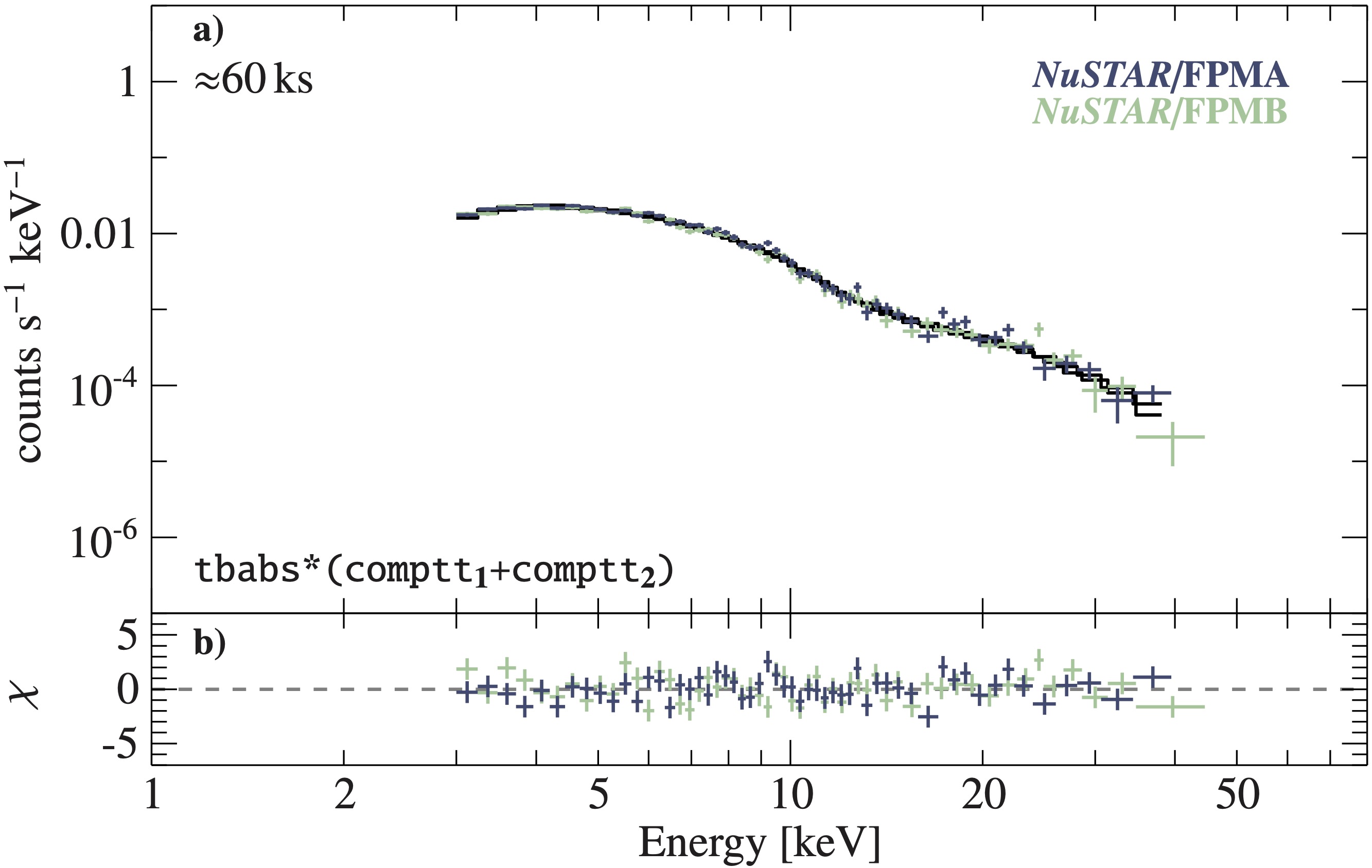} &
        \vspace{0pt} \includegraphics[width=0.48\textwidth]{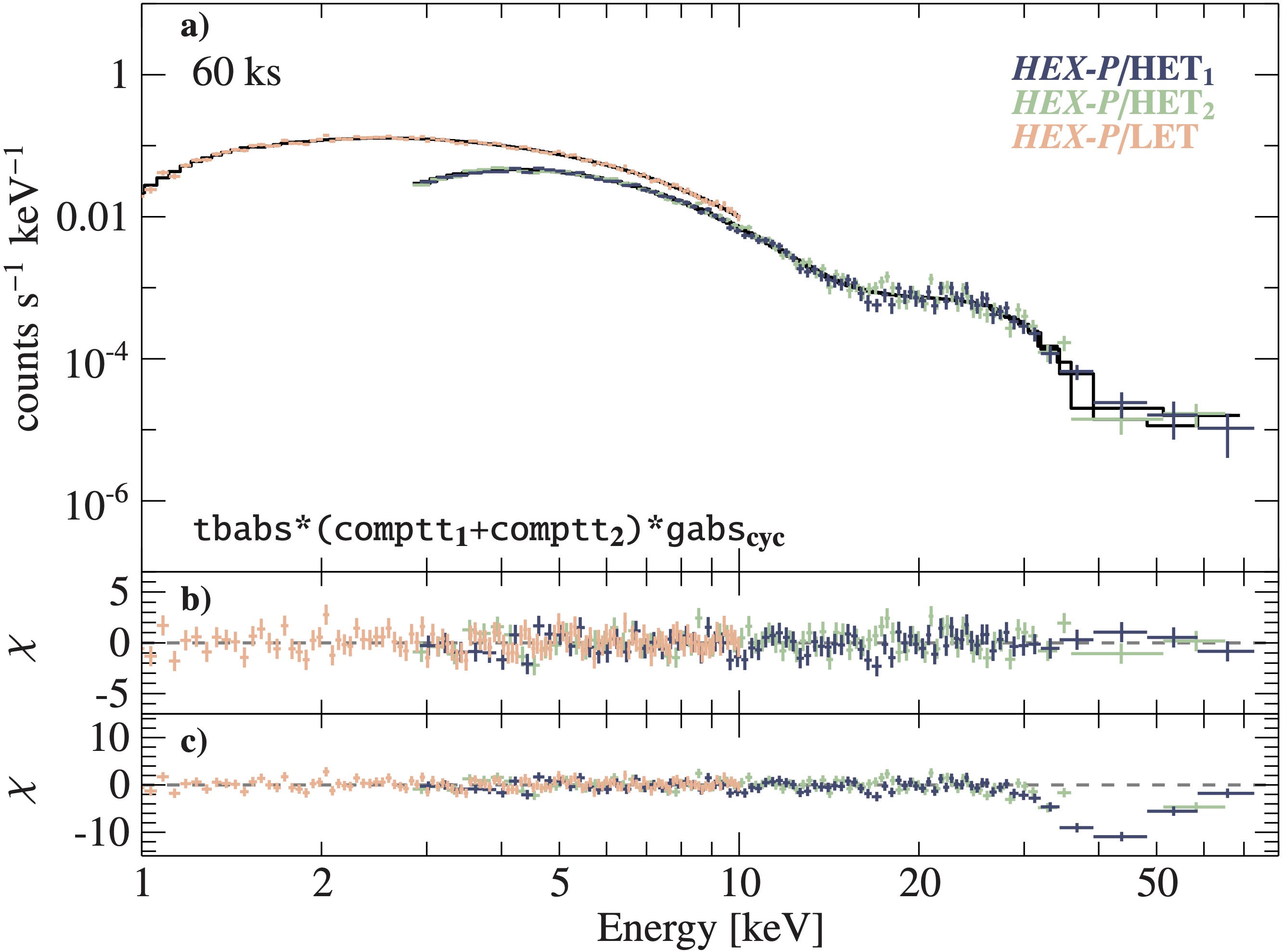}
    \end{tabular}
    \hfill
    \caption{\nustar observation (left, a) and simulated \hexp (right, a) spectra of a 60$\,\mathrm{ks}$ observation of the Be~X-ray GX\,304$-$1 in quiescence and corresponding best-fit models. Both panels b show residuals for the best-fit models. For the simulated \hexp spectra, we also show the residuals for the case when the cyclotron line strength is set to zero (left, c).
    The high-energy cyclotron line can be constrained from the \hexp data, with the centroid energy
    $46.5^{+4.1}_{-2.7}\,\mathrm{keV}$.
    }
    \label{fig:gxlowl}
\end{figure*}

We compare the \hexp spectra against a \nustar observation of the same exposure. 
We first fit both, the simulated \hexp and existing \nustar observations, using a model that includes two independent Comptonized components \citep[\textsc{comptt};][]{Titarchuk1994}, \textsc{tbabs*(comptt$_1$ + comptt$_2$)}.
This model, with or without a multiplicative Gaussian-like cyclotron line (\textsc{gabs}), is commonly used to describe the two-component spectra of low-luminosity states \citep[see, e.g.,][]{Tsygankov2019, Lutovinov2021, Doroshenko2021}.
Similarly to earlier analyses \citep{Tsygankov2019a}, we obtain a good description of the \nustar data with this model, with
$\chi^2_\mathrm{red}=136.93/107=1.28$
for the best fit.
For the simulated \hexp data, the same model consistent of two absorbed Comptonized components provides
a formally satisfactory fit ($\chi^2_\mathrm{red}=322.64/275=1.17$), however, 
the residuals are flat only at low and intermediate energies, but indicate a
dip at around 40--60\,keV, where the cyclotron line is expected from the underlying physical model (see Figure~\ref{fig:gxmod}).
The best fit for a model which includes an additional Gaussian absorption line at high energies to describe the cyclotron line, provides the line's centroid energy of 
$46.5^{+4.1}_{-2.7}\,\mathrm{keV}$
, width
$8.2^{+3.1}_{-2.1}\,\mathrm{keV}$
, and strength
$54^{+41}_{-21}\,\mathrm{keV}$,
with the
$\chi^2_\mathrm{red}=261.23/276=0.95$.
The centroid energy corresponds (within its uncertainties) to the redshifted cyclotron line from the physical model, which is expected to be at ${\approx}60/(1+z_\mathrm{g})\,\mathrm{keV}\approx48\,\mathrm{keV}$, assuming the standard NS parameters to estimate the redshift, $z_\mathrm{g}=0.24$.
Figure~\ref{fig:gxlowl} shows the resulting spectra and the corresponding best-fit models for the real \nustar and the simulated \hexp observations.

This example shows that even at this exposure, which is relatively low for the quiescent state of Be X-ray binaries, we can constrain the cyclotron line energy and thus the surface magnetic field to within ${\sim}20\%$.
This capability is crucial for systems which have never been observed at high-luminosity states, that is, for which cyclotron lines have never been observed in the spectra. This class of Be X-ray binaries accreting in quiescence is actively growing, e.g., through \nustar follow-up of Be X-ray binary candidates found during the first eROSITA All Sky Survey \citep[see, e.g.,][]{Doroshenko2022}. More similar discoveries are expected as deeper X-ray survey data from eROSITA become available.
The spectral shape of the latter sources, in particular, the location of the high-energy excess associated with the energies below the cyclotron resonance, suggests a similar magnetic field strength as in GX\,304$-$1.
The search for high-energy cyclotron lines, which makes it possible to unambiguously constrain the surface magnetic field of an accreting neutron star, requires superior sensitivity at hard X-rays, as will be provided by \hexp.

For a discussion of NS science that can be done with \hexp for B-fields in excess of ${\sim}10^{14}\,$G (i.e., magnetars), we refer the reader to Alford et al. 2023, (\textit{in prep.}).

\subsubsection{Constraining super-critical accretion and the critical luminosity via cyclotron line evolution in extragalactic sources}

A correlation between cyclotron line energy and X-ray luminosity has been reported for a handful of X-ray pulsars \citep[see][and above]{Staubert2019}. The dependence of the line energy on luminosity can be attributed to changes in the height of the accretion column and can be positive or negative for low and high X-ray luminosity, respectively \citep{Becker2012}. However, in a couple of sources a secondary effect has been reported: for equal levels of luminosity, the energy of the cyclotron line can be different (up to 10\% change) between the rise and the decay of an outburst or between different outbursts (e.g., V\,0332$+$53, \citealt{Cusumano2016} and SMC~X-2, \citealt{Jaisawal2023}).
\citet{Doroshenko2017} proposed that this effect is likely caused by a change of the emission region geometry (e.g., different combinations of height and width of the accretion column), while an alternative explanation is that it is due to accretion-induced decay of the NS's magnetic field \citep{{Cusumano2016,Jaisawal2023}}.

Given this complex observational behavior, it is critical to detect CRSFs in more transient systems and to study their evolution over luminous outbursts. 
Although a significant population of HMXBs is found in the Milky Way, interpreting results from observing its members is often hampered by their uncertain distance and large foreground absorption. In that sense nearby galaxies like the star-forming Large and Small Magellanic Clouds (LMC and SMC) offer a unique laboratory to complement our studies of luminous Galactic HMXBs. Sources in these galaxies have well determined distances of $\sim$50\,kpc (LMC) or $\sim$60\,kpc (SMC) and low Galactic foreground absorption (${\sim}10^{20}\,\mathrm{cm}^{-2}$) making them ideal targets for spectral and temporal studies during major outbursts. 
Based on past observations and recent statistics, an outburst that peaks above $2\times10^{38}$\,erg\,s$^{-1}$ occurs in the Magellanic Clouds every few years
\citep[e.g.][]{Vasilopoulos2014,Koliopanos2018,Maitra2018,Vasilopoulos2020}.
Such outbursts are brighter than the Eddington luminosity for a typical NS, more precisely, they are brighter than the critical luminosity\footnote{The critical luminosity is sometimes called the "local Eddington limit" which is descriptive but formally not well defined.}, $L_{\rm crit}{\sim}10^{37}\,\mathrm{erg}\,\mathrm{s}^{-1}$, for a typical accretion column (see \S\ref{subsec:hmxbback}). They are therefore called super-Eddington or super-critical outbursts.

\begin{figure*}
    \centering
    \includegraphics[width=0.485\columnwidth]{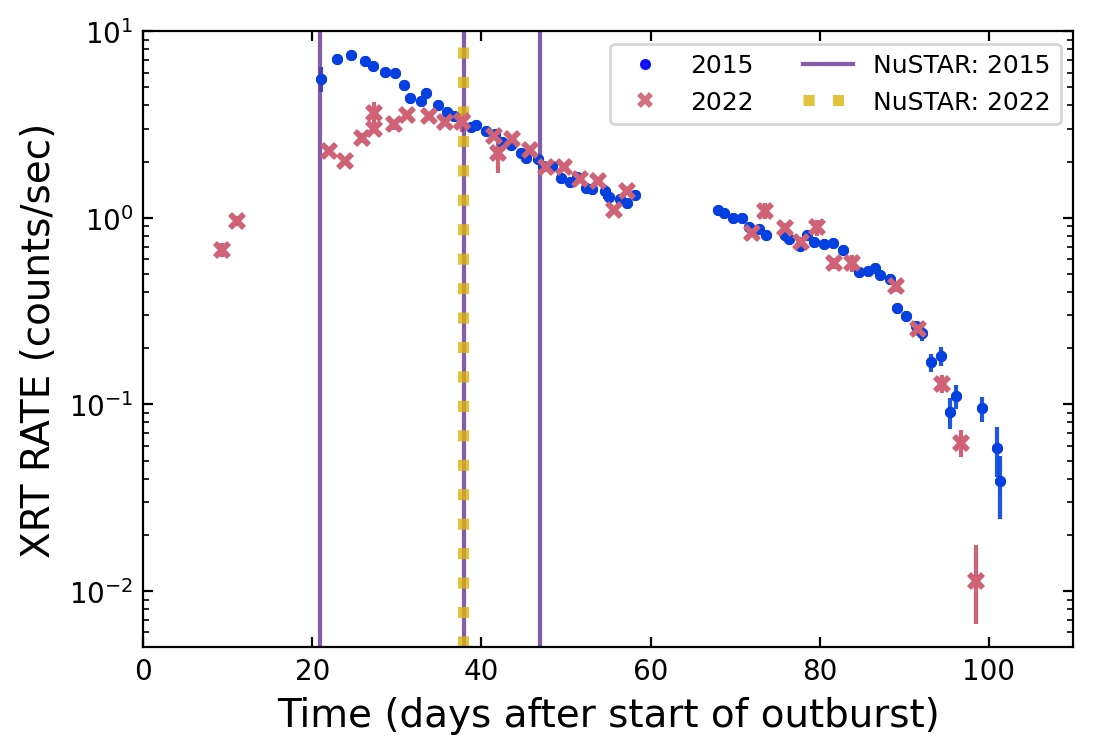}
    \includegraphics[width=0.47\columnwidth]{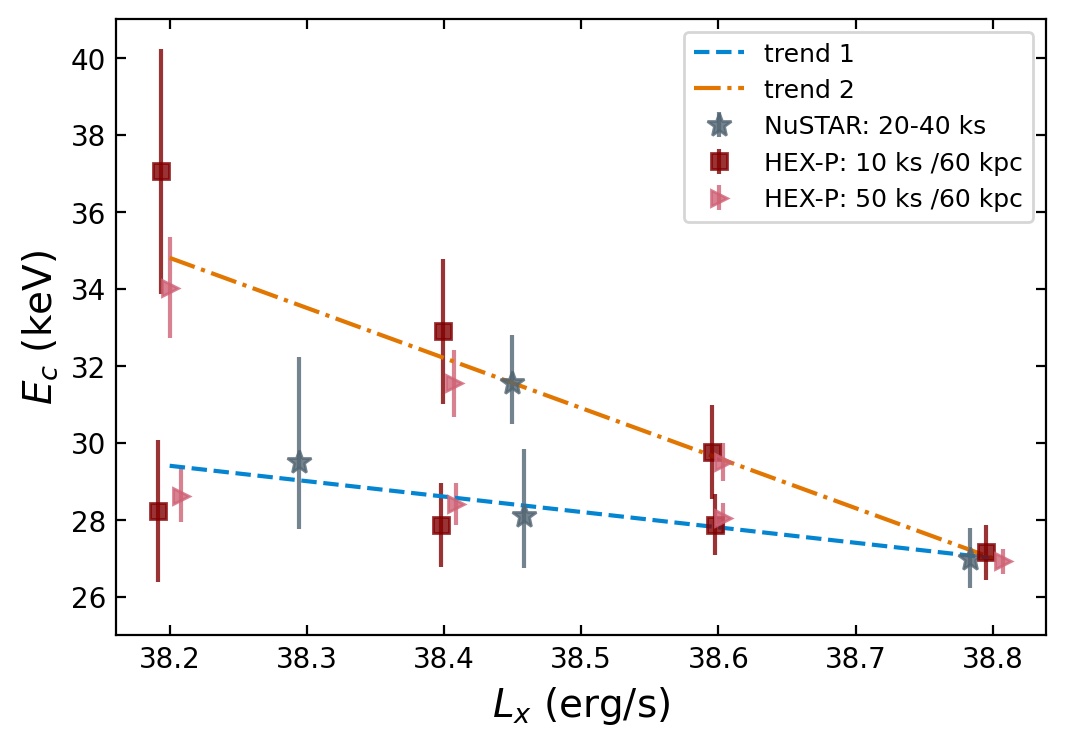}
   \caption{\emph{Left:} The 2015 and 2022 super-critical outbursts of SMC~X-2 observed by \swift/XRT. The vertical lines mark the epochs of four \nustar observations of the system. \emph{Right:} The cyclotron line dependence on luminosity. The star symbols mark the measured CRSF values for SMC~X-2 for the \nustar observations \citep{Jaisawal2023}. Note that the \nustar points on the trend 1 line are from the 2015 outburst, while the point on the trend 2 line was measured during the 2022 outburst; hence different cyclotron line energies can be measured even when the source is at nearly the same luminosity during outburst indicating an inherent difference between outbursts. We demonstrate \hexp's capabilities to discern between different possible trends in cyclotron line energy as a function of luminosity in these types of systems in as little at 10~ks, though greatly improved for 50 ks exposures.}
    \label{fig:smcx2}
\end{figure*}

SMC~X-2 is a good example to demonstrate \hexp's capabilities to detect and follow the evolution of a CRSF during a bright outburst. This BeXRB pulsar has exhibited two outbursts which reached luminosities above the Eddington limit 
over the last decade, one in 2015 and another in 2022. Both outbursts were followed up with \swift/XRT, as well as with \nustar ToOs (three observations in 2015 and one in 2022), covering a broad range in luminosity of $(2-6)\times10^{38}\,\mathrm{erg}\,\mathrm{s}^{-1}$ (see Fig.~\ref{fig:smcx2}, left panel). When comparing the CRSF energy for the two outbursts at the same luminosity level, the line energy was about 2\,keV higher during the 2022 outburst than observed previously in 2015 (see Fig. \ref{fig:smcx2}, right panel). This corresponds to a difference in magnetic field strength of $B\sim2\times10^{11}$\,G, akin to the difference reported by \citet{Cusumano2016} for V\,0332$+$53. 

In order to test \hexp's capabilities to detect CRSF features for extragalactic sources and determine if the data would be sufficient to distinguish trends with luminosity, we performed several simulations based on the spectral properties of SMC~X-2 as measured by \nustar \citep{Jaisawal2023}. 
For all simulations we used both HET and LET detectors. For the continuum we used an absorbed \citep[\textsc{tbabs} with $N_{\rm H}=10^{20}\ {\rm cm}^{-2}$;][]{Wilms2000} cut-off power-law model (i.e. \textsc{cutoffpl}) and a cyclotron absorption feature with the \textsc{gabs} model. We limited our simulations to the super-critical regime and a bolometric luminosity of 1.5-6$\times10^{38}$\,\lumcgs  which yields an absorbed flux range of 2-8$\times10^{-10}$\,\fluxcgs (1--50 keV). 
We simulated spectra where the CRSF energy varied with luminosity assuming two arbitrary inverse linear energy dependencies (trend 1 and trend 2 in Fig.~\ref{fig:smcx2}). We then fit the spectra and derived the best-fit parameters for the line with corresponding uncertainties. 
The results are shown in Figure~\ref{fig:smcx2} (right panel). 
\hexp can constrain the CRSF parameters (here the central line energy) at least two times better than \nustar and with an exposure time (10\,ks) that is only between half and a fourth that of \nustar, as well as discern between different trend inputs.
For a longer exposure time of 50\,ks the line parameters can be measured with unparalleled accuracy at the distance of the SMC. Thus, \hexp will be able to detect and follow cyclotron line evolution for super-critical outbursts in the Magellanic Clouds and thereby test competing physical models \citep[i.e. similar to V\,0332$+$53][]{Cusumano2016,Vybornov2018}. 
Note that the flux regime for which these trends are simulated (see Table~S1) and the science cases of the previous subsections suggest that trends in CRSF energy can also be probed at sub-critical through super-critical luminosities in galactic sources.
For science that can be executed with \hexp regarding pulsations and cyclotron lines in extragalactic ultraluminous X-ray (ULX) sources which reach even higher luminosities of $>100\times$ their Eddington limit, we refer the reader to Bachetti et al.\ ($accepted$).

In addition, it is expected that accreting XRPs crossing the critical luminosity should show a reversal of the cyclotron line energy versus luminosity correlation. To date only marginal evidence has been observed for such a trend reversal and in only a couple of sources \citep{Doroshenko2017, Malacaria2022}. As mentioned above, the Magellanic Clouds host comparatively many transient accreting pulsars displaying high-luminosity outbursts. 
\hexp's ability to constrain the cyclotron line energy with unprecedented accuracy across orders of magnitude in luminosity (and thus over distinct accretion regimes) will provide us with the opportunity to trace more and well defined trend reversals. This will allow us to constrain the critical luminosity for a given source which, in turn, places constraints on physical parameters of the observed system, such as the NS magnetic field strength or source distance \citep{Becker2012}.

\section{Conclusion}

We have presented open questions about NSs and accretion onto strongly magnetized sources, and demonstrated \hexp's unique ability to address these questions. The broad X-ray passband, improved sensitivity, and low X-ray background make \hexp ideally suited to understand accretion in X-ray soft sources, down into low accretion rate regimes. 
In particular for LMXBs, we have shown that \hexp observations will discriminate between competing continuum emission models that diverge above 30\,keV; an energy band inaccessible to current focusing X-ray telescopes for these soft spectrum sources. Additionally, leveraging the improved sensitivity and broad X-ray passband, \hexp will achieve tighter constraints on NS radius measurements through reflection modeling. These measurements are complementary and independent to other methodologies, and will narrow the allowed region on the NS mass-radius plane for viable EoS models for ultra-dense, cold matter, providing fundamental physics information in a regime inaccessible to terrestrial laboratories. 

For the case of NSs in HMXBs, \hexp will vastly improve our ability to: 
1) detect multiple cyclotron line features in a single observation, aiding in our understanding of the magnetic field strength in different regions of the accretion column and the poorly known physical mechanisms behind the formation of higher harmonic features; 
2) obtain detailed phase-resolved spectra to track the dependence of  CRSFs on pulse phase in order to explore the geometric configuration of the accretion column emission in unprecedented details;
3) constrain the surface magnetic field strength of accreting NSs at the lowest accretion regime and characterize the continuum spectral formation to provide updated physically-motivated spectral models for future observations; and 
4) identify CRSFs in extragalactic sources and follow the evolution of their line energy as a function of luminosity in order to distinguish between competing theories regarding changing emission region geometry or accretion-induced decay of the NS $B$-field.

\hexp will provide a new avenue for testing magnetic field configuration, emission pattern, and accretion column physics close to the surface of NSs, as well as enhance our understanding of extreme accretion physics in both LMXBs and HMXBs.

\section*{Author Contributions}

R.M.L.\ is responsible for the creation of the manuscript, authoring the abstract, \S1--2, \S4, editing \S3, graphic rendering of Figure~\ref{fig:XRB}, designing and executing the simulations for continuum modeling (\S2.2.1) and determining \rin and spin for an accreting NS LMXB (\S2.2.2).
A.W.S.\ wrote scripts to conduct the simulations for determining \rin and spin for an accreting NS LMXB (\S2.2.2).
C.M.\ authored \S3.1 and contributed to define the science cases of \S3.
P.P.\ conducted the simulations and authored  \S3.2.1.
F.F.\ conducted the simulations for CRSF phase dependence through phase resolved spectroscopy as described in \S3.2.2.
E.S.L.\ conducted the simulations and authored \S3.2.3, as well as created the figure for cyclotron line creation (Fig.~\ref{fig:cycf}).
G.V.\ conducted the simulations and authored  \S3.2.4. 
K.P.\ contributed to the science cases of  \S3.
J.W.\ contributed to the science cases of \S3.
All authors contributed to the final editing of the manuscript.

\section*{Funding}

The material is based upon work supported by NASA under award number 80GSFC21M0002.
G.V.\ acknowledges support by the Hellenic Foundation for Research and Innovation, (H.F.R.I.) through the project ASTRAPE (Project ID 7802). The work of D.S.\ was carried out at the Jet Propulsion Laboratory, California Institute of Technology, under a contract with NASA.
E.S.L.\ and J.W.\ acknowledge partial funding under Deutsche Forschungsgemeinschaft grant WI 1860/11-2 and Deutsches Zentrum f\"ur Luft- und Raumfahrt grant 50 QR 2202.

\section*{Acknowledgments}

R.M.L.\ and A.W.S.\ would like to thank Dr.\ Kazuhiko Shinki for providing consultation on statistical methods of \S 2.2.2. The authors thank the reviewers for their detailed comments that enhanced the science cases presented within the manuscript. 

\section*{Supplemental Material}
Here we provide the input values used for creating the various {\it HEX-P} simulations that are presented in the accompanying manuscript. Additionally, we specify the flux levels of the simulations since the introduction mentions a pile-up limit for the LET in units of mCrab. 

\setcounter{table}{0}
\renewcommand{\thetable}{S\arabic{table}}

\begin{table}[h!]
\centering
\caption{Flux level for each {\it HEX-P} simulation based on different source cases.}
\vspace{5pt}
\label{tab:flux}
\begin{tabular}{llcc}
 \hline
 Source & Type & $F_{\rm{2-10\,keV}}$ (mCrab) & $F_{\rm{2-80\,keV}}$  (erg cm$^{-2}$ s$^{-1}$)\\
 \hline
 4U 1735$-$44 & LMXB, atoll & 61 & $2.0\times10^{-9}$ \\
 Cygnus X-2 & LMXB, Z & 763 & $2.1\times10^{-8}$\\
 Cen X-3 & HMXB & 36 & $1.0\times10^{-8}$ \\
 Vela X-1 & HMXB & 120 & $6.7\times10^{-9}$\\
 GX 304$-$1 & HMXB, Be XRB & 0.4 & $1.2\times10^{-11}$\\
 SMC X-2 & HMXB, extragalactic & 2-8.3 & 2-8$\times10^{-10}$\\
 \hline
\end{tabular}

\end{table}

\begin{table}[h!]
\centering
\caption{Input spectral parameters for the two continuum models used to demonstrate {\it HEX-P}'s capabilities to distinguish continuum shapes in NS LMXBs in section 2.2.1. The int\_type parameter of {\sc nthcomp} in Model 2 is set to 0 so that the seed photons come from a single-temperature blackbody.}
\vspace{5pt}
\label{tab:4U1735}
\begin{tabular}{lcc|lcc}
 \hline
 \multicolumn{3}{c}{Model 1} & \multicolumn{3}{c}{Model 2}\\
 Model & Parameter & Value & Model & Parameter & Value\\
 \hline
 {\sc tbabs} &
 $N_\text{H}$ ($10^{22}\,\text{cm}^{-2}$) & 0.4  &  {\sc tbabs} &
 $N_\text{H}$ ($10^{22}\,\text{cm}^{-2}$) & 0.4\\

{\sc diskbb} & kT (keV) & 1.26 & {\sc diskbb} & kT (keV) & 0.68\\
 & norm & 28 &  & norm & 350\\
{\sc bbody} & kT (keV) & 2.43 & {\sc nthcomp} & $\Gamma$ & 1.97\\
& norm & $1.02\times10^{-2}$ &  & $kT_{\rm e}$ (keV) & 3.13\\
{\sc pow} & $\Gamma$ & 2.57 &  & $kT_{\rm bb}$ (keV) & 1.07\\
& norm & 0.45 &  & norm & $5.9\times10^{-2}$\\

 \hline
\end{tabular}

\end{table}

\begin{table}[h!]
\centering
\caption{Spectral parameters for the Vela X-1 simulations, see also Figure~9. The model can be written as \texttt{tbabs(2)*((powerlaw(1)*fdcut(1))*gabs(1)*gabs(2)+} \texttt{+egauss(1)+egauss(3)+egauss(10)+egauss(11)+egauss(12)+egauss(13)+} \texttt{+egauss(100))*(constant(10)*tbabs(1)+(1-constant(10)))}.}
\vspace{5pt}
\label{tab:velax1pars}
\begin{tabular}{lll}
 \hline
 Parameter & Value & Unit \\
 \hline
 $N_\text{H,1}$ & 0.37 & $10^{22}\,\text{cm}^{-2}$ \\
 Normalization & 0.33 & ph\,keV$^{-1}$\,cm$^{-2}$\,s${-1}$ at 1\,keV \\
 $\Gamma$ & 1.08 &  \\
 $E_\text{fold}$ & 26.1 & keV \\
 $E_\text{cut}$ & 10.2 & keV \\
 $E_\text{CRSF,2}$ & 56.8 & keV \\
 $\sigma_\text{CRSF,2}$ & 8.75 & keV \\
 $d_\text{CRSF,2}$ & 21.7 &  \\
 $E_\text{CRSF,1}$ & 24.3 & keV \\
 $\sigma_\text{CRSF,2}$ & 4.37 & keV \\
 $d_\text{CRSF,2}$ & 0.49 &  \\
 $A_{\text{Fe K}\alpha}$ & 5.83 & $10^{-3}$ ph\,s$^{-1}$\,cm$^{-2}$ \\
 $E_{\text{Fe K}\alpha}$ & 6.49 & keV \\
 $\sigma_{\text{Fe K}\alpha}$ & 0.08 & keV \\
 $A_{\text{Fe K}\beta}$ & 2.45 & eV \\
 $E_{\text{Fe K}\beta}$ & 7.14 & keV \\
 $\sigma_{\text{Fe K}\beta}$ & 0.09 & keV \\
 $A_{\text{Ne IX}\alpha}$ & 3.58 & $10^{-3}$ ph\,s$^{-1}$\,cm$^{-2}$ \\
 $E_{\text{Ne IX}\alpha}$ & 0.93 & keV \\
 $\sigma_{\text{Ne IX}\alpha}$ & 0.10 & keV \\
 $A_{\text{Mg L}\alpha}$ & 1.95 & $10^{-3}$ ph\,s$^{-1}$\,cm$^{-2}$ \\
 $E_{\text{Mg L}\alpha}$ & 1.34 & keV \\
 $\sigma_{\text{Mg L}\alpha}$ & 0.10 & keV \\
 $A_{\text{Si K}\alpha}$ & 5.74 & $10^{-4}$ ph\,s$^{-1}$\,cm$^{-2}$ \\
 $E_{\text{Si K}\alpha}$ & 1.83 & keV \\
 $\sigma_{\text{Si K}\alpha}$ & 15.5 & eV \\
 $A_{\text{S K}\alpha}$ & 4.97 & $10^{-4}$ ph\,s$^{-1}$\,cm$^{-2}$ \\
 $E_{\text{S K}\alpha}$ & 2.44 & keV \\
 $\sigma_{\text{S K}\alpha}$ & 0.05 & keV \\
 Cov. Frac. & 0.50 &  \\
 $N_\text{H,2}$ & 32.1 & $10^{22}\,\text{cm}^{-2}$ \\
 $A_{\text{10keV}}$ & -1.25 & $10^{-2}$ ph\,s$^{-1}$\,cm$^{-2}$ \\
 $E_{\text{10keV}}$ & 9.13 & keV \\
 $\sigma_{\text{10keV}}$ & 2.34 & keV \\
 \hline
\end{tabular}

\end{table}

\newpage

\bibliographystyle{Frontiers-Harvard} 

\bibliography{references}

\begin{thebibliography}{158}
\providecommand{\natexlab}[1]{#1}
\expandafter\ifx\csname urlstyle\endcsname\relax
  \providecommand{\doi}[1]{doi:\discretionary{}{}{}#1}\else
  \providecommand{\doi}{doi:\discretionary{}{}{}\begingroup \urlstyle{rm}\Url}\fi
\providecommand{\selectlanguage}[1]{\relax}
\providecommand{\bibAnnoteFile}[1]{%
  \IfFileExists{#1}{\begin{quotation}\noindent\textsc{Key:} #1\\
  \textsc{Annotation:}\ \input{#1}\end{quotation}}{}}
\providecommand{\bibAnnote}[2]{%
  \begin{quotation}\noindent\textsc{Key:} #1\\
  \textsc{Annotation:}\ #2\end{quotation}}

\bibitem[{{Abbott} et~al.(2019){Abbott}, {Abbott}, {Abbott}, {Acernese}, {Ackley}, {Adams} et~al.}]{abbott19}
{Abbott}, B.~P., {Abbott}, R., {Abbott}, T.~D., {Acernese}, F., {Ackley}, K., {Adams}, C., et~al. (2019).
\newblock {Properties of the Binary Neutron Star Merger GW170817}.
\newblock \emph{Physical Review X} 9, 011001.
\newblock \doi{10.1103/PhysRevX.9.011001}
\bibAnnoteFile{abbott19}

\bibitem[{{Aftab} et~al.(2019){Aftab}, {Paul}, and {Kretschmar}}]{aftab2019}
{Aftab}, N., {Paul}, B., and {Kretschmar}, P. (2019).
\newblock {X-Ray Reprocessing: Through the Eclipse Spectra of High-mass X-Ray Binaries with XMM-Newton}.
\newblock \emph{\apjs} 243, 29.
\newblock \doi{10.3847/1538-4365/ab2a77}
\bibAnnoteFile{aftab2019}

\bibitem[{{Alexander} and {Meszaros}(1991)}]{Alexander1991}
{Alexander}, S.~G. and {Meszaros}, P. (1991).
\newblock {Cyclotron Harmonics in Accreting Pulsars and Gamma-Ray Bursters: Effect of Two-Photon Processes}.
\newblock \emph{\apj} 372, 565.
\newblock \doi{10.1086/170001}
\bibAnnoteFile{Alexander1991}

\bibitem[{{Arnaud}(1996)}]{arnaud96}
{Arnaud}, K.~A. (1996).
\newblock {XSPEC: The First Ten Years}.
\newblock In \emph{Astronomical Data Analysis Software and Systems V}, eds. G.~H. {Jacoby} and J.~{Barnes} (San Francisco: Astron.\ Soc.\ Pacific), vol. 101 of \emph{Astron.\ Soc.\ Pacific Conf.\ Ser.}, 17
\bibAnnoteFile{arnaud96}

\bibitem[{{Bala} et~al.(2020){Bala}, {Bhattacharya}, {Staubert}, and {Maitra}}]{Bala2020}
{Bala}, S., {Bhattacharya}, D., {Staubert}, R., and {Maitra}, C. (2020).
\newblock {Time evolution of cyclotron line of Her X-1: a detailed statistical analysis including new AstroSat data}.
\newblock \emph{\mnras} 497, 1029--1042.
\newblock \doi{10.1093/mnras/staa1988}
\bibAnnoteFile{Bala2020}

\bibitem[{{Bardeen} et~al.(1972){Bardeen}, {Press}, and {Teukolsky}}]{bardeen72}
{Bardeen}, J.~M., {Press}, W.~H., and {Teukolsky}, S.~A. (1972).
\newblock {Rotating Black Holes: Locally Nonrotating Frames, Energy Extraction, and Scalar Synchrotron Radiation}.
\newblock \emph{\apj} 178, 347--370.
\newblock \doi{10.1086/151796}
\bibAnnoteFile{bardeen72}

\bibitem[{{Barret}(2001)}]{barret01}
{Barret}, D. (2001).
\newblock {The broad band x-ray/hard x-ray spectra of accreting neutron stars}.
\newblock \emph{Adv.\ Space Res.} 28, 307--321.
\newblock \doi{10.1016/S0273-1177(01)00414-8}
\bibAnnoteFile{barret01}

\bibitem[{{Barret} et~al.(2000){Barret}, {Olive}, {Boirin}, {Done}, {Skinner}, and {Grindlay}}]{barret00}
{Barret}, D., {Olive}, J.~F., {Boirin}, L., {Done}, C., {Skinner}, G.~K., and {Grindlay}, J.~E. (2000).
\newblock {Hard X-Ray Emission from Low-Mass X-Ray Binaries}.
\newblock \emph{\apj} 533, 329--351.
\newblock \doi{10.1086/308651}
\bibAnnoteFile{barret00}

\bibitem[{{Basko} and {Sunyaev}(1976)}]{Basko1976}
{Basko}, M.~M. and {Sunyaev}, R.~A. (1976).
\newblock {The limiting luminosity of accreting neutron stars with magnetic fields.}
\newblock \emph{\mnras} 175, 395--417.
\newblock \doi{10.1093/mnras/175.2.395}
\bibAnnoteFile{Basko1976}

\bibitem[{{Becker} et~al.(2012){Becker}, {Klochkov}, {Sch{\"o}nherr}, {Nishimura}, {Ferrigno}, {Caballero} et~al.}]{Becker2012}
{Becker}, P.~A., {Klochkov}, D., {Sch{\"o}nherr}, G., {Nishimura}, O., {Ferrigno}, C., {Caballero}, I., et~al. (2012).
\newblock {Spectral formation in accreting X-ray pulsars: bimodal variation of the cyclotron energy with luminosity}.
\newblock \emph{\aap} 544, A123.
\newblock \doi{10.1051/0004-6361/201219065}
\bibAnnoteFile{Becker2012}

\bibitem[{{Becker} and {Wolff}(2022)}]{Becker2022}
{Becker}, P.~A. and {Wolff}, M.~T. (2022).
\newblock {A Generalized Analytical Model for Thermal and Bulk Comptonization in Accretion-powered X-Ray Pulsars}.
\newblock \emph{\apj} 939, 67.
\newblock \doi{10.3847/1538-4357/ac8d95}
\bibAnnoteFile{Becker2022}

\bibitem[{{Bozzo} et~al.(2022){Bozzo}, {Romano}, {Ferrigno}, and {Oskinova}}]{bozzo22}
{Bozzo}, E., {Romano}, P., {Ferrigno}, C., and {Oskinova}, L. (2022).
\newblock {The symbiotic X-ray binaries Sct X-1, 4U 1700+24, and IGR J17329-2731}.
\newblock \emph{\mnras} 513, 42--54.
\newblock \doi{10.1093/mnras/stac907}
\bibAnnoteFile{bozzo22}

\bibitem[{{Caballero} and {Wilms}(2012)}]{caballero12}
{Caballero}, I. and {Wilms}, J. (2012).
\newblock {X-ray pulsars: a review.}
\newblock \emph{\memsai} 83, 230.
\newblock \doi{10.48550/arXiv.1206.3124}
\bibAnnoteFile{caballero12}

\bibitem[{{Cackett} et~al.(2009){Cackett}, {Altamirano}, {Patruno}, {Miller}, {Reynolds}, {Linares} et~al.}]{cackett09}
{Cackett}, E.~M., {Altamirano}, D., {Patruno}, A., {Miller}, J.~M., {Reynolds}, M., {Linares}, M., et~al. (2009).
\newblock {Broad Relativistic Iron Emission Line Observed in SAX J1808.4$-$3658}.
\newblock \emph{\apjl} 694, L21--L25.
\newblock \doi{10.1088/0004-637X/694/1/L21}
\bibAnnoteFile{cackett09}

\bibitem[{{Cackett} et~al.(2008){Cackett}, {Miller}, {Bhattacharyya}, {Grindlay}, {Homan}, {van der Klis} et~al.}]{cackett08}
{Cackett}, E.~M., {Miller}, J.~M., {Bhattacharyya}, S., {Grindlay}, J.~E., {Homan}, J., {van der Klis}, M., et~al. (2008).
\newblock {Relativistic Iron Emission Lines in Neutron Star Low-Mass X-Ray Binaries as Probes of Neutron Star Radii}.
\newblock \emph{\apj} 674, 415--420.
\newblock \doi{10.1086/524936}
\bibAnnoteFile{cackett08}

\bibitem[{{Caiazzo} and {Heyl}(2021)}]{Caiazzo2021}
{Caiazzo}, I. and {Heyl}, J. (2021).
\newblock {Polarization of accreting X-ray pulsars - II. Hercules X-1}.
\newblock \emph{\mnras} 501, 129--136.
\newblock \doi{10.1093/mnras/staa3429}
\bibAnnoteFile{Caiazzo2021}

\bibitem[{{Casares} et~al.(2010){Casares}, {Gonz{\'a}lez Hern{\'a}ndez}, {Israelian}, and {Rebolo}}]{casares10}
{Casares}, J., {Gonz{\'a}lez Hern{\'a}ndez}, J.~I., {Israelian}, G., and {Rebolo}, R. (2010).
\newblock {On the mass of the neutron star in Cyg X-2}.
\newblock \emph{\mnras} 401, 2517--2520.
\newblock \doi{10.1111/j.1365-2966.2009.15828.x}
\bibAnnoteFile{casares10}

\bibitem[{{Chodil} et~al.(1967){Chodil}, {Mark}, {Rodrigues}, {Seward}, and {Swift}}]{Chodil}
{Chodil}, G., {Mark}, H., {Rodrigues}, R., {Seward}, F.~D., and {Swift}, C.~D. (1967).
\newblock {X-Ray Intensities and Spectra from Several Cosmic Sources}.
\newblock \emph{\apj} 150, 57.
\newblock \doi{10.1086/149312}
\bibAnnoteFile{Chodil}

\bibitem[{{Church} and {Baluci{\'n}ska-Church}(2001)}]{church01}
{Church}, M.~J. and {Baluci{\'n}ska-Church}, M. (2001).
\newblock {Results of a LMXB survey: Variation in the height of the neutron star blackbody emission region}.
\newblock \emph{\aap} 369, 915--924.
\newblock \doi{10.1051/0004-6361:20010150}
\bibAnnoteFile{church01}

\bibitem[{{Cocchi} et~al.(2023){Cocchi}, {Gnarini}, {Fabiani}, {Ursini}, {Poutanen}, {Capitanio} et~al.}]{Cocchi23}
{Cocchi}, M., {Gnarini}, A., {Fabiani}, S., {Ursini}, F., {Poutanen}, J., {Capitanio}, F., et~al. (2023).
\newblock {Discovery of strongly variable X-ray polarization in the neutron star low-mass X-ray binary transient XTE J1701{\ensuremath{-}}462}.
\newblock \emph{\aap} 674, L10.
\newblock \doi{10.1051/0004-6361/202346275}
\bibAnnoteFile{Cocchi23}

\bibitem[{{Cusumano} et~al.(2016){Cusumano}, {La Parola}, {D'A{\`\i}}, {Segreto}, {Tagliaferri}, {Barthelmy} et~al.}]{Cusumano2016}
{Cusumano}, G., {La Parola}, V., {D'A{\`\i}}, A., {Segreto}, A., {Tagliaferri}, G., {Barthelmy}, S.~D., et~al. (2016).
\newblock {An unexpected drop in the magnetic field of the X-ray pulsar V0332+53 after the bright outburst occurred in 2015}.
\newblock \emph{\mnras} 460, L99--L103.
\newblock \doi{10.1093/mnrasl/slw084}
\bibAnnoteFile{Cusumano2016}

\bibitem[{{Dauser} et~al.(2013){Dauser}, {Garc\'ia}, {Wilms}, {B{\"o}ck}, {Brenneman}, {Falanga} et~al.}]{Dauser2013}
{Dauser}, T., {Garc\'ia}, J., {Wilms}, J., {B{\"o}ck}, M., {Brenneman}, L.~W., {Falanga}, M., et~al. (2013).
\newblock {Irradiation of an accretion disc by a jet: general properties and implications for spin measurements of black holes}.
\newblock \emph{\mnras} 430, 1694--1708.
\newblock \doi{10.1093/mnras/sts710}
\bibAnnoteFile{Dauser2013}

\bibitem[{{Davies} and {Pringle}(1981)}]{DaviesPringle1981}
{Davies}, R.~E. and {Pringle}, J.~E. (1981).
\newblock {Spindown of neutron stars in close binary systems - II.}
\newblock \emph{\mnras} 196, 209--224.
\newblock \doi{10.1093/mnras/196.2.209}
\bibAnnoteFile{DaviesPringle1981}

\bibitem[{{Degenaar} et~al.(2018){Degenaar}, {Ballantyne}, {Belloni}, {Chakraborty}, {Chen}, {Ji} et~al.}]{Degenaar18}
{Degenaar}, N., {Ballantyne}, D.~R., {Belloni}, T., {Chakraborty}, M., {Chen}, Y.-P., {Ji}, L., et~al. (2018).
\newblock {Accretion Disks and Coronae in the X-Ray Flashlight}.
\newblock \emph{Space~Sci.~Rev.} 214, 15.
\newblock \doi{10.1007/s11214-017-0448-3}
\bibAnnoteFile{Degenaar18}

\bibitem[{{Di Salvo} et~al.(1998){Di Salvo}, {Burderi}, {Robba}, and {Guainazzi}}]{DiSalvo1998}
{Di Salvo}, T., {Burderi}, L., {Robba}, N.~R., and {Guainazzi}, M. (1998).
\newblock {The Two-Component X-Ray Broadband Spectrum of X Persei Observed by BeppoSAX}.
\newblock \emph{\apj} 509, 897--903.
\newblock \doi{10.1086/306525}
\bibAnnoteFile{DiSalvo1998}

\bibitem[{{Diez} et~al.(2023){Diez}, {Grinberg}, {F{\"u}rst}, {El Mellah}, {Zhou}, {Santangelo} et~al.}]{Diez2023}
{Diez}, C.~M., {Grinberg}, V., {F{\"u}rst}, F., {El Mellah}, I., {Zhou}, M., {Santangelo}, A., et~al. (2023).
\newblock {Observing the onset of the accretion wake in Vela X-1}.
\newblock \emph{\aap} 674, A147.
\newblock \doi{10.1051/0004-6361/202245708}
\bibAnnoteFile{Diez2023}

\bibitem[{{Diez} et~al.(2022){Diez}, {Grinberg}, {F{\"u}rst}, {Sokolova-Lapa}, {Santangelo}, {Wilms} et~al.}]{Diez2022}
{Diez}, C.~M., {Grinberg}, V., {F{\"u}rst}, F., {Sokolova-Lapa}, E., {Santangelo}, A., {Wilms}, J., et~al. (2022).
\newblock {Continuum, cyclotron line, and absorption variability in the high-mass X-ray binary Vela X-1}.
\newblock \emph{\aap} 660, A19.
\newblock \doi{10.1051/0004-6361/202141751}
\bibAnnoteFile{Diez2022}

\bibitem[{{Doroshenko} et~al.(2020){Doroshenko}, {Piraino}, {Doroshenko}, and {Santangelo}}]{Doroshenko2020}
{Doroshenko}, R., {Piraino}, S., {Doroshenko}, V., and {Santangelo}, A. (2020).
\newblock {Revisiting BeppoSAX and NuSTAR observations of KS 1947+300 and the missing cyclotron line}.
\newblock \emph{\mnras} 493, 3442--3448.
\newblock \doi{10.1093/mnras/staa490}
\bibAnnoteFile{Doroshenko2020}

\bibitem[{{Doroshenko} et~al.(2012){Doroshenko}, {Santangelo}, {Kreykenbohm}, and {Doroshenko}}]{Doroshenko2012}
{Doroshenko}, V., {Santangelo}, A., {Kreykenbohm}, I., and {Doroshenko}, R. (2012).
\newblock {The hard X-ray emission of X Persei}.
\newblock \emph{\aap} 540, L1.
\newblock \doi{10.1051/0004-6361/201218878}
\bibAnnoteFile{Doroshenko2012}

\bibitem[{{Doroshenko} et~al.(2021){Doroshenko}, {Santangelo}, {Tsygankov}, and {Ji}}]{Doroshenko2021}
{Doroshenko}, V., {Santangelo}, A., {Tsygankov}, S.~S., and {Ji}, L. (2021).
\newblock {SGR 0755{\ensuremath{-}}2933: a new high-mass X-ray binary with the wrong name}.
\newblock \emph{\aap} 647, A165.
\newblock \doi{10.1051/0004-6361/202039785}
\bibAnnoteFile{Doroshenko2021}

\bibitem[{{Doroshenko} et~al.(2022){Doroshenko}, {Staubert}, {Maitra}, {Rau}, {Haberl}, {Santangelo} et~al.}]{Doroshenko2022}
{Doroshenko}, V., {Staubert}, R., {Maitra}, C., {Rau}, A., {Haberl}, F., {Santangelo}, A., et~al. (2022).
\newblock {SRGA J124404.1-632232/SRGU J124403.8-632231: New X-ray pulsar discovered in the all-sky survey by the SRG}.
\newblock \emph{\aap} 661, A21.
\newblock \doi{10.1051/0004-6361/202141147}
\bibAnnoteFile{Doroshenko2022}

\bibitem[{{Doroshenko} et~al.(2017){Doroshenko}, {Tsygankov}, {Mushtukov}, {Lutovinov}, {Santangelo}, {Suleimanov} et~al.}]{Doroshenko2017}
{Doroshenko}, V., {Tsygankov}, S.~S., {Mushtukov}, A.~A., {Lutovinov}, A.~A., {Santangelo}, A., {Suleimanov}, V.~F., et~al. (2017).
\newblock {Luminosity dependence of the cyclotron line and evidence for the accretion regime transition in V 0332+53}.
\newblock \emph{\mnras} 466, 2143--2150.
\newblock \doi{10.1093/mnras/stw3236}
\bibAnnoteFile{Doroshenko2017}

\bibitem[{{Eraerds} et~al.(2021){Eraerds}, {Antonelli}, {Davis}, {Hall}, {Hetherington}, {Holland} et~al.}]{Eraerds2021}
{Eraerds}, T., {Antonelli}, V., {Davis}, C., {Hall}, D., {Hetherington}, O., {Holland}, A., et~al. (2021).
\newblock {Enhanced simulations on the Athena/Wide Field Imager instrumental background}.
\newblock \emph{Journal of Astronomical Telescopes, Instruments, and Systems} 7, 034001.
\newblock \doi{10.1117/1.JATIS.7.3.034001}
\bibAnnoteFile{Eraerds2021}

\bibitem[{{Fabian} et~al.(1989){Fabian}, {Rees}, {Stella}, and {White}}]{Fabian1989}
{Fabian}, A.~C., {Rees}, M.~J., {Stella}, L., and {White}, N.~E. (1989).
\newblock {X-ray fluorescence from the inner disc in Cygnus X-1}.
\newblock \emph{\mnras} 238, 729--736
\bibAnnoteFile{Fabian1989}

\bibitem[{{Farinelli} et~al.(2023){Farinelli}, {Fabiani}, {Poutanen}, {Ursini}, {Ferrigno}, {Bianchi} et~al.}]{Farinelli23}
{Farinelli}, R., {Fabiani}, S., {Poutanen}, J., {Ursini}, F., {Ferrigno}, C., {Bianchi}, S., et~al. (2023).
\newblock {Accretion geometry of the neutron star low mass X-ray binary Cyg X-2 from X-ray polarization measurements}.
\newblock \emph{\mnras} 519, 3681--3690.
\newblock \doi{10.1093/mnras/stac3726}
\bibAnnoteFile{Farinelli23}

\bibitem[{{Farinelli} et~al.(2016){Farinelli}, {Ferrigno}, {Bozzo}, and {Becker}}]{Farinelli16}
{Farinelli}, R., {Ferrigno}, C., {Bozzo}, E., and {Becker}, P.~A. (2016).
\newblock {A new model for the X-ray continuum of the magnetized accreting pulsars}.
\newblock \emph{\aap} 591, A29.
\newblock \doi{10.1051/0004-6361/201527257}
\bibAnnoteFile{Farinelli16}

\bibitem[{{Fiocchi} et~al.(2019){Fiocchi}, {Bazzano}, {Bruni}, {Ludlam}, {Natalucci}, {Onori} et~al.}]{fiocchi19}
{Fiocchi}, M., {Bazzano}, A., {Bruni}, G., {Ludlam}, R., {Natalucci}, L., {Onori}, F., et~al. (2019).
\newblock {Quasi-simultaneous INTEGRAL, SWIFT, and NuSTAR Observations of the New X-Ray Clocked Burster 1RXS J180408.9-342058}.
\newblock \emph{\apj} 887, 30.
\newblock \doi{10.3847/1538-4357/ab4d59}
\bibAnnoteFile{fiocchi19}

\bibitem[{{Forsblom} et~al.(2023){Forsblom}, {Poutanen}, {Tsygankov}, {Bachetti}, {Di Marco}, {Doroshenko} et~al.}]{Forsblom2023}
{Forsblom}, S.~V., {Poutanen}, J., {Tsygankov}, S.~S., {Bachetti}, M., {Di Marco}, A., {Doroshenko}, V., et~al. (2023).
\newblock {IXPE Observations of the Quintessential Wind-accreting X-Ray Pulsar Vela X-1}.
\newblock \emph{\apjl} 947, L20.
\newblock \doi{10.3847/2041-8213/acc391}
\bibAnnoteFile{Forsblom2023}

\bibitem[{{F{\"u}rst} et~al.(2018){F{\"u}rst}, {Falkner}, {Marcu-Cheatham}, {Grefenstette}, {Tomsick}, {Pottschmidt} et~al.}]{Furst2018}
{F{\"u}rst}, F., {Falkner}, S., {Marcu-Cheatham}, D., {Grefenstette}, B., {Tomsick}, J., {Pottschmidt}, K., et~al. (2018).
\newblock {Multiple cyclotron line-forming regions in GX 301$-$2}.
\newblock \emph{\aap} 620, A153.
\newblock \doi{10.1051/0004-6361/201732132}
\bibAnnoteFile{Furst2018}

\bibitem[{{F{\"u}rst} et~al.(2014){F{\"u}rst}, {Pottschmidt}, {Wilms}, {Tomsick}, {Bachetti}, {Boggs} et~al.}]{fuerst14}
{F{\"u}rst}, F., {Pottschmidt}, K., {Wilms}, J., {Tomsick}, J.~A., {Bachetti}, M., {Boggs}, S.~E., et~al. (2014).
\newblock {NuSTAR Discovery of a Luminosity Dependent Cyclotron Line Energy in Vela X-1}.
\newblock \emph{\apj} 780, 133.
\newblock \doi{10.1088/0004-637X/780/2/133}
\bibAnnoteFile{fuerst14}

\bibitem[{{Galloway} et~al.(2008){Galloway}, {Muno}, {Hartman}, {Psaltis}, and {Chakrabarty}}]{galloway2008}
{Galloway}, D.~K., {Muno}, M.~P., {Hartman}, J.~M., {Psaltis}, D., and {Chakrabarty}, D. (2008).
\newblock {Thermonuclear (Type I) X-Ray Bursts Observed by the Rossi X-Ray Timing Explorer}.
\newblock \emph{\apjs} 179, 360--422.
\newblock \doi{10.1086/592044}
\bibAnnoteFile{galloway2008}

\bibitem[{{Garc{\'{\i}}a} and {Kallman}(2010)}]{Garcia2010}
{Garc{\'{\i}}a}, J. and {Kallman}, T.~R. (2010).
\newblock {X-ray Reflected Spectra from Accretion Disk Models. I. Constant Density Atmospheres}.
\newblock \emph{\apj} 718, 695--706.
\newblock \doi{10.1088/0004-637X/718/2/695}
\bibAnnoteFile{Garcia2010}

\bibitem[{{Garc{\'\i}a} et~al.(2022){Garc{\'\i}a}, {Dauser}, {Ludlam}, {Parker}, {Fabian}, {Harrison} et~al.}]{Garcia22}
{Garc{\'\i}a}, J.~A., {Dauser}, T., {Ludlam}, R., {Parker}, M., {Fabian}, A., {Harrison}, F.~A., et~al. (2022).
\newblock {Relativistic X-Ray Reflection Models for Accreting Neutron Stars}.
\newblock \emph{\apj} 926, 13.
\newblock \doi{10.3847/1538-4357/ac3cb7}
\bibAnnoteFile{Garcia22}

\bibitem[{{Gehrels} et~al.(2004){Gehrels}, {Chincarini}, {Giommi}, {Mason}, {Nousek}, {Wells} et~al.}]{Gehrels04}
{Gehrels}, N., {Chincarini}, G., {Giommi}, P., {Mason}, K.~O., {Nousek}, J.~A., {Wells}, A.~A., et~al. (2004).
\newblock {The Swift Gamma-Ray Burst Mission}.
\newblock \emph{\apj} 611, 1005--1020.
\newblock \doi{10.1086/422091}
\bibAnnoteFile{Gehrels04}

\bibitem[{{Gendreau} et~al.(2012){Gendreau}, {Arzoumanian}, and {Okajima}}]{Gendreau12}
{Gendreau}, K.~C., {Arzoumanian}, Z., and {Okajima}, T. (2012).
\newblock {The Neutron star Interior Composition ExploreR (NICER): an Explorer mission of opportunity for soft x-ray timing spectroscopy}.
\newblock In \emph{Space Telescopes and Instrumentation 2012: Ultraviolet to Gamma Ray}, eds. T.~{Takahashi}, S.~S. {Murray}, and J.-W.~A. {den Herder} (Bellingham, WA: SPIE), vol. 8443 of \emph{Society of Photo-Optical Instrumentation Engineers (SPIE) Conference Series}, 844313.
\newblock \doi{10.1117/12.926396}
\bibAnnoteFile{Gendreau12}

\bibitem[{{Giacconi} et~al.(1971){Giacconi}, {Gursky}, {Kellogg}, {Schreier}, and {Tananbaum}}]{Giacconi}
{Giacconi}, R., {Gursky}, H., {Kellogg}, E., {Schreier}, E., and {Tananbaum}, H. (1971).
\newblock {Discovery of Periodic X-Ray Pulsations in Centaurus X-3 from UHURU}.
\newblock \emph{\apjl} 167, L67.
\newblock \doi{10.1086/180762}
\bibAnnoteFile{Giacconi}

\bibitem[{{Harrison} et~al.(2013){Harrison}, {Craig}, {Christensen}, {Hailey}, {Zhang}, {Boggs} et~al.}]{harrison13}
{Harrison}, F.~A., {Craig}, W.~W., {Christensen}, F.~E., {Hailey}, C.~J., {Zhang}, W.~W., {Boggs}, S.~E., et~al. (2013).
\newblock {The Nuclear Spectroscopic Telescope Array (NuSTAR) High-energy X-Ray Mission}.
\newblock \emph{\apj} 770, 103.
\newblock \doi{10.1088/0004-637X/770/2/103}
\bibAnnoteFile{harrison13}

\bibitem[{{Hasinger} and {van der Klis}(1989)}]{hasinger89}
{Hasinger}, G. and {van der Klis}, M. (1989).
\newblock {Two patterns of correlated X-ray timing and spectral behaviour in low-mass X-ray binaries.}
\newblock \emph{\aap} 225, 79--96
\bibAnnoteFile{hasinger89}

\bibitem[{{Heindl} et~al.(2004){Heindl}, {Rothschild}, {Coburn}, {Staubert}, {Wilms}, {Kreykenbohm} et~al.}]{Heindl2004}
{Heindl}, W.~A., {Rothschild}, R.~E., {Coburn}, W., {Staubert}, R., {Wilms}, J., {Kreykenbohm}, I., et~al. (2004).
\newblock {Timing and Spectroscopy of Accreting X-ray Pulsars: the State of Cyclotron Line Studies}.
\newblock In \emph{X-ray Timing 2003: Rossi and Beyond}, eds. P.~{Kaaret}, F.~K. {Lamb}, and J.~H. {Swank}. vol. 714 of \emph{American Institute of Physics Conference Series}, 323--330.
\newblock \doi{10.1063/1.1781049}
\bibAnnoteFile{Heindl2004}

\bibitem[{{Homan} et~al.(2010){Homan}, {van der Klis}, {Fridriksson}, {Remillard}, {Wijnands}, {M{\'e}ndez} et~al.}]{homan10}
{Homan}, J., {van der Klis}, M., {Fridriksson}, J.~K., {Remillard}, R.~A., {Wijnands}, R., {M{\'e}ndez}, M., et~al. (2010).
\newblock {XTE J1701$-$462 and Its Implications for the Nature of Subclasses in Low-magnetic-field Neutron Star Low-mass X-ray Binaries}.
\newblock \emph{\apj} 719, 201--212.
\newblock \doi{10.1088/0004-637X/719/1/201}
\bibAnnoteFile{homan10}

\bibitem[{{Houck} and {Denicola}(2000)}]{Houck}
{Houck}, J.~C. and {Denicola}, L.~A. (2000).
\newblock {ISIS: An Interactive Spectral Interpretation System for High Resolution X-Ray Spectroscopy}.
\newblock In \emph{Astronomical Data Analysis Software and Systems IX}, eds. N.~{Manset}, C.~{Veillet}, and D.~{Crabtree} (San Francisco: Astron.\ Soc.\ Pacific), vol. 216 of \emph{Astron.\ Soc.\ Pacific Conf.\ Ser.}, 591
\bibAnnoteFile{Houck}

\bibitem[{{Iaria} et~al.(2005){Iaria}, {Di Salvo}, {Robba}, {Burderi}, {Lavagetto}, and {Riggio}}]{iaria2005}
{Iaria}, R., {Di Salvo}, T., {Robba}, N.~R., {Burderi}, L., {Lavagetto}, G., and {Riggio}, A. (2005).
\newblock {Resolving the Fe \textsc{xxv} Triplet with Chandra in Centaurus X-3}.
\newblock \emph{\apjl} 634, L161--L164.
\newblock \doi{10.1086/499040}
\bibAnnoteFile{iaria2005}

\bibitem[{{Ibragimov} and {Poutanen}(2009)}]{ibragimov2009}
{Ibragimov}, A. and {Poutanen}, J. (2009).
\newblock {Accreting millisecond pulsar SAX J1808.4$-$3658 during its 2002 outburst: evidence for a receding disc}.
\newblock \emph{\mnras} 400, 492--508.
\newblock \doi{10.1111/j.1365-2966.2009.15477.x}
\bibAnnoteFile{ibragimov2009}

\bibitem[{{Iwakiri} et~al.(2019){Iwakiri}, {Pottschmidt}, {Falkner}, {Hemphill}, {F{\"u}rst}, {Nishimura} et~al.}]{Iwakiri2019}
{Iwakiri}, W.~B., {Pottschmidt}, K., {Falkner}, S., {Hemphill}, P.~B., {F{\"u}rst}, F., {Nishimura}, O., et~al. (2019).
\newblock {Spectral and Timing Analysis of the Accretion-powered Pulsar 4U 1626$-$67 Observed with Suzaku and NuSTAR}.
\newblock \emph{\apj} 878, 121.
\newblock \doi{10.3847/1538-4357/ab1f87}
\bibAnnoteFile{Iwakiri2019}

\bibitem[{{Jaisawal} et~al.(2016){Jaisawal}, {Naik}, and {Epili}}]{Jaisawal2016}
{Jaisawal}, G.~K., {Naik}, S., and {Epili}, P. (2016).
\newblock {Suzaku view of the Be/X-ray binary pulsar GX 304$-$1 during Type I X-ray outbursts}.
\newblock \emph{\mnras} 457, 2749--2760.
\newblock \doi{10.1093/mnras/stw085}
\bibAnnoteFile{Jaisawal2016}

\bibitem[{{Jaisawal} et~al.(2023){Jaisawal}, {Vasilopoulos}, {Naik}, {Maitra}, {Malacaria}, {Chhotaray} et~al.}]{Jaisawal2023}
{Jaisawal}, G.~K., {Vasilopoulos}, G., {Naik}, S., {Maitra}, C., {Malacaria}, C., {Chhotaray}, B., et~al. (2023).
\newblock {On the cyclotron absorption line and evidence of the spectral transition in SMC X-2 during 2022 giant outburst}.
\newblock \emph{\mnras} 521, 3951--3961.
\newblock \doi{10.1093/mnras/stad781}
\bibAnnoteFile{Jaisawal2023}

\bibitem[{{Jansen} et~al.(2001){Jansen}, {Lumb}, {Altieri}, {Clavel}, {Ehle}, {Erd} et~al.}]{xmm}
{Jansen}, F., {Lumb}, D., {Altieri}, B., {Clavel}, J., {Ehle}, M., {Erd}, C., et~al. (2001).
\newblock {XMM-Newton observatory. I. The spacecraft and operations}.
\newblock \emph{\aap} 365, L1--L6.
\newblock \doi{10.1051/0004-6361:20000036}
\bibAnnoteFile{xmm}

\bibitem[{{Jayasurya} et~al.(2023){Jayasurya}, {Agrawal}, and {Chatterjee}}]{Jayasurya23}
{Jayasurya}, K.~M., {Agrawal}, V.~K., and {Chatterjee}, R. (2023).
\newblock {Detection of significant X-ray polarization from transient NS-LMXB XTE J1701-462 with IXPE and its implication on the coronal geometry}.
\newblock \emph{\mnras} 525, 4657--4662.
\newblock \doi{10.1093/mnras/stad2601}
\bibAnnoteFile{Jayasurya23}

\bibitem[{{King} et~al.(2016){King}, {Tomsick}, {Miller}, {Chenevez}, {Barret}, {Boggs} et~al.}]{king16}
{King}, A.~L., {Tomsick}, J.~A., {Miller}, J.~M., {Chenevez}, J., {Barret}, D., {Boggs}, S.~E., et~al. (2016).
\newblock {Measuring a Truncated Disk in Aquila X-1}.
\newblock \emph{\apjl} 819, L29.
\newblock \doi{10.3847/2041-8205/819/2/L29}
\bibAnnoteFile{king16}

\bibitem[{{Klochkov} et~al.(2008){Klochkov}, {Santangelo}, {Staubert}, and {Ferrigno}}]{Klochkov2008}
{Klochkov}, D., {Santangelo}, A., {Staubert}, R., and {Ferrigno}, C. (2008).
\newblock {Giant outburst of EXO 2030+375: pulse-phase resolved analysis of INTEGRAL data}.
\newblock \emph{\aap} 491, 833--840.
\newblock \doi{10.1051/0004-6361:200810673}
\bibAnnoteFile{Klochkov2008}

\bibitem[{{Klochkov} et~al.(2011){Klochkov}, {Staubert}, {Santangelo}, {Rothschild}, and {Ferrigno}}]{Klochkov2011}
{Klochkov}, D., {Staubert}, R., {Santangelo}, A., {Rothschild}, R.~E., and {Ferrigno}, C. (2011).
\newblock {Pulse-amplitude-resolved spectroscopy of bright accreting pulsars: indication of two accretion regimes}.
\newblock \emph{\aap} 532, A126.
\newblock \doi{10.1051/0004-6361/201116800}
\bibAnnoteFile{Klochkov2011}

\bibitem[{{Koliopanos} and {Vasilopoulos}(2018)}]{Koliopanos2018}
{Koliopanos}, F. and {Vasilopoulos}, G. (2018).
\newblock {Accreting, highly magnetized neutron stars at the Eddington limit: a study of the 2016 outburst of SMC X-3}.
\newblock \emph{\aap} 614, A23.
\newblock \doi{10.1051/0004-6361/201731623}
\bibAnnoteFile{Koliopanos2018}

\bibitem[{{Kong} et~al.(2022){Kong}, {Zhang}, {Zhang}, {Ji}, {Doroshenko}, {Santangelo} et~al.}]{Kong2022}
{Kong}, L.-D., {Zhang}, S., {Zhang}, S.-N., {Ji}, L., {Doroshenko}, V., {Santangelo}, A., et~al. (2022).
\newblock {Insight-HXMT Discovery of the Highest-energy CRSF from the First Galactic Ultraluminous X-Ray Pulsar Swift J0243.6+6124}.
\newblock \emph{\apjl} 933, L3.
\newblock \doi{10.3847/2041-8213/ac7711}
\bibAnnoteFile{Kong2022}

\bibitem[{{Kretschmar} et~al.(2021){Kretschmar}, {El Mellah}, {Mart{\'\i}nez-N{\'u}{\~n}ez}, {F{\"u}rst}, {Grinberg}, {Sander} et~al.}]{kretschmar21}
{Kretschmar}, P., {El Mellah}, I., {Mart{\'\i}nez-N{\'u}{\~n}ez}, S., {F{\"u}rst}, F., {Grinberg}, V., {Sander}, A.~A.~C., et~al. (2021).
\newblock {Revisiting the archetypical wind accretor Vela X-1 in depth. Case study of a well-known X-ray binary and the limits of our knowledge}.
\newblock \emph{\aap} 652, A95.
\newblock \doi{10.1051/0004-6361/202040272}
\bibAnnoteFile{kretschmar21}

\bibitem[{{Kreykenbohm} et~al.(2002){Kreykenbohm}, {Coburn}, {Wilms}, {Kretschmar}, {Staubert}, {Heindl} et~al.}]{kreykenbohm2002}
{Kreykenbohm}, I., {Coburn}, W., {Wilms}, J., {Kretschmar}, P., {Staubert}, R., {Heindl}, W.~A., et~al. (2002).
\newblock {Confirmation of two cyclotron lines in Vela X-1}.
\newblock \emph{\aap} 395, 129--140.
\newblock \doi{10.1051/0004-6361:20021181}
\bibAnnoteFile{kreykenbohm2002}

\bibitem[{{Kreykenbohm} et~al.(1999){Kreykenbohm}, {Kretschmar}, {Wilms}, {Staubert}, {Kendziorra}, {Gruber} et~al.}]{kreykenbohm99}
{Kreykenbohm}, I., {Kretschmar}, P., {Wilms}, J., {Staubert}, R., {Kendziorra}, E., {Gruber}, D.~E., et~al. (1999).
\newblock {VELA X-1 as seen by RXTE}.
\newblock \emph{\aap} 341, 141--150.
\newblock \doi{10.48550/arXiv.astro-ph/9810282}
\bibAnnoteFile{kreykenbohm99}

\bibitem[{{Kuulkers} et~al.(1997){Kuulkers}, {van der Klis}, {Oosterbroek}, {van Paradijs}, and {Lewin}}]{kuulkers97}
{Kuulkers}, E., {van der Klis}, M., {Oosterbroek}, T., {van Paradijs}, J., and {Lewin}, W.~H.~G. (1997).
\newblock {GX17+2: X-ray spectral and timing behaviour of a bursting Z source}.
\newblock \emph{\mnras} 287, 495--514.
\newblock \doi{10.1093/mnras/287.3.495}
\bibAnnoteFile{kuulkers97}

\bibitem[{{Kylafis} et~al.(2021){Kylafis}, {Tr{\"u}mper}, and {Loudas}}]{Kylafis2021}
{Kylafis}, N.~D., {Tr{\"u}mper}, J.~E., and {Loudas}, N.~A. (2021).
\newblock {Cyclotron line formation by reflection on the surface of a magnetic neutron star}.
\newblock \emph{\aap} 655, A39.
\newblock \doi{10.1051/0004-6361/202039361}
\bibAnnoteFile{Kylafis2021}

\bibitem[{{Lattimer}(2011)}]{Lattimer2011}
{Lattimer}, J.~M. (2011).
\newblock {Neutron stars and the dense matter equation of state}.
\newblock \emph{\apss} 336, 67--74.
\newblock \doi{10.1007/s10509-010-0529-1}
\bibAnnoteFile{Lattimer2011}

\bibitem[{{Lattimer} and {Prakash}(2001)}]{Lattimer01}
{Lattimer}, J.~M. and {Prakash}, M. (2001).
\newblock {Neutron Star Structure and the Equation of State}.
\newblock \emph{\apj} 550, 426--442.
\newblock \doi{10.1086/319702}
\bibAnnoteFile{Lattimer01}

\bibitem[{{Lattimer} and {Prakash}(2004)}]{Lattimer04}
{Lattimer}, J.~M. and {Prakash}, M. (2004).
\newblock {The Physics of Neutron Stars}.
\newblock \emph{Science} 304, 536--542.
\newblock \doi{10.1126/science.1090720}
\bibAnnoteFile{Lattimer04}

\bibitem[{{Lin} et~al.(2007){Lin}, {Remillard}, and {Homan}}]{Lin07}
{Lin}, D., {Remillard}, R.~A., and {Homan}, J. (2007).
\newblock {Evaluating Spectral Models and the X-Ray States of Neutron Star X-Ray Transients}.
\newblock \emph{\apj} 667, 1073--1086.
\newblock \doi{10.1086/521181}
\bibAnnoteFile{Lin07}

\bibitem[{{Lin} et~al.(2010){Lin}, {Remillard}, and {Homan}}]{Lin10}
{Lin}, D., {Remillard}, R.~A., and {Homan}, J. (2010).
\newblock {Suzaku and BeppoSAX X-ray Spectra of the Persistently Accreting Neutron-star Binary 4U 1705$-$44}.
\newblock \emph{\apj} 719, 1350--1361.
\newblock \doi{10.1088/0004-637X/719/2/1350}
\bibAnnoteFile{Lin10}

\bibitem[{{Liu} et~al.(2020){Liu}, {Tao}, {Zhang}, {Li}, {Ge}, {Qu} et~al.}]{Liu2020}
{Liu}, B.-S., {Tao}, L., {Zhang}, S.-N., {Li}, X.-D., {Ge}, M.-Y., {Qu}, J.-L., et~al. (2020).
\newblock {A Peculiar Cyclotron Line near 16 keV Detected in the 2015 Outburst of 4U 0115+63?}
\newblock \emph{\apj} 900, 41.
\newblock \doi{10.3847/1538-4357/aba4a5}
\bibAnnoteFile{Liu2020}

\bibitem[{{Liu} et~al.(2022){Liu}, {Wang}, {Chen}, {Ding}, {Lu}, {Song} et~al.}]{liu22}
{Liu}, Q., {Wang}, W., {Chen}, X., {Ding}, Y.~Z., {Lu}, F.~J., {Song}, L.~M., et~al. (2022).
\newblock {Variations of cyclotron resonant scattering features in Vela X-1 revealed with Insight-HXMT}.
\newblock \emph{\mnras} 514, 2805--2814.
\newblock \doi{10.1093/mnras/stac1520}
\bibAnnoteFile{liu22}

\bibitem[{{Longair}(2011)}]{Longair2011}
{Longair}, M.~S. (2011).
\newblock \emph{{High Energy Astrophysics}} (Cambridge: Cambridge Univ.\ Press)
\bibAnnoteFile{Longair2011}

\bibitem[{{Ludlam} et~al.(2020){Ludlam}, {Cackett}, {Garc{\'\i}a}, {Miller}, {Bult}, {Strohmayer} et~al.}]{ludlam20}
{Ludlam}, R.~M., {Cackett}, E.~M., {Garc{\'\i}a}, J.~A., {Miller}, J.~M., {Bult}, P.~M., {Strohmayer}, T.~E., et~al. (2020).
\newblock {NICER-NuSTAR Observations of the Neutron Star Low-mass X-Ray Binary 4U 1735$-$44}.
\newblock \emph{\apj} 895, 45.
\newblock \doi{10.3847/1538-4357/ab89a6}
\bibAnnoteFile{ludlam20}

\bibitem[{{Ludlam} et~al.(2022){Ludlam}, {Cackett}, {Garc{\'\i}a}, {Miller}, {Stevens}, {Fabian} et~al.}]{Ludlam22}
{Ludlam}, R.~M., {Cackett}, E.~M., {Garc{\'\i}a}, J.~A., {Miller}, J.~M., {Stevens}, A.~L., {Fabian}, A.~C., et~al. (2022).
\newblock {Radius Constraints from Reflection Modeling of Cygnus X-2 with NuSTAR and NICER}.
\newblock \emph{\apj} 927, 112.
\newblock \doi{10.3847/1538-4357/ac5028}
\bibAnnoteFile{Ludlam22}

\bibitem[{{Ludlam} et~al.(2021){Ludlam}, {Jaodand}, {Garc{\'\i}a}, {Degenaar}, {Tomsick}, {Cackett} et~al.}]{Ludlam21}
{Ludlam}, R.~M., {Jaodand}, A.~D., {Garc{\'\i}a}, J.~A., {Degenaar}, N., {Tomsick}, J.~A., {Cackett}, E.~M., et~al. (2021).
\newblock {Simultaneous NICER and NuSTAR Observations of the Ultracompact X-Ray Binary 4U 1543-624}.
\newblock \emph{\apj} 911, 123.
\newblock \doi{10.3847/1538-4357/abedb0}
\bibAnnoteFile{Ludlam21}

\bibitem[{{Ludlam} et~al.(2018){Ludlam}, {Miller}, {Arzoumanian}, {Bult}, {Cackett}, {Chakrabarty} et~al.}]{ludlam18}
{Ludlam}, R.~M., {Miller}, J.~M., {Arzoumanian}, Z., {Bult}, P.~M., {Cackett}, E.~M., {Chakrabarty}, D., et~al. (2018).
\newblock {Detection of Reflection Features in the Neutron Star Low-mass X-Ray Binary Serpens X-1 with NICER}.
\newblock \emph{\apjl} 858, L5.
\newblock \doi{10.3847/2041-8213/aabee6}
\bibAnnoteFile{ludlam18}

\bibitem[{{Ludlam} et~al.(2017{\natexlab{a}}){Ludlam}, {Miller}, {Bachetti}, {Barret}, {Bostrom}, {Cackett} et~al.}]{ludlam17a}
{Ludlam}, R.~M., {Miller}, J.~M., {Bachetti}, M., {Barret}, D., {Bostrom}, A.~C., {Cackett}, E.~M., et~al. (2017{\natexlab{a}}).
\newblock {A Hard Look at the Neutron Stars and Accretion Disks in 4U 1636$-$53, GX 17+2, and 4U 1705$-$44 with NuStar}.
\newblock \emph{\apj} 836, 140.
\newblock \doi{10.3847/1538-4357/836/1/140}
\bibAnnoteFile{ludlam17a}

\bibitem[{{Ludlam} et~al.(2016){Ludlam}, {Miller}, {Cackett}, {Fabian}, {Bachetti}, {Parker} et~al.}]{ludlam16}
{Ludlam}, R.~M., {Miller}, J.~M., {Cackett}, E.~M., {Fabian}, A.~C., {Bachetti}, M., {Parker}, M.~L., et~al. (2016).
\newblock {NuSTAR and XMM-Newton Observations of the Neutron Star X-Ray Binary 1RXS J180408.9-34205}.
\newblock \emph{\apj} 824, 37.
\newblock \doi{10.3847/0004-637X/824/1/37}
\bibAnnoteFile{ludlam16}

\bibitem[{{Ludlam} et~al.(2017{\natexlab{b}}){Ludlam}, {Miller}, {Degenaar}, {Sanna}, {Cackett}, {Altamirano} et~al.}]{ludlam17}
{Ludlam}, R.~M., {Miller}, J.~M., {Degenaar}, N., {Sanna}, A., {Cackett}, E.~M., {Altamirano}, D., et~al. (2017{\natexlab{b}}).
\newblock {Truncation of the Accretion Disk at One-third of the Eddington Limit in the Neutron Star Low-mass X-Ray Binary Aquila X-1}.
\newblock \emph{\apj} 847, 135.
\newblock \doi{10.3847/1538-4357/aa8b1b}
\bibAnnoteFile{ludlam17}

\bibitem[{{Lutovinov} et~al.(2021){Lutovinov}, {Tsygankov}, {Molkov}, {Doroshenko}, {Mushtukov}, {Arefiev} et~al.}]{Lutovinov2021}
{Lutovinov}, A., {Tsygankov}, S., {Molkov}, S., {Doroshenko}, V., {Mushtukov}, A., {Arefiev}, V., et~al. (2021).
\newblock {SRG/ART-XC and NuSTAR Observations of the X-Ray pulsar GRO J1008$-$57 in the Lowest Luminosity State}.
\newblock \emph{\apj} 912, 17.
\newblock \doi{10.3847/1538-4357/abec43}
\bibAnnoteFile{Lutovinov2021}

\bibitem[{{Madej} et~al.(2014){Madej}, {Garc{\'\i}a}, {Jonker}, {Parker}, {Ross}, {Fabian} et~al.}]{madej14}
{Madej}, O.~K., {Garc{\'\i}a}, J., {Jonker}, P.~G., {Parker}, M.~L., {Ross}, R., {Fabian}, A.~C., et~al. (2014).
\newblock {X-ray reflection in oxygen-rich accretion discs of ultracompact X-ray binaries}.
\newblock \emph{\mnras} 442, 1157--1165.
\newblock \doi{10.1093/mnras/stu884}
\bibAnnoteFile{madej14}

\bibitem[{{Maitra} and {Paul}(2013)}]{Maitra2013}
{Maitra}, C. and {Paul}, B. (2013).
\newblock {Pulse-phase-resolved Spectroscopy of Vela X-1 with Suzaku}.
\newblock \emph{\apj} 763, 79.
\newblock \doi{10.1088/0004-637X/763/2/79}
\bibAnnoteFile{Maitra2013}

\bibitem[{{Maitra} et~al.(2018){Maitra}, {Paul}, {Haberl}, and {Vasilopoulos}}]{Maitra2018}
{Maitra}, C., {Paul}, B., {Haberl}, F., and {Vasilopoulos}, G. (2018).
\newblock {Detection of a cyclotron line in SXP 15.3 during its 2017 outburst}.
\newblock \emph{\mnras} 480, L136--L140.
\newblock \doi{10.1093/mnrasl/sly141}
\bibAnnoteFile{Maitra2018}

\bibitem[{{Makishima} et~al.(1999){Makishima}, {Mihara}, {Nagase}, and {Tanaka}}]{makishima1999}
{Makishima}, K., {Mihara}, T., {Nagase}, F., and {Tanaka}, Y. (1999).
\newblock {Cyclotron Resonance Effects in Two Binary X-Ray Pulsars and the Evolution of Neutron Star Magnetic Fields}.
\newblock \emph{\apj} 525, 978--994.
\newblock \doi{10.1086/307912}
\bibAnnoteFile{makishima1999}

\bibitem[{{Malacaria} et~al.(2022){Malacaria}, {Bhargava}, {Coley}, {Ducci}, {Pradhan}, {Ballhausen} et~al.}]{Malacaria2022}
{Malacaria}, C., {Bhargava}, Y., {Coley}, J.~B., {Ducci}, L., {Pradhan}, P., {Ballhausen}, R., et~al. (2022).
\newblock {Accreting on the Edge: A Luminosity-dependent Cyclotron Line in the Be/X-Ray Binary 2S 1553-542 Accompanied by Accretion Regimes Transition}.
\newblock \emph{\apj} 927, 194.
\newblock \doi{10.3847/1538-4357/ac524f}
\bibAnnoteFile{Malacaria2022}

\bibitem[{{Malacaria} et~al.(2023{\natexlab{a}}){Malacaria}, {Ducci}, {Falanga}, {Altamirano}, {Bozzo}, {Guillot} et~al.}]{Malacaria2023}
{Malacaria}, C., {Ducci}, L., {Falanga}, M., {Altamirano}, D., {Bozzo}, E., {Guillot}, S., et~al. (2023{\natexlab{a}}).
\newblock {The unaltered pulsar: GRO J1750$-$27, a supercritical X-ray neutron star that does not blink an eye}.
\newblock \emph{\aap} 669, A38.
\newblock \doi{10.1051/0004-6361/202245123}
\bibAnnoteFile{Malacaria2023}

\bibitem[{{Malacaria} et~al.(2023{\natexlab{b}}){Malacaria}, {Heyl}, {Doroshenko}, {Tsygankov}, {Poutanen}, {Forsblom} et~al.}]{Malacaria23b}
{Malacaria}, C., {Heyl}, J., {Doroshenko}, V., {Tsygankov}, S.~S., {Poutanen}, J., {Forsblom}, S.~V., et~al. (2023{\natexlab{b}}).
\newblock {A polarimetrically oriented X-ray stare at the accreting pulsar EXO 2030+375}.
\newblock \emph{\aap} 675, A29.
\newblock \doi{10.1051/0004-6361/202346581}
\bibAnnoteFile{Malacaria23b}

\bibitem[{{Malacaria} et~al.(2015){Malacaria}, {Klochkov}, {Santangelo}, and {Staubert}}]{Malacaria2015}
{Malacaria}, C., {Klochkov}, D., {Santangelo}, A., and {Staubert}, R. (2015).
\newblock {Luminosity-dependent spectral and timing properties of the accreting pulsar GX 304$-$1 measured with INTEGRAL}.
\newblock \emph{\aap} 581, A121.
\newblock \doi{10.1051/0004-6361/201526417}
\bibAnnoteFile{Malacaria2015}

\bibitem[{{Marino} et~al.(2023){Marino}, {Russell}, {Del Santo}, {Beri}, {Sanna}, {Coti Zelati} et~al.}]{marino23}
{Marino}, A., {Russell}, T.~D., {Del Santo}, M., {Beri}, A., {Sanna}, A., {Coti Zelati}, F., et~al. (2023).
\newblock {The accretion/ejection link in the neutron star X-ray binary 4U 1820-30 I: A boundary layer-jet coupling?}
\newblock \emph{arXiv e-prints} , arXiv:2307.16566\doi{10.48550/arXiv.2307.16566}
\bibAnnoteFile{marino23}

\bibitem[{{Meidinger} et~al.(2020){Meidinger}, {Albrecht}, {Beitler}, {Bonholzer}, {Emberger}, {Frank} et~al.}]{wfi}
{Meidinger}, N., {Albrecht}, S., {Beitler}, C., {Bonholzer}, M., {Emberger}, V., {Frank}, J., et~al. (2020).
\newblock {Development status of the wide field imager instrument for Athena}.
\newblock In \emph{Space Telescopes and Instrumentation 2020: Ultraviolet to Gamma Ray}, eds. J.-W.~A. {den Herder}, S.~{Nikzad}, and K.~{Nakazawa}. vol. 11444 of \emph{Society of Photo-Optical Instrumentation Engineers (SPIE) Conference Series}, 114440T.
\newblock \doi{10.1117/12.2560507}
\bibAnnoteFile{wfi}

\bibitem[{{Meszaros}(1992)}]{Meszaros1992}
{Meszaros}, P. (1992).
\newblock \emph{{High-energy radiation from magnetized neutron stars}} (Chicago: Univ.\ Chicago Press)
\bibAnnoteFile{Meszaros1992}

\bibitem[{{Meszaros} et~al.(1988){Meszaros}, {Novick}, {Szentgyorgyi}, {Chanan}, and {Weisskopf}}]{Meszaros88}
{Meszaros}, P., {Novick}, R., {Szentgyorgyi}, A., {Chanan}, G.~A., and {Weisskopf}, M.~C. (1988).
\newblock {Astrophysical Implications and Observational Prospects of X-Ray Polarimetry}.
\newblock \emph{\apj} 324, 1056.
\newblock \doi{10.1086/165962}
\bibAnnoteFile{Meszaros88}

\bibitem[{{Mihara}(1995)}]{M95}
{Mihara}, T. (1995).
\newblock \emph{{Observational study of X-ray spectra of binary pulsars with Ginga}}.
\newblock Ph.D. thesis, , Dept.~of Physics, Univ.~of Tokyo (M95), (1995)
\bibAnnoteFile{M95}

\bibitem[{{Mihara} et~al.(2010){Mihara}, {Yamamoto}, {Sugizaki}, and {Yamaoka}}]{Mihara2010}
{Mihara}, T., {Yamamoto}, T., {Sugizaki}, M., and {Yamaoka}, K. (2010).
\newblock {Discovery of the cyclotron line at 51 keV from GX 304$-$1}.
\newblock ATEL 2796
\bibAnnoteFile{Mihara2010}

\bibitem[{{Miller} et~al.(2011){Miller}, {Maitra}, {Cackett}, {Bhattacharyya}, and {Strohmayer}}]{miller2011}
{Miller}, J.~M., {Maitra}, D., {Cackett}, E.~M., {Bhattacharyya}, S., and {Strohmayer}, T.~E. (2011).
\newblock {A Fast X-ray Disk Wind in the Transient Pulsar IGR J17480$-$2446 in Terzan 5}.
\newblock \emph{\apjl} 731, L7.
\newblock \doi{10.1088/2041-8205/731/1/L7}
\bibAnnoteFile{miller2011}

\bibitem[{{Miller} et~al.(2019){Miller}, {Lamb}, {Dittmann}, {Bogdanov}, {Arzoumanian}, {Gendreau} et~al.}]{miller19}
{Miller}, M.~C., {Lamb}, F.~K., {Dittmann}, A.~J., {Bogdanov}, S., {Arzoumanian}, Z., {Gendreau}, K.~C., et~al. (2019).
\newblock {PSR J0030+0451 Mass and Radius from NICER Data and Implications for the Properties of Neutron Star Matter}.
\newblock \emph{\apjl} 887, L24.
\newblock \doi{10.3847/2041-8213/ab50c5}
\bibAnnoteFile{miller19}

\bibitem[{{Miller} et~al.(2021){Miller}, {Lamb}, {Dittmann}, {Bogdanov}, {Arzoumanian}, {Gendreau} et~al.}]{miller21}
{Miller}, M.~C., {Lamb}, F.~K., {Dittmann}, A.~J., {Bogdanov}, S., {Arzoumanian}, Z., {Gendreau}, K.~C., et~al. (2021).
\newblock {The Radius of PSR J0740+6620 from NICER and XMM-Newton Data}.
\newblock \emph{\apjl} 918, L28.
\newblock \doi{10.3847/2041-8213/ac089b}
\bibAnnoteFile{miller21}

\bibitem[{{Mitsuda} et~al.(1989){Mitsuda}, {Inoue}, {Nakamura}, and {Tanaka}}]{mitsuda89}
{Mitsuda}, K., {Inoue}, H., {Nakamura}, N., and {Tanaka}, Y. (1989).
\newblock {Luminosity-related changes of the energy spectrum of X 1608$-$522.}
\newblock \emph{\pasj} 41, 97--111
\bibAnnoteFile{mitsuda89}

\bibitem[{{Mondal} et~al.(2018){Mondal}, {Dewangan}, {Pahari}, and {Raychaudhuri}}]{mondal18}
{Mondal}, A.~S., {Dewangan}, G.~C., {Pahari}, M., and {Raychaudhuri}, B. (2018).
\newblock {NuSTAR view of the Z-type neutron star low-mass X-ray binary Cygnus X-2}.
\newblock \emph{\mnras} 474, 2064--2072.
\newblock \doi{10.1093/mnras/stx2931}
\bibAnnoteFile{mondal18}

\bibitem[{{M{\"u}ller} et~al.(2013){M{\"u}ller}, {Ferrigno}, {K{\"u}hnel}, {Sch{\"o}nherr}, {Becker}, {Wolff} et~al.}]{mueller2013}
{M{\"u}ller}, S., {Ferrigno}, C., {K{\"u}hnel}, M., {Sch{\"o}nherr}, G., {Becker}, P.~A., {Wolff}, M.~T., et~al. (2013).
\newblock {No anticorrelation between cyclotron line energy and X-ray flux in 4U 0115+634}.
\newblock \emph{\aap} 551, A6.
\newblock \doi{10.1051/0004-6361/201220359}
\bibAnnoteFile{mueller2013}

\bibitem[{{Mushtukov} and {Tsygankov}(2022)}]{Mushtukov2022}
{Mushtukov}, A. and {Tsygankov}, S. (2022).
\newblock {Accreting strongly magnetised neutron stars: X-ray Pulsars}.
\newblock \emph{arXiv e-prints} , arXiv:2204.14185\doi{10.48550/arXiv.2204.14185}
\bibAnnoteFile{Mushtukov2022}

\bibitem[{{Mushtukov} et~al.(2021){Mushtukov}, {Suleimanov}, {Tsygankov}, and {Portegies Zwart}}]{Mushtukov2021}
{Mushtukov}, A.~A., {Suleimanov}, V.~F., {Tsygankov}, S.~S., and {Portegies Zwart}, S. (2021).
\newblock {Spectrum formation in X-ray pulsars at very low mass accretion rate: Monte Carlo approach}.
\newblock \emph{\mnras} 503, 5193--5203.
\newblock \doi{10.1093/mnras/stab811}
\bibAnnoteFile{Mushtukov2021}

\bibitem[{{Mushtukov} et~al.(2015){Mushtukov}, {Suleimanov}, {Tsygankov}, and {Poutanen}}]{Mushtukov2015}
{Mushtukov}, A.~A., {Suleimanov}, V.~F., {Tsygankov}, S.~S., and {Poutanen}, J. (2015).
\newblock {The critical accretion luminosity for magnetized neutron stars}.
\newblock \emph{\mnras} 447, 1847--1856.
\newblock \doi{10.1093/mnras/stu2484}
\bibAnnoteFile{Mushtukov2015}

\bibitem[{{Naik} and {Paul}(2012)}]{naik2012}
{Naik}, S. and {Paul}, B. (2012).
\newblock {Investigation of variability of iron emission lines in Centaurus X-3}.
\newblock \emph{Bulletin of the Astronomical Society of India} 40, 503
\bibAnnoteFile{naik2012}

\bibitem[{{Nandra} et~al.(2013){Nandra}, {Barret}, {Barcons}, {Fabian}, {den Herder}, {Piro} et~al.}]{Nandra13}
{Nandra}, K., {Barret}, D., {Barcons}, X., {Fabian}, A., {den Herder}, J.-W., {Piro}, L., et~al. (2013).
\newblock {The Hot and Energetic Universe: A White Paper presenting the science theme motivating the Athena+ mission}.
\newblock \emph{arXiv e-prints} , arXiv:1306.2307\doi{10.48550/arXiv.1306.2307}
\bibAnnoteFile{Nandra13}

\bibitem[{Nishimura(2003)}]{Nishimura2003}
Nishimura, O. (2003).
\newblock {The Influence of a Dipole Magnetic Field on the Structures of Cyclotron Lines}.
\newblock \emph{Publications of the Astronomical Society of Japan} 55, 849--857.
\newblock \doi{10.1093/pasj/55.4.849}
\bibAnnoteFile{Nishimura2003}

\bibitem[{{Nishimura}(2011)}]{Nishimura2011}
{Nishimura}, O. (2011).
\newblock {Superposition of Cyclotron Lines in Accreting X-Ray Pulsars. I. Long Spin Period}.
\newblock \emph{\apj} 730, 106.
\newblock \doi{10.1088/0004-637X/730/2/106}
\bibAnnoteFile{Nishimura2011}

\bibitem[{{Nishimura}(2014)}]{Nishimura2014}
{Nishimura}, O. (2014).
\newblock {Variations of Cyclotron Line Energy with Luminosity in Accreting X-Ray Pulsars}.
\newblock \emph{\apj} 781, 30.
\newblock \doi{10.1088/0004-637X/781/1/30}
\bibAnnoteFile{Nishimura2014}

\bibitem[{{Oppenheimer} and {Volkoff}(1939)}]{Oppenheimer1939}
{Oppenheimer}, J.~R. and {Volkoff}, G.~M. (1939).
\newblock {On Massive Neutron Cores}.
\newblock \emph{Physical Review} 55, 374--381.
\newblock \doi{10.1103/PhysRev.55.374}
\bibAnnoteFile{Oppenheimer1939}

\bibitem[{{Orlandini} et~al.(2012){Orlandini}, {Frontera}, {Masetti}, {Sguera}, and {Sidoli}}]{Orlandini2012}
{Orlandini}, M., {Frontera}, F., {Masetti}, N., {Sguera}, V., and {Sidoli}, L. (2012).
\newblock {BeppoSAX Observations of the X-Ray Pulsar MAXI J1409$-$619 in Low State: Discovery of Cyclotron Resonance Features}.
\newblock \emph{\apj} 748, 86.
\newblock \doi{10.1088/0004-637X/748/2/86}
\bibAnnoteFile{Orlandini2012}

\bibitem[{{Orosz} and {Kuulkers}(1999)}]{orosz99}
{Orosz}, J.~A. and {Kuulkers}, E. (1999).
\newblock {The optical light curves of Cygnus X-2 (V1341 Cyg) and the mass of its neutron star}.
\newblock \emph{\mnras} 305, 132--142.
\newblock \doi{10.1046/j.1365-8711.1999.t01-1-02420.x}
\bibAnnoteFile{orosz99}

\bibitem[{{Parikh} et~al.(2017){Parikh}, {Wijnands}, {Degenaar}, {Altamirano}, {Patruno}, {Gusinskaia} et~al.}]{parikh17}
{Parikh}, A.~S., {Wijnands}, R., {Degenaar}, N., {Altamirano}, D., {Patruno}, A., {Gusinskaia}, N.~V., et~al. (2017).
\newblock {Very hard states in neutron star low-mass X-ray binaries}.
\newblock \emph{\mnras} 468, 3979--3984.
\newblock \doi{10.1093/mnras/stx747}
\bibAnnoteFile{parikh17}

\bibitem[{{Popham} and {Sunyaev}(2001)}]{popham2001}
{Popham}, R. and {Sunyaev}, R. (2001).
\newblock {Accretion Disk Boundary Layers around Neutron Stars: X-Ray Production in Low-Mass X-Ray Binaries}.
\newblock \emph{\apj} 547, 355--383.
\newblock \doi{10.1086/318336}
\bibAnnoteFile{popham2001}

\bibitem[{{Pottschmidt} et~al.(2005){Pottschmidt}, {Kreykenbohm}, {Wilms}, {Coburn}, {Rothschild}, {Kretschmar} et~al.}]{Pottschmidt2005}
{Pottschmidt}, K., {Kreykenbohm}, I., {Wilms}, J., {Coburn}, W., {Rothschild}, R.~E., {Kretschmar}, P., et~al. (2005).
\newblock {RXTE Discovery of Multiple Cyclotron Lines during the 2004 December Outburst of V0332+53}.
\newblock \emph{\apjl} 634, L97--L100.
\newblock \doi{10.1086/498689}
\bibAnnoteFile{Pottschmidt2005}

\bibitem[{{Poutanen} et~al.(2013){Poutanen}, {Mushtukov}, {Suleimanov}, {Tsygankov}, {Nagirner}, {Doroshenko} et~al.}]{Poutanen2013}
{Poutanen}, J., {Mushtukov}, A.~A., {Suleimanov}, V.~F., {Tsygankov}, S.~S., {Nagirner}, D.~I., {Doroshenko}, V., et~al. (2013).
\newblock {A Reflection Model for the Cyclotron Lines in the Spectra of X-Ray Pulsars}.
\newblock \emph{\apj} 777, 115.
\newblock \doi{10.1088/0004-637X/777/2/115}
\bibAnnoteFile{Poutanen2013}

\bibitem[{Pradhan et~al.(2021)Pradhan, Paul, Bozzo, Maitra, and Paul}]{pradhan2021}
Pradhan, P., Paul, B., Bozzo, E., Maitra, C., and Paul, B.~C. (2021).
\newblock {Comprehensive broad-band study of accreting neutron stars with Suzaku: Is there a bi-modality in the X-ray spectrum?}
\newblock \emph{Monthly Notices of the Royal Astronomical Society} 502, 1163--1190.
\newblock \doi{10.1093/mnras/stab024}
\bibAnnoteFile{pradhan2021}

\bibitem[{{Raaijmakers} et~al.(2021){Raaijmakers}, {Greif}, {Hebeler}, {Hinderer}, {Nissanke}, {Schwenk} et~al.}]{Raaijmakers21}
{Raaijmakers}, G., {Greif}, S.~K., {Hebeler}, K., {Hinderer}, T., {Nissanke}, S., {Schwenk}, A., et~al. (2021).
\newblock {Constraints on the Dense Matter Equation of State and Neutron Star Properties from NICER's Mass-Radius Estimate of PSR J0740+6620 and Multimessenger Observations}.
\newblock \emph{\apjl} 918, L29.
\newblock \doi{10.3847/2041-8213/ac089a}
\bibAnnoteFile{Raaijmakers21}

\bibitem[{{Riley} et~al.(2019){Riley}, {Watts}, {Bogdanov}, {Ray}, {Ludlam}, {Guillot} et~al.}]{Riley19}
{Riley}, T.~E., {Watts}, A.~L., {Bogdanov}, S., {Ray}, P.~S., {Ludlam}, R.~M., {Guillot}, S., et~al. (2019).
\newblock {A NICER View of PSR J0030+0451: Millisecond Pulsar Parameter Estimation}.
\newblock \emph{\apjl} 887, L21.
\newblock \doi{10.3847/2041-8213/ab481c}
\bibAnnoteFile{Riley19}

\bibitem[{{Riley} et~al.(2021){Riley}, {Watts}, {Ray}, {Bogdanov}, {Guillot}, {Morsink} et~al.}]{Riley21}
{Riley}, T.~E., {Watts}, A.~L., {Ray}, P.~S., {Bogdanov}, S., {Guillot}, S., {Morsink}, S.~M., et~al. (2021).
\newblock {A NICER View of the Massive Pulsar PSR J0740+6620 Informed by Radio Timing and XMM-Newton Spectroscopy}.
\newblock \emph{\apjl} 918, L27.
\newblock \doi{10.3847/2041-8213/ac0a81}
\bibAnnoteFile{Riley21}

\bibitem[{{Rodes-Roca} et~al.(2009){Rodes-Roca}, {Torrej{\'o}n}, {Kreykenbohm}, {Mart{\'\i}nez N{\'u}{\~n}ez}, {Camero-Arranz}, and {Bernab{\'e}u}}]{Rodes2009}
{Rodes-Roca}, J.~J., {Torrej{\'o}n}, J.~M., {Kreykenbohm}, I., {Mart{\'\i}nez N{\'u}{\~n}ez}, S., {Camero-Arranz}, A., and {Bernab{\'e}u}, G. (2009).
\newblock {The first cyclotron harmonic of 4U 1538$-$52}.
\newblock \emph{\aap} 508, 395--400.
\newblock \doi{10.1051/0004-6361/200912815}
\bibAnnoteFile{Rodes2009}

\bibitem[{{Ross} and {Fabian}(2005)}]{Ross2005}
{Ross}, R.~R. and {Fabian}, A.~C. (2005).
\newblock {A comprehensive range of X-ray ionized-reflection models}.
\newblock \emph{\mnras} 358, 211--216.
\newblock \doi{10.1111/j.1365-2966.2005.08797.x}
\bibAnnoteFile{Ross2005}

\bibitem[{{Rothschild} et~al.(2017){Rothschild}, {K{\"u}hnel}, {Pottschmidt}, {Hemphill}, {Postnov}, {Gornostaev} et~al.}]{Rothschild2017}
{Rothschild}, R.~E., {K{\"u}hnel}, M., {Pottschmidt}, K., {Hemphill}, P., {Postnov}, K., {Gornostaev}, M., et~al. (2017).
\newblock {Discovery and modelling of a flattening of the positive cyclotron line/luminosity relation in GX 304$-$1 with RXTE}.
\newblock \emph{MNRAS} 466, 2752--2779.
\newblock \doi{10.1093/mnras/stw3222}
\bibAnnoteFile{Rothschild2017}

\bibitem[{{Rouco Escorial} et~al.(2018){Rouco Escorial}, {van den Eijnden}, and {Wijnands}}]{Rouco2018}
{Rouco Escorial}, A., {van den Eijnden}, J., and {Wijnands}, R. (2018).
\newblock {Discovery of accretion-driven pulsations in the prolonged low X-ray luminosity state of the Be/X-ray transient GX 304$-$1}.
\newblock \emph{\aap} 620, L13.
\newblock \doi{10.1051/0004-6361/201834572}
\bibAnnoteFile{Rouco2018}

\bibitem[{{Sanjurjo-Ferr{\'\i}n} et~al.(2021){Sanjurjo-Ferr{\'\i}n}, {Torrej{\'o}n}, {Postnov}, {Oskinova}, {Rodes-Roca}, and {Bernabeu}}]{ferrin2021}
{Sanjurjo-Ferr{\'\i}n}, G., {Torrej{\'o}n}, J.~M., {Postnov}, K., {Oskinova}, L., {Rodes-Roca}, J.~J., and {Bernabeu}, G. (2021).
\newblock {X-ray variability of the HMXB Cen X-3: evidence for inhomogeneous accretion flows}.
\newblock \emph{\mnras} 501, 5892--5909.
\newblock \doi{10.1093/mnras/staa3953}
\bibAnnoteFile{ferrin2021}

\bibitem[{{Sch{\"o}nherr} et~al.(2007){Sch{\"o}nherr}, {Wilms}, {Kretschmar}, {Kreykenbohm}, {Santangelo}, {Rothschild} et~al.}]{Sch2007}
{Sch{\"o}nherr}, G., {Wilms}, J., {Kretschmar}, P., {Kreykenbohm}, I., {Santangelo}, A., {Rothschild}, R.~E., et~al. (2007).
\newblock {A model for cyclotron resonance scattering features}.
\newblock \emph{\aap} 472, 353--365.
\newblock \doi{10.1051/0004-6361:20077218}
\bibAnnoteFile{Sch2007}

\bibitem[{{Schreier} et~al.(1972){Schreier}, {Levinson}, {Gursky}, {Kellogg}, {Tananbaum}, and {Giacconi}}]{Schreier}
{Schreier}, E., {Levinson}, R., {Gursky}, H., {Kellogg}, E., {Tananbaum}, H., and {Giacconi}, R. (1972).
\newblock {Evidence for the Binary Nature of Centaurus X-3 from UHURU X-Ray Observations.}
\newblock \emph{\apjl} 172, L79.
\newblock \doi{10.1086/180896}
\bibAnnoteFile{Schreier}

\bibitem[{{Schwarm} et~al.(2017){Schwarm}, {Ballhausen}, {Falkner}, {Sch{\"o}nherr}, {Pottschmidt}, {Wolff} et~al.}]{Schwarm2017}
{Schwarm}, F.~W., {Ballhausen}, R., {Falkner}, S., {Sch{\"o}nherr}, G., {Pottschmidt}, K., {Wolff}, M.~T., et~al. (2017).
\newblock {Cyclotron resonant scattering feature simulations. II. Description of the CRSF simulation process}.
\newblock \emph{\aap} 601, A99.
\newblock \doi{10.1051/0004-6361/201630250}
\bibAnnoteFile{Schwarm2017}

\bibitem[{{Sibgatullin} and {Sunyaev}(1998)}]{sibgatullin98}
{Sibgatullin}, N.~R. and {Sunyaev}, R.~A. (1998).
\newblock {Disk accretion in the gravitational field of a rapidly rotating neutron star with a rotationally induced quadrupole mass distribution}.
\newblock \emph{Astronomy Letters} 24, 774--787.
\newblock \doi{10.48550/arXiv.astro-ph/9811028}
\bibAnnoteFile{sibgatullin98}

\bibitem[{{Sokolova-Lapa} et~al.(2021){Sokolova-Lapa}, {Gornostaev}, {Wilms}, {Ballhausen}, {Falkner}, {Postnov} et~al.}]{Sokolova-Lapa2021}
{Sokolova-Lapa}, E., {Gornostaev}, M., {Wilms}, J., {Ballhausen}, R., {Falkner}, S., {Postnov}, K., et~al. (2021).
\newblock {X-ray emission from magnetized neutron star atmospheres at low mass-accretion rates. I. Phase-averaged spectrum}.
\newblock \emph{\aap} 651, A12.
\newblock \doi{10.1051/0004-6361/202040228}
\bibAnnoteFile{Sokolova-Lapa2021}

\bibitem[{{Sokolova-Lapa} et~al.(2023){Sokolova-Lapa}, {Stierhof}, {Dauser}, and {Wilms}}]{Sokolova-Lapa2023}
{Sokolova-Lapa}, E., {Stierhof}, J., {Dauser}, T., and {Wilms}, J. (2023).
\newblock {Vacuum polarization alters the spectra of accreting X-ray pulsars}.
\newblock \emph{\aap} 674, L2.
\newblock \doi{10.1051/0004-6361/202346265}
\bibAnnoteFile{Sokolova-Lapa2023}

\bibitem[{{Staubert} et~al.(2020){Staubert}, {Ducci}, {Ji}, {F{\"u}rst}, {Wilms}, {Rothschild} et~al.}]{Staubert2020}
{Staubert}, R., {Ducci}, L., {Ji}, L., {F{\"u}rst}, F., {Wilms}, J., {Rothschild}, R.~E., et~al. (2020).
\newblock {Cyclotron line energy in Hercules X-1: stable after the decay}.
\newblock \emph{\aap} 642, A196.
\newblock \doi{10.1051/0004-6361/202038855}
\bibAnnoteFile{Staubert2020}

\bibitem[{{Staubert} et~al.(2014){Staubert}, {Klochkov}, {Wilms}, {Postnov}, {Shakura}, {Rothschild} et~al.}]{Staubert2014}
{Staubert}, R., {Klochkov}, D., {Wilms}, J., {Postnov}, K., {Shakura}, N.~I., {Rothschild}, R.~E., et~al. (2014).
\newblock {Long-term change in the cyclotron line energy in Hercules X-1}.
\newblock \emph{\aap} 572, A119.
\newblock \doi{10.1051/0004-6361/201424203}
\bibAnnoteFile{Staubert2014}

\bibitem[{{Staubert} et~al.(2019){Staubert}, {Tr{\"u}mper}, {Kendziorra}, {Klochkov}, {Postnov}, {Kretschmar} et~al.}]{Staubert2019}
{Staubert}, R., {Tr{\"u}mper}, J., {Kendziorra}, E., {Klochkov}, D., {Postnov}, K., {Kretschmar}, P., et~al. (2019).
\newblock {Cyclotron lines in highly magnetized neutron stars}.
\newblock \emph{\aap} 622, A61.
\newblock \doi{10.1051/0004-6361/201834479}
\bibAnnoteFile{Staubert2019}

\bibitem[{{Sunyaev} et~al.(1991){Sunyaev}, {Arefev}, {Borozdin}, {Gilfanov}, {Efremov}, {Kaniovskii} et~al.}]{Sunyaev1991}
{Sunyaev}, R.~A., {Arefev}, V.~A., {Borozdin}, K.~N., {Gilfanov}, M.~R., {Efremov}, V.~V., {Kaniovskii}, A.~S., et~al. (1991).
\newblock {Broadband X-Ray Spectra of Black-Hole Candidates X-Ray Pulsars and Low-Mass Binary X-Ray Systems - KVANT Module Results}.
\newblock \emph{Soviet Astronomy Letters} 17, 409
\bibAnnoteFile{Sunyaev1991}

\bibitem[{{Tanaka}(1986)}]{tanaka86}
{Tanaka}, Y. (1986).
\newblock {Observations of Compact X-Ray Sources}.
\newblock In \emph{IAU Colloq. 89: Radiation Hydrodynamics in Stars and Compact Objects}, eds. D.~{Mihalas} and K.-H.~A. {Winkler} (Berlin, Heidelberg: Springer), vol. 255. 198.
\newblock \doi{10.1007/3-540-16764-1_12}
\bibAnnoteFile{tanaka86}

\bibitem[{{The LIGO Scientific Collaboration} et~al.(2020){The LIGO Scientific Collaboration}, {The Virgo Collaboration}, {Abbott}, {Abbott}, {Abraham}, {Acernese} et~al.}]{ligo20}
{The LIGO Scientific Collaboration}, {The Virgo Collaboration}, B.~P., {Abbott}, {Abbott}, R., {Abbott}, T.~D., {Abraham}, S., {Acernese}, F., et~al. (2020).
\newblock {GW190425: Observation of a Compact Binary Coalescence with Total Mass {\ensuremath{\sim}} 3.4 M$_{{\ensuremath{\odot}}}$}.
\newblock \emph{\apjl} 892, L3.
\newblock \doi{10.3847/2041-8213/ab75f5}
\bibAnnoteFile{ligo20}

\bibitem[{{Titarchuk}(1994)}]{Titarchuk1994}
{Titarchuk}, L. (1994).
\newblock {Generalized Comptonization Models and Application to the Recent High-Energy Observations}.
\newblock \emph{\apj} 434, 570.
\newblock \doi{10.1086/174760}
\bibAnnoteFile{Titarchuk1994}

\bibitem[{{Tolman}(1934)}]{Tolman1934}
{Tolman}, R.~C. (1934).
\newblock \emph{{Relativity, Thermodynamics, and Cosmology}} (Oxford: Clarendon Press)
\bibAnnoteFile{Tolman1934}

\bibitem[{{Tolman}(1939)}]{Tolman1939}
{Tolman}, R.~C. (1939).
\newblock {Static Solutions of Einstein's Field Equations for Spheres of Fluid}.
\newblock \emph{Physical Review} 55, 364--373.
\newblock \doi{10.1103/PhysRev.55.364}
\bibAnnoteFile{Tolman1939}

\bibitem[{{Tomar} et~al.(2021){Tomar}, {Pradhan}, and {Paul}}]{tomar2021}
{Tomar}, G., {Pradhan}, P., and {Paul}, B. (2021).
\newblock {New measurements of the cyclotron line energy in Cen X-3}.
\newblock \emph{\mnras} 500, 3454--3461.
\newblock \doi{10.1093/mnras/staa3477}
\bibAnnoteFile{tomar2021}

\bibitem[{{Tsygankov} et~al.(2019{\natexlab{a}}){Tsygankov}, {Doroshenko}, {Mushtukov}, {Suleimanov}, {Lutovinov}, and {Poutanen}}]{Tsygankov2019}
{Tsygankov}, S.~S., {Doroshenko}, V., {Mushtukov}, A.~A., {Suleimanov}, V.~F., {Lutovinov}, A.~A., and {Poutanen}, J. (2019{\natexlab{a}}).
\newblock {Cyclotron emission, absorption, and the two faces of X-ray pulsar A 0535+262}.
\newblock \emph{\mnras} 487, L30--L34.
\newblock \doi{10.1093/mnrasl/slz079}
\bibAnnoteFile{Tsygankov2019}

\bibitem[{{Tsygankov} et~al.(2022){Tsygankov}, {Doroshenko}, {Poutanen}, {Heyl}, {Mushtukov}, {Caiazzo} et~al.}]{Tsygankov2022}
{Tsygankov}, S.~S., {Doroshenko}, V., {Poutanen}, J., {Heyl}, J., {Mushtukov}, A.~A., {Caiazzo}, I., et~al. (2022).
\newblock {The X-Ray Polarimetry View of the Accreting Pulsar Cen X-3}.
\newblock \emph{\apjl} 941, L14.
\newblock \doi{10.3847/2041-8213/aca486}
\bibAnnoteFile{Tsygankov2022}

\bibitem[{{Tsygankov} et~al.(2019{\natexlab{b}}){Tsygankov}, {Rouco Escorial}, {Suleimanov}, {Mushtukov}, {Doroshenko}, {Lutovinov} et~al.}]{Tsygankov2019a}
{Tsygankov}, S.~S., {Rouco Escorial}, A., {Suleimanov}, V.~F., {Mushtukov}, A.~A., {Doroshenko}, V., {Lutovinov}, A.~A., et~al. (2019{\natexlab{b}}).
\newblock {Dramatic spectral transition of X-ray pulsar GX 304$-$1 in low luminous state}.
\newblock \emph{\mnras} 483, L144--L148.
\newblock \doi{10.1093/mnrasl/sly236}
\bibAnnoteFile{Tsygankov2019a}

\bibitem[{{Tugay} and {Vasylenko}(2009)}]{tugay2009}
{Tugay}, A.~V. and {Vasylenko}, A.~A. (2009).
\newblock {XMM-Newton Observations of X-ray Pulsar Cen X-3}.
\newblock In \emph{Young Scientists 16th Proceedings}, eds. V.~Y. {Choliy} and G.~{Ivashchenko}. 58--61
\bibAnnoteFile{tugay2009}

\bibitem[{{Ursini} et~al.(2023){Ursini}, {Farinelli}, {Gnarini}, {Poutanen}, {Bianchi}, {Capitanio} et~al.}]{Ursini23}
{Ursini}, F., {Farinelli}, R., {Gnarini}, A., {Poutanen}, J., {Bianchi}, S., {Capitanio}, F., et~al. (2023).
\newblock {X-ray polarimetry and spectroscopy of the neutron star low-mass X-ray binary GX 9+9: An in-depth study with IXPE and NuSTAR}.
\newblock \emph{\aap} 676, A20.
\newblock \doi{10.1051/0004-6361/202346541}
\bibAnnoteFile{Ursini23}

\bibitem[{{van der Klis}(2005)}]{vanderklis05}
{van der Klis}, M. (2005).
\newblock {Timing Neutron Stars}.
\newblock In \emph{The Electromagnetic Spectrum of Neutron Stars}. vol. 210 of \emph{NATO Advanced Study Institute (ASI) Series B}, 283
\bibAnnoteFile{vanderklis05}

\bibitem[{{Vasilopoulos} et~al.(2014){Vasilopoulos}, {Haberl}, {Sturm}, {Maggi}, and {Udalski}}]{Vasilopoulos2014}
{Vasilopoulos}, G., {Haberl}, F., {Sturm}, R., {Maggi}, P., and {Udalski}, A. (2014).
\newblock {Spectral and temporal properties of RX J0520.5-6932 (LXP 8.04) during a type-I outburst}.
\newblock \emph{\aap} 567, A129.
\newblock \doi{10.1051/0004-6361/201423934}
\bibAnnoteFile{Vasilopoulos2014}

\bibitem[{{Vasilopoulos} et~al.(2020){Vasilopoulos}, {Ray}, {Gendreau}, {Jenke}, {Jaisawal}, {Wilson-Hodge} et~al.}]{Vasilopoulos2020}
{Vasilopoulos}, G., {Ray}, P.~S., {Gendreau}, K.~C., {Jenke}, P.~A., {Jaisawal}, G.~K., {Wilson-Hodge}, C.~A., et~al. (2020).
\newblock {The 2019 super-Eddington outburst of RX J0209.6-7427: detection of pulsations and constraints on the magnetic field strength}.
\newblock \emph{\mnras} 494, 5350--5359.
\newblock \doi{10.1093/mnras/staa991}
\bibAnnoteFile{Vasilopoulos2020}

\bibitem[{{Vybornov} et~al.(2018){Vybornov}, {Doroshenko}, {Staubert}, and {Santangelo}}]{Vybornov2018}
{Vybornov}, V., {Doroshenko}, V., {Staubert}, R., and {Santangelo}, A. (2018).
\newblock {Changes in the cyclotron line energy on short and long timescales in V 0332+53}.
\newblock \emph{\aap} 610, A88.
\newblock \doi{10.1051/0004-6361/201731750}
\bibAnnoteFile{Vybornov2018}

\bibitem[{{Vybornov} et~al.(2017){Vybornov}, {Klochkov}, {Gornostaev}, {Postnov}, {Sokolova-Lapa}, {Staubert} et~al.}]{Vybornov2017}
{Vybornov}, V., {Klochkov}, D., {Gornostaev}, M., {Postnov}, K., {Sokolova-Lapa}, E., {Staubert}, R., et~al. (2017).
\newblock {Luminosity-dependent changes of the cyclotron line energy and spectral hardness in Cep X-4}.
\newblock \emph{A\&A} 601, A126
\bibAnnoteFile{Vybornov2017}

\bibitem[{{White} et~al.(1988){White}, {Stella}, and {Parmar}}]{white88}
{White}, N.~E., {Stella}, L., and {Parmar}, A.~N. (1988).
\newblock {The X-Ray Spectral Properties of Accretion Disks in X-Ray Binaries}.
\newblock \emph{\apj} 324, 363.
\newblock \doi{10.1086/165901}
\bibAnnoteFile{white88}

\bibitem[{{Wilms} et~al.(2000){Wilms}, {Allen}, and {McCray}}]{Wilms2000}
{Wilms}, J., {Allen}, A., and {McCray}, R. (2000).
\newblock {On the Absorption of X-Rays in the Interstellar Medium}.
\newblock \emph{\apj} 542, 914--924
\bibAnnoteFile{Wilms2000}

\bibitem[{{Wolff} et~al.(2019){Wolff}, {Becker}, {Coley}, {F\"urst}, {Guillot}, {Harding} et~al.}]{Wolff2019}
{Wolff}, M., {Becker}, P.~A., {Coley}, J., {F\"urst}, F., {Guillot}, S., {Harding}, A.~K., et~al. (2019).
\newblock {The Physics of Accretion Onto Highly Magnetized Neutron Stars}.
\newblock \emph{\baas} 51, 386.
\newblock \doi{10.48550/arXiv.1904.00108}
\bibAnnoteFile{Wolff2019}

\bibitem[{{Yang} et~al.(2023){Yang}, {Wang}, {Liu}, {Chen}, {Wu}, {Tian} et~al.}]{Yang2023}
{Yang}, W., {Wang}, W., {Liu}, Q., {Chen}, X., {Wu}, H.~J., {Tian}, P.~F., et~al. (2023).
\newblock {Discovery of two cyclotron resonance scattering features in X-ray pulsar Cen X-3 by Insight-HXMT}.
\newblock \emph{\mnras} 519, 5402--5409.
\newblock \doi{10.1093/mnras/stad048}
\bibAnnoteFile{Yang2023}

\end{thebibliography}

\end{document}